\documentclass{aa}
\usepackage{subfig}
\usepackage{textcomp} 
\usepackage{rotating} 
\usepackage[dvips]{color}  
\usepackage{graphics}
\usepackage{epsfig}
\usepackage{amsmath}
\usepackage{amssymb}
\usepackage{natbib}
\usepackage{threeparttable} 
\usepackage{longtable}
\usepackage[breaklinks]{hyperref}  
\usepackage{breakurl}
\usepackage[varg]{txfonts}

\def\kms{\mbox{${\rm km}\,{\rm s}^{-1}$}}

\begin{document}
\hyphenation{brems-strah-lung}
\title{The X-shooter Spectral  Library (XSL): I. DR1. Near-ultraviolet through optical spectra from the first year of
  the survey 
  \thanks{Table 3 and Table B.1 are also available in electronic form at the CDS.}}

\author{Yan-Ping Chen\inst{1, \thanks{yanping@astro.rug.nl,  chenyp.astro@gmail.com}},  
S. C. Trager\inst{1}, R. F. Peletier\inst{1}, A. Lan\c{c}on\inst{2}, \\
A. Vazdekis\inst{3, 4}, Ph. Prugniel\inst{5}, D.R. Silva\inst{6}, and A. Gonneau\inst{2} }

\institute
{Kapteyn Astronomical Institute, University of Groningen,
  P.O. BOX 800, 9700 AV Groningen, The Netherlands
\and
Observatoire Astronomique, 11, rue de l'Universit\'e, F-67000
  Strasbourg, France
\and
Instituto de Astrof\'\i sica de Canarias (IAC), E-38200 La Laguna, Tenerife, Spain
\and
Departamento de Astrof\'\i sica, Universidad de La Laguna, E-38205, Tenerife, Spain
\and
Universit\'e de Lyon, Universit\'e Lyon 1, 69622 Villeurbanne, France; CRAL, Observatoire de Lyon, CNRS UMR 5574,
             69561 Saint-Genis Laval, France
\and
National Optical Astronomy Observatory, 950 North Cherry Avenue, Tucson, AZ, 85719 USA
}         

\offprints{Yan-Ping Chen,
\email{chenyp.astro@gmail.com}}

\date{ }
\date{Received  / Accepted }

\titlerunning{XSL. I. First year optical--NUV spectra}

\authorrunning{Chen et al.}


\abstract
{
We present the first release of XSL, the X-shooter Spectral Library.
This release contains 237 stars.
The spectra in this release span a wavelength range of 3000--10200 \AA\
and have been observed at a resolving power of 
$R \equiv \lambda / \Delta\lambda \sim 10000$.
The spectra were
obtained at ESO's 8-m Very Large Telescope (VLT). The sample contains
O -- M, long-period variable (LPV), C and S stars.
The spectra are flux-calibrated and telluric-corrected.
We describe a new technique for the telluric correction. 
 The wavelength
coverage, spectral resolution, and spectral type of this library make it
well suited to stellar population synthesis of galaxies and clusters,
kinematical investigation of stellar systems, and the study of the physics
of cool stars. 
}

{} 
\keywords{stars: abundances -- stars: fundamental parameters -- stars: AGB and post-AGB 
        -- stars: atmospheres -- galaxies: stellar content}

\maketitle

%

\section{Introduction}\label{}
Spectral libraries play an important role in different fields of astrophysics. 
In particular they serve as a reference for the classification and automatic 
analysis of large stellar spectroscopic surveys and are fundamental ingredients 
for models of stellar populations used to study the evolution of galaxies.

Much of what we know about the formation, evolution, and current state
of galaxies comes
through studies of their starlight.  In distant galaxies,
where the ability to study their stellar populations star-by-star is
compromised by crowding and blending due to poor resolution, we must
resort to studying their integrated light.  This typically means
comparing colors or spectra to models of simple or composite stellar
population models \citep[e.g.,][]{IAPmodel,Buzzoni94, Worthey94,Leitherer96,   
Pegase,Starburst99,BC03, Pegase-HR,Granada05,Maraston05,Vazdekis10,
 Conroy12a}.  These comparisons
give insight into a galaxy's evolution process: stellar population
analysis can reveal the detailed chemical composition and star-formation 
history of a galaxy \citep[e.g.,][]{Gonzalez93, Davies93,
Trager00,Thomas05,Yamada06,Koleva13,Conroy13}.

Modern stellar population models consist of three primary ingredients
\citep[e.g.,][]{Vazdekis10}: 
stellar isochrones that represent the location in the luminosity--effective 
temperature plane (and, as a consequence, also surface gravity--effective 
temperature) of stars with different masses, the same initial chemical 
composition, and age;
an initial mass function that populates the isochrones with a
sensible number of stars; and a stellar library.  A spectral library
is a collection of stellar spectra that share similar wavelength coverage
and spectral resolution. The spectra change
as a function of effective temperature ($T _{\mathrm{eff}}$), surface gravity
($\log g$), and metallicity ([Fe/H]). To reproduce galaxy
spectra as precisely as possible, one requires a comprehensive stellar
spectral library that covers the entire desired parameter space of $T
_{\mathrm{eff}}$, $\log g$, and [Fe/H].  Moreover, extended wavelength
coverage is strongly desirable, because different stellar phases
contribute their light in different bands. For instance, cool giants
contribute more light than warmer faint giant stars in the
near-infrared, while the situation is reversed  in the optical
\citep{Frogel88, Maraston05}.  Asymptotic giant branch (AGB) stars
dominate the light of intermediate-aged stellar populations in the
near-infrared but are unimportant in the optical \citep{CB91,
  Worthey94, Maraston98}.  Detecting their presence requires
\emph{broad wavelength coverage} in both the target and model spectra.

Stellar spectral libraries can be classified into empirical and
theoretical libraries, depending on how libraries are obtained. Both
theoretical and empirical libraries have improved in recent years. 
Widely-used theoretical libraries in stellar population models
include those of \citet{Kurucz93,Coelho05,Martins05,Munari05, Coelho07,Gustafsson08}, 
and \citet{Allard11}. Theoretical libraries have the advantage of
(nearly) unlimited resolution and selectable abundance patterns, which include not
only scaled-solar abundances but also non-solar patterns.
Unfortunately, theoretical libraries suffer from systematic
uncertainties, as they rely on model atmospheres and require a
reliable list of atomic and molecular line wavelengths and opacities
\citep{Coelho05}.  Empirical stellar libraries, on the other hand,
have the advantage of being drawn from real, observed stars and
therefore do not suffer this limitation. However, they frequently have
relatively low resolution (with a few exceptions; see below) and are typically
unable to reproduce the indices measured in giant elliptical galaxies,
because they are based on local stars with Milky Way disk abundance patterns
\citep{Reynier89, Worthey92}.
Table~\ref{tablib}
lists several previous empirical stellar libraries and their principal
features. In the optical, there are the (among others) Lick/IDS
\citep{Worthey97}, MILES \citep{miles06}, ELODIE
\citep{Prugniel01,Prugniel04,Prugniel07}, STELIB \citep{LeBorgne03},
NGSL \citep{Gregg06}, and the \citet{Pickles98} libraries.  Building stellar libraries
in the near-IR is a challenging task, but pioneering work has been
done by \citet{Lancon92,LW2000} and \citet{LM2002} (LW2000, LM2002);
\citet{MQ08}; and \citet[IRTF-SpeX]{Rayner09}. However, spectral libraries with
extended wavelength coverage at moderate resolution are still largely
missing.

\begin{table*}
\caption{A selection of previous empirical stellar libraries.}\label{tablib}
\begin{center}
\begin{tabular}{lrrrl}\hline
Library&Resolution&Spectral range&Number &Reference\\
&\multicolumn{1}{c}{R=$\lambda/\Delta\lambda$}& \multicolumn{1}{c}{(nm)}& \multicolumn{1}{c}{of stars}&\\\hline 
STELIB    & 2000  & 320--930 & 249 & \citet{LeBorgne03}\\
ELODIE    & 10000 & 390--680 & 1388 & \citet{Prugniel01,Prugniel04}\\
          &       &         &      & \citet{Prugniel07}\\
INDO-US   & 5000  & 346--946 & 1237& \citet{Valdes04}\\
MILES     & 2000  & 352--750 & 985 & \citet{miles06}\\
IRTF-SpeX & 2000  & 800--5000& 210 & \citet{Rayner09}\\
NGSL      & 1000  & 167--1025& 374 & \citet{Gregg06}\\
UVES-POP  & 80000 & 307--1030& 300 & \citet{Bagnulo03}\\
LW2000    & 1100  & 500--2500& 100 & \citet{LW2000}\\\hline
\end{tabular}
\end{center}
\end{table*}

With the  high-efficiency, broad-wavelength coverage
and high/moderate resolution of the X-shooter spectrograph at ESO's VLT
\citep{XShooter}, we can now fill the gap between high-resolution 
theoretical stellar libraries and low-resolution 
empirical stellar libraries. To this end, we developed
the X-shooter Spectral Library (XSL, PI: Trager), which is a survey of
$\gtrsim700$ stars that cover the entire Hertzsprung--Russell (HR) diagram,
including both cool (M dwarfs, M giants, C stars, long-period
variables, etc.) and hot stars (up to late O stars) with wavelength
coverage from $300$--$2480\, \mathrm{nm}$. This includes the entire
near-ultraviolet (NUV) to near-infrared (NIR) range at $R\sim8000$--11000.  

Here, we present the first two of six periods of XSL data (from ESO Periods
84 and 85).  We concentrate on the NUV--optical data (3000--10200 \AA) from 
the UVB and VIS arms of X-shooter in this paper and leave the
NIR arm data for a forthcoming paper.

\section{Sample selection and observation}

The targets of XSL are selected from several empirical stellar libraries 
and supplementary literature sources. We take stars from Lick/IDS,
MILES, and NGSL to cover $T _{\mathrm{eff}}$, $\log g$, and [Fe/H] as
uniformly as possible. However, these libraries mostly lack the cool,
bright stars, which are important in the near-infrared.  For this
purpose, we select AGB and long-period variable (LPV) stars from LW2000
and IRTF-SpeX with a declination $<35\,^{\circ}$ marked M, C or
S-stars. The Long-period variable stars are also collected from \citet[][for the Large Magellanic Cloud]{Hughes90}
and the \citet[][for the Small Magellanic Cloud]{Cioni03}. Red supergiant stars are taken from the
lists of LW2000 and \citet{Levesque05,Levesque07}. To cover metal-rich
stars with abundances similar to giant elliptical galaxies, we also
include Galactic Bulge giants from the samples of \cite{Blanco84} and
\cite{Groenewegen05}. 

\begin{figure}
   \includegraphics[angle=90,scale=0.35]
   {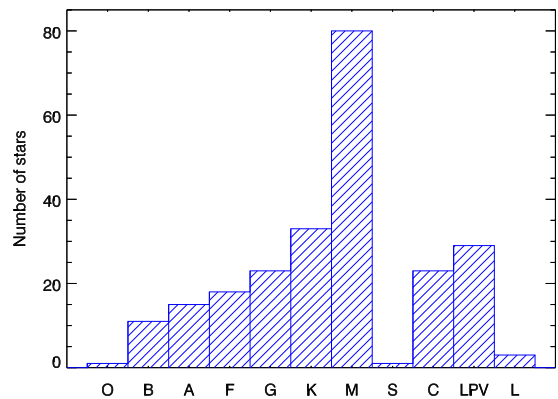}
   \caption{Distribution of spectral types in XSL observed in Periods
     84 and 85 (excluding telluric calibrators). Spectral types were
     retrieved from SIMBAD or based on educated guesses from the
     source libraries or atmospheric parameters.}
   \label{sp-type-hist}
\end{figure}    

\begin{figure}
   \includegraphics[angle=90,scale=0.37]
   {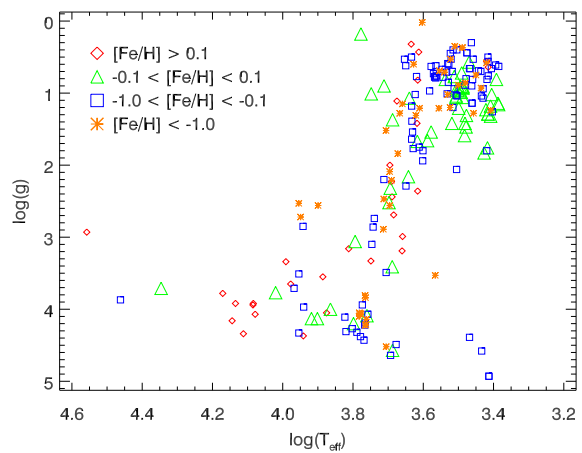}
   \caption{
     HR diagram of the 219 XSL stars (O -- M, LPV, S) with calculated
     $T_{\mathrm{eff}}$, $\log g$ and [Fe/H], where [Fe/H] is
     presented in different colors.}
   \label{sp-param}
\end{figure}    

As of the end of March 2011 (i.e., the completion of the Period 85 observing campaign),
 258 observations of 237 individual stars from
the XSL input catalog had been completed. Figure \ref{sp-type-hist}
shows the distribution of stellar types of these XSL stars.  In Figure
\ref{sp-param} we show those sample stars with calculated stellar
parameters in an HR diagram; these parameters were computed using
ULySS \citep{Koleva09,Wu11} and will be described in more detail in a
forthcoming paper.

\subsection{Observations}
The observations described here were carried out in Periods 84 and 85
using X-shooter.
The X-shooter spectrograph was built by a consortium of 11 institutes in Denmark,
France, Italy and the Netherlands, and ESO.  It is a
single-target echelle spectrograph, which maximizes the sensitivity
over a broad wavelength by splitting the spectra into three different
arms (UVB, VIS, and NIR). 
Depending on wavelength and slit width, X-shooter yields a resolving power of $R = 4000$--14000. 
A unique
capability of X-shooter is that it collects spectra in the wavelength
range from the near-ultraviolet to the near-infrared through its three
arms simultaneously.  This property is extremely useful for observing
variable stars, especially very cool stars, like AGB stars, whose
spectra may vary substantially during their pulsation cycles.

\begin{table*}
\centering
  \caption{X-shooter observing modes for XSL in Periods 84 and
    85.}\label{tabobs}
  \begin{tabular}{llrrrll}
    \hline
    Mode&Arm&Slit&$\lambda$ (nm)&$R$&Readout (bright)&Readout
    (faint)\\ \hline
    Nod&UVB&$0\farcs5\times11\arcsec$&300--600&9100&400k/1pt/hg&100k/1pt/hg\\
    Nod&VIS&$0\farcs7\times11\arcsec$&600--1020&11000&400k/1pt/hg&100k/1pt/hg/2$\times$2\\
    Nod&NIR&$0\farcs6\times11\arcsec$&1000-2480&8100&&\\
    Stare (P84)&UVB&$5\arcsec\times11\arcsec$&300--600& &100k/1pt/hg/2$\times$2&100k/1pt/hg/2$\times$2\\
    Stare (P85)&UVB&&&&400k/1pt/hg&100k/1pt/hg\\
    Stare (P84)&VIS&$5\arcsec\times11\arcsec$&600--1020& &100k/1pt/hg/2$\times$2&100k/1pt/hg/2$\times$2\\
    Stare (P85)&VIS&&&&400k/1pt/hg&100k/1pt/hg\\
    Stare&NIR&$5\arcsec\times11\arcsec$&1000-2480& &&\\
    \hline
  \end{tabular}
  
\end{table*}

X-shooter offers multiple spectroscopic observation modes; we used the
longslit SLIT mode for all observations. Three observing strategies
are supported in SLIT mode: classical ``staring'' observations, A--B
``nodding'' along the slit for improved sky subtraction, and on--off
target--sky switching (``offset''). Almost all XSL stars were observed
in nodding mode with a narrow slit, yielding an intermediate resolving
power of $\sim10000$ and good sky subtraction; nearly all were
also observed in staring mode with a wide slit for flux
calibration. The observed modes used, slit widths, and CCD readout
modes are given in Table~\ref{tabobs}.  Wide-slit observations were
carried out (except on the brightest stars, for which no wide-slit
observations were made, since they would saturate the detectors) just
before nodding observations. The CCD binning and readout speeds were
altered depending on the brightness of a given target: faint stars ($K
\ge 8$ or $V \ge 10$) were generally binned in the VIS arm 
and unbinned in the UVB arm with slow readout speeds in
nodding mode. Fast readout speeds were used for
brighter stars (Table~\ref{tabobs}\footnote{See \url{http://www.hq.eso.org/sci
/facilities/paranal/instruments/xshooter/doc/VLT-MAN-ESO-14650-4942_P91P92.pdf}
for more information}).

Along with the program stars, we observed hot stars (mostly B-type) to be 
used as telluric standard stars. These were observed using the same narrow-slit setup
to preserve the spectral resolution. Flux standard stars taken as part of
the normal ESO X-shooter calibration program were collected from the
archive; these were observed using a wide slit
($5.0^{\prime\prime}\times11^{\prime\prime}$) to sample as much of the
total flux as possible.

\section{Data reduction}

Basic data reduction was performed with the public release of the
X-shooter pipeline version 1.5.0, following the standard steps
described in the X-shooter pipeline manual\footnote{See
  \url{http://www.eso.org/sci/software/pipelines/}\label{manual}} up
to the production of 2D spectra. This includes bias and/or dark
correction, flat-fielding, geometric correction, wavelength
calibration and sometimes sky subtraction. A master bias constructed
from a set of at least five exposures was used to remove the bias level of the
CCD and to correct for large-scale electronic noise patterns. 
A master flat field constructed from a set of at least five exposures
was used to correct the pixel-to-pixel variations in the CCD 
response and the non-homogeneous illumination by the telescope.
Wavelength and spatial scale calibration are performed with 
observations of a Th-Ar lamp in the UVB and VIS arm.  


\subsection{Difficulties encountered and proposed solutions}\label{extra:calib}

A number of issues were discovered during the pipeline
reduction process, which required further processing steps.  We give details about
the most significant of these steps here.
 
\subsubsection{Issues with nodding mode observations}

\begin{figure}
   \includegraphics[angle=0,scale=0.35]
   {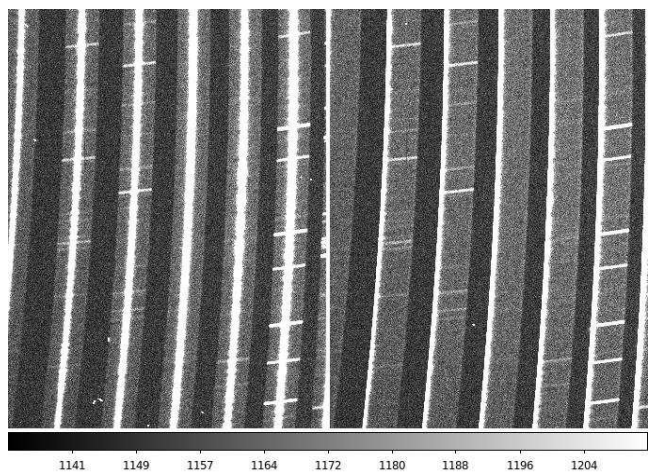}
   \caption{Data observed in nodding mode in the VIS arm.  Here, the
     star is sitting in the center of the slit in the first exposure
     (left panel), while the star is almost out of the slit in the
     second exposure (right panel).}
   \label{off_frame}
\end{figure}    

Most spectra were reduced using the pipeline recipes that correspond to
the observation mode used: spectra observed in staring mode were
reduced by the pipeline recipe ``xsh\_scired\_slit\_stare", and
spectra observed in nodding mode were reduced by the pipeline recipe
``xsh\_scired\_slit\_nod". However, in the case of nodding-mode
observations, the first exposure in an A--B pair was occasionally
centered in the slit, and the ``throw'' to the B image (which is fixed
angular difference plus a random extra ``jitter'') was large enough to
put the star at the end or even off of the slit in the B image. We
show an example in Figure~\ref{off_frame}. In these cases, we reduce
the well-centered slit in ``offset'' mode using the second frame as
the ``sky'' frame.  This does an excellent job of sky subtraction for
these frames, but it means the effective exposure time for these stars
is half that anticipated.

\subsubsection{Background subtraction issues in UVB spectra}

\begin{figure}
   \includegraphics[angle=0,scale=0.43]
   {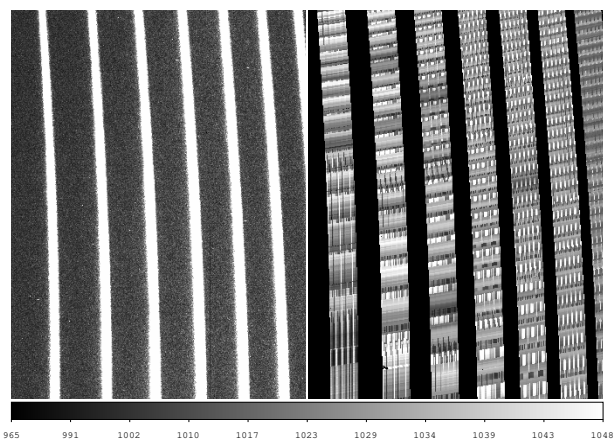}
   \caption{Example of background (``sky'') modeled by the pipeline
     from data observed in stare mode in the UVB arm, in which sky
     lines are nearly absent. Left panel: Original raw frame. Right
     panel: Sky frame generated by the pipeline scaled to the same
     intensity as the left panel.  Note the very poor background model.}
   \label{bd-sky-pip}
\end{figure}    

When the sky lines are very weak in stare mode, which is often the case
in short-exposure frames, such as in the UVB arm, the background
model constructed by the pipeline can fail catastrophically,
particularly if weak pattern noise is not properly subtracted by the
master bias. 
This arises from electronic noise from the CCD readout amplifiers and 
varies in position (and amplitude) on each frame in an unpredictable manner.
The derived master bias frame contains an average of the electronic noise 
features of each input bias frame, and neither the individual bias frames nor 
the average master bias frame had the same noise pattern as the science (star) frame.  
When the pipeline attempts to remove the background (``sky'') in the science (star) frame, 
it generates features in the ``sky'' that attempt to replicate the noise pattern of the 
image after corretion by the corrupted master bias frame.
Figure \ref{bd-sky-pip} illustrates an extreme case of
this: the left-hand panel is the observed star, and the right-hand
panel is the background modeled by the pipeline, where the modeling
has clearly failed. In such cases, we turn off sky subtraction in the
pipeline. 
Instead, we estimate the sky using pixels on either side of the stellar 
spectrum in the rectified 2D spectra of each spectrograph order, 
as would be done for observations in stare mode.

\begin{figure*}
  \centering
  \hfill
  \begin{minipage}[c]{.45\textwidth}
     \subfloat{\includegraphics[angle=90,scale=0.35]{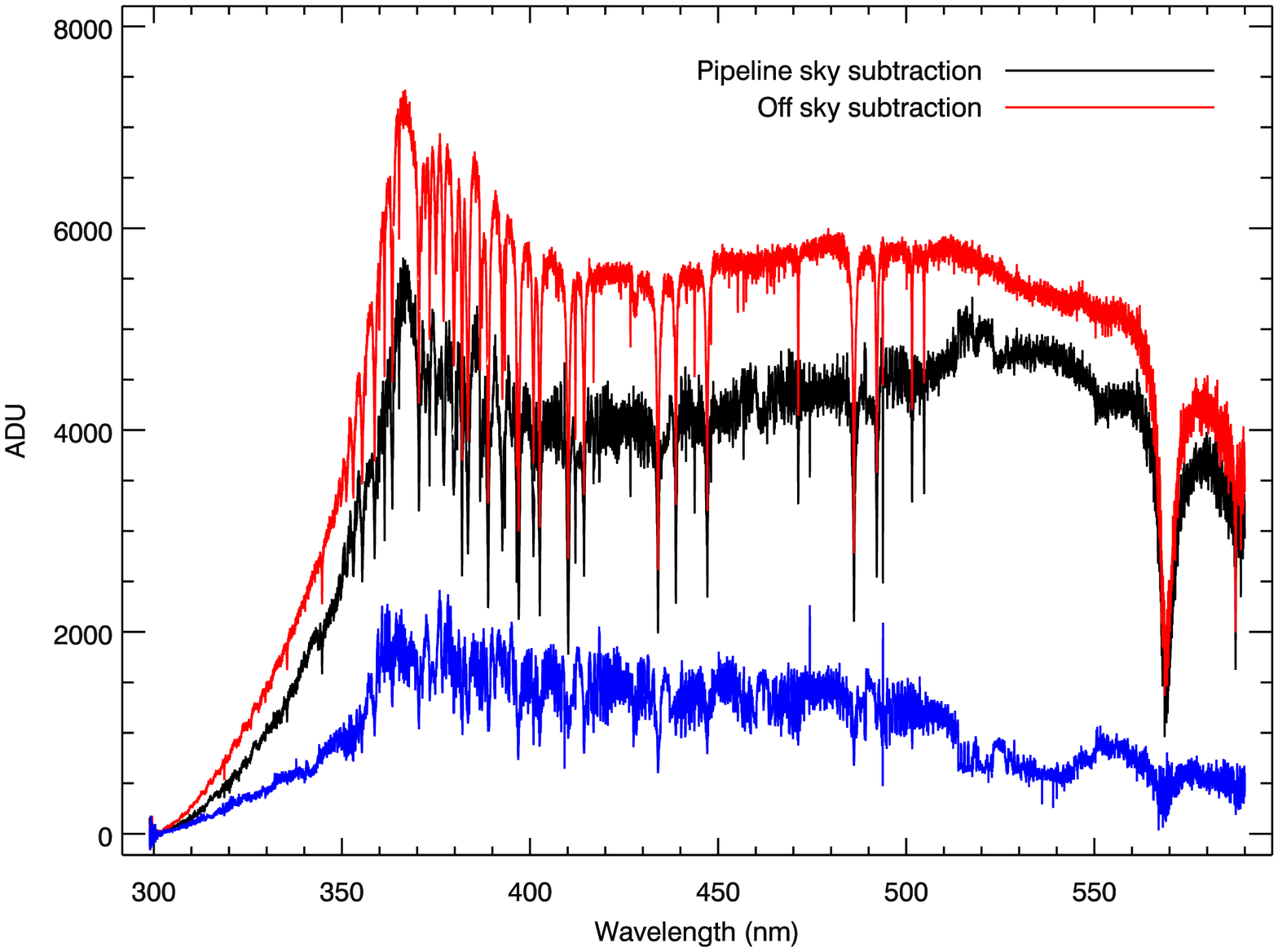}}\quad 
  \end{minipage}
  \hfill
  \begin{minipage}[c]{.45\textwidth}
    \subfloat{\includegraphics[angle=90,scale=0.33]{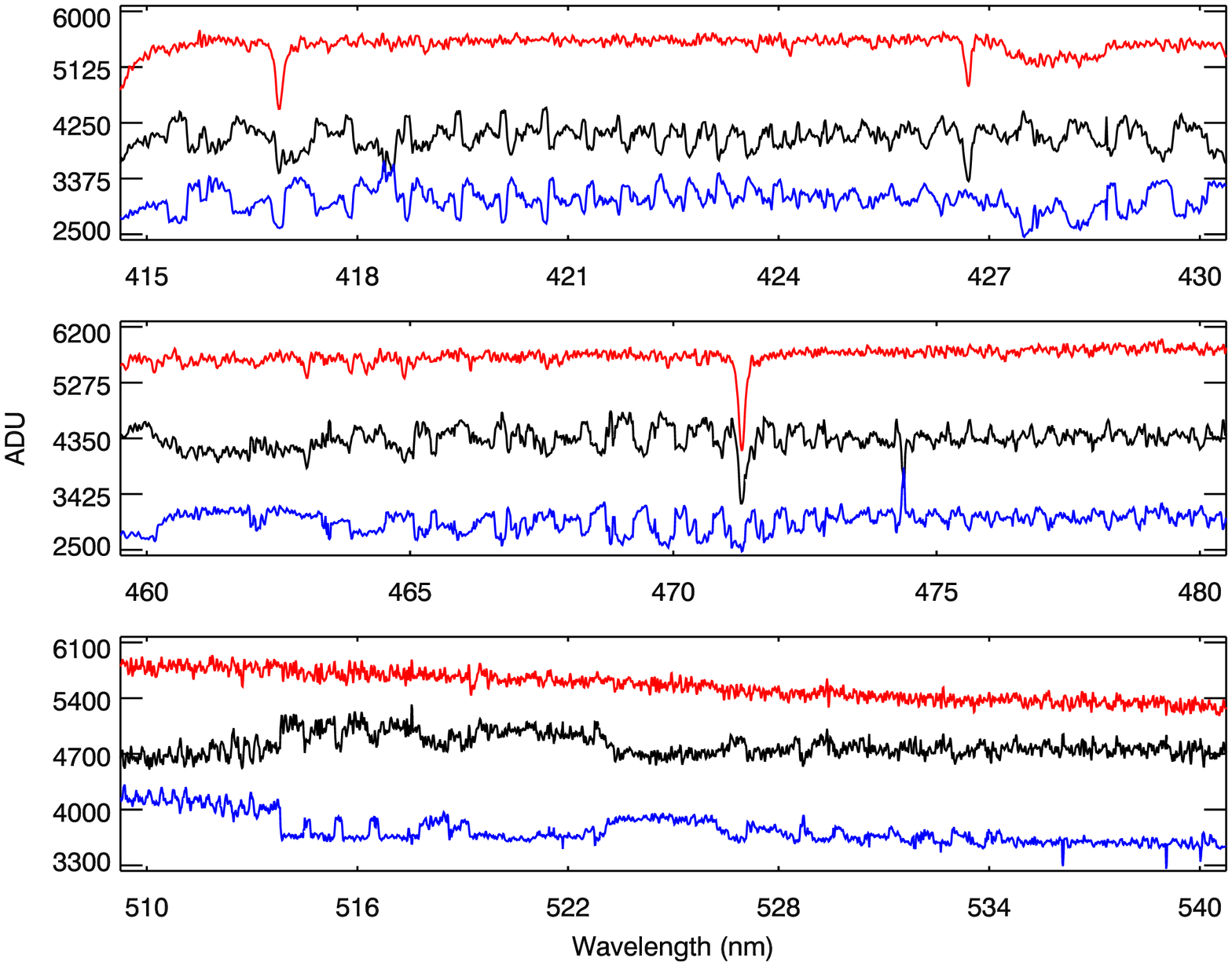}}
  \end{minipage}
   \caption{Extracted 1D spectra from the UVB arm, using the pipeline
     sky model (black) and a background extracted directly from
     the observation (red), when the sky emission lines are weak (or
     not observed).  The blue spectrum shows the difference between the two
     methods.  In the right panels, we zoom into three small
     wavelength regions, so that the sky modeling errors of the
     pipeline are clear.}
   \label{bd-sky-1d}
\end{figure*}

The final sky-subtracted 1D spectrum is derived by subtracting the 1D
sky spectrum from the 1D spectrum.  In Figure \ref{bd-sky-1d}, we
compare the two methods, showing that the in-place background
estimation gives a clear improvement.

\subsubsection{Extra cosmic ray cleaning}

\begin{figure*}
  \centering
  \begin{minipage}[c]{.45\textwidth}
  	\centering
         \subfloat[]{\label{bd-crh-1d}\includegraphics[angle=90,scale=0.35]{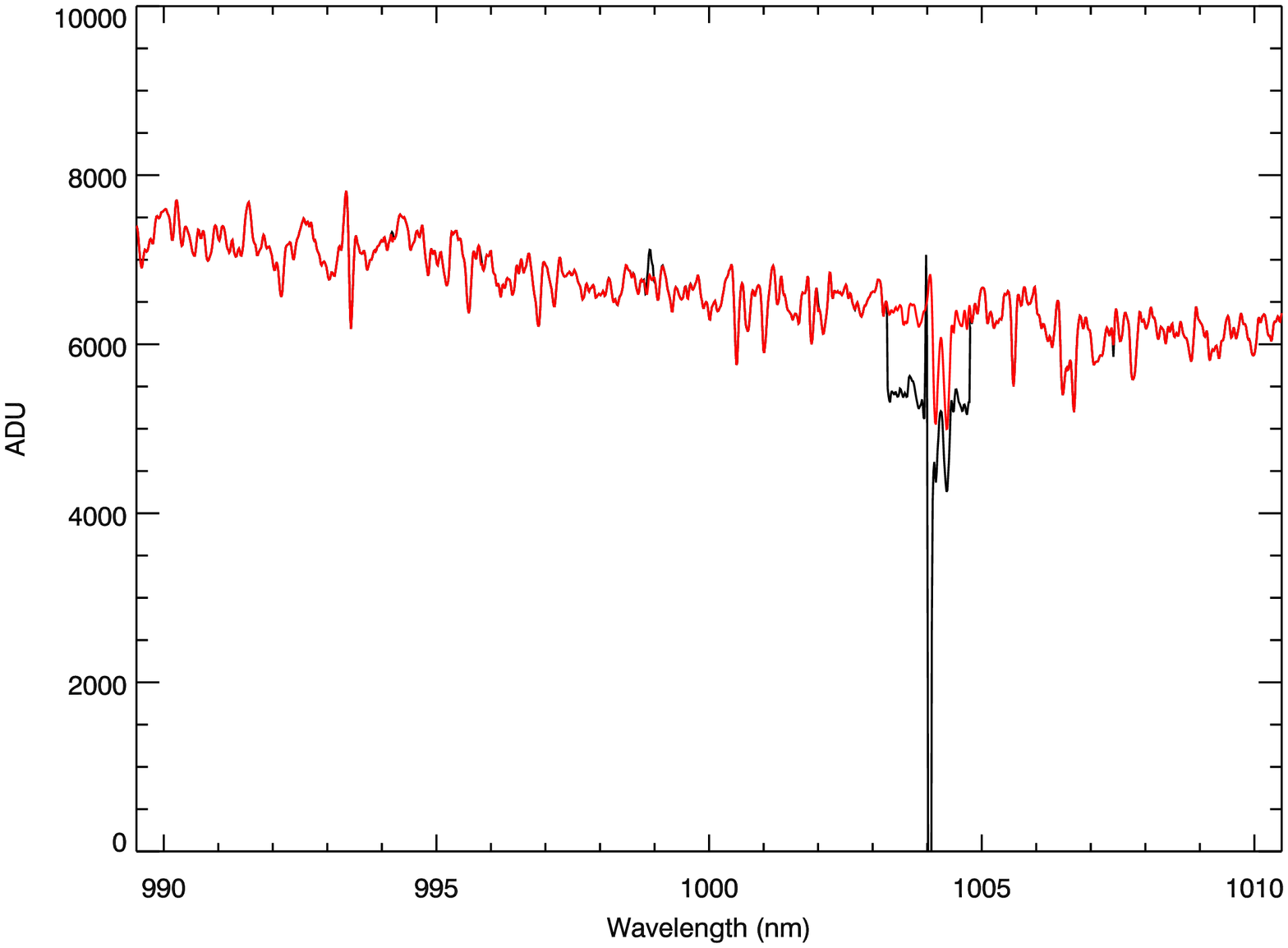}}
	 
  \end{minipage}
  \begin{minipage}[c]{.45\textwidth}
	\centering    	  
         \subfloat[]{\label{bd-crh-raw}\includegraphics[angle=0,scale=0.28]{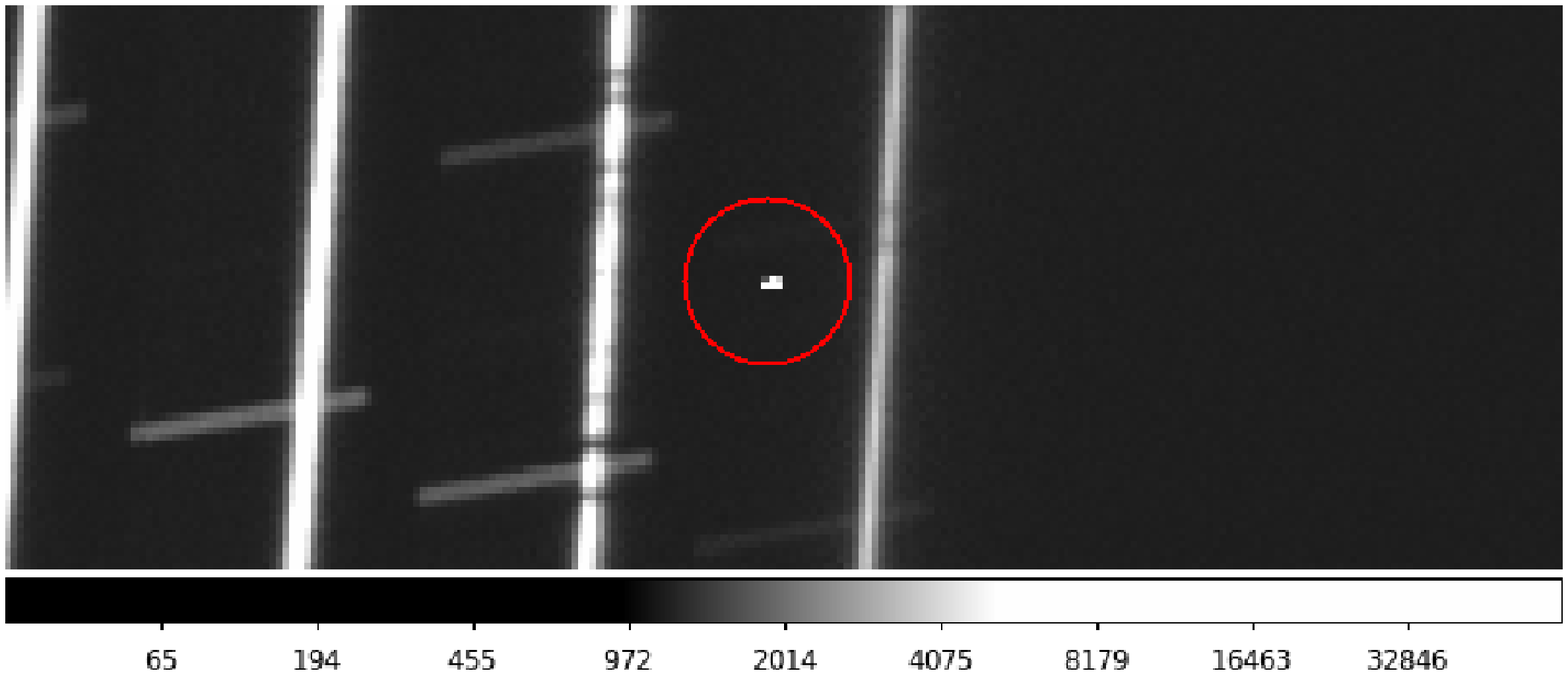}}
	 
         \subfloat[]{\label{bd-crh-2d}\includegraphics[angle=0,scale=0.28]{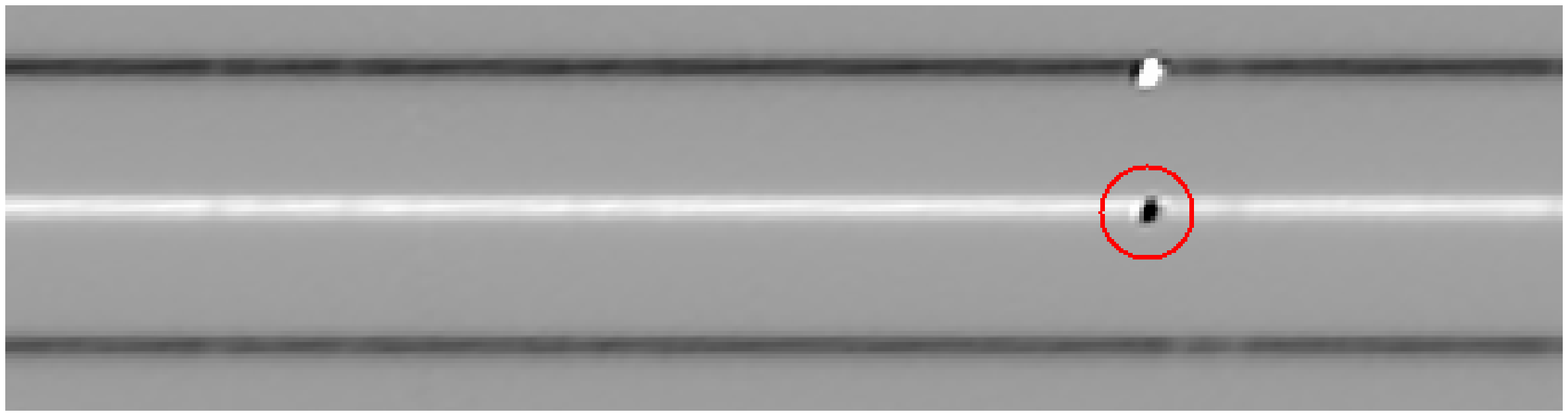}}
	 
  \end{minipage}
  \caption{Panel (a): Black: extracted 1D spectrum from the pipeline;
     red: extracted 1D spectra after raw frames corrected
    using the algorithm of \protect \cite{vanDokkum01}.
    Panel (b): Raw image of ``B" frame zoomed in on the CRH feature
    (red circle).  Panel(c): Pipeline-corrected 2D spectrum zoomed in
    on the CRH feature (red circle).}
  \label{bd-crh}
\end{figure*}

The X-shooter pipeline (ver. 1.5.0) removes the cosmic ray hits (CRHs)
for multiple input images ($n \ge 3$) in nodding and offset
modes by computing a median of these images and applying a
sigma-clipping. If only two raw images are used as input, as is the case for
most of our nodding and offset mode exposures, proper CRH cleaning is
not performed by the pipeline. This becomes an issue when a CRH in
image ``A" is sitting on the same position as the star in image
``B''; this results in an artificial emission or absorption line in the
final 1D spectrum.  We present the case of exposures of the star
[M2002] SMC 55188 in the VIS arm in Figure~\ref{bd-crh-1d} to
illustrate the problem.  We see an artificial feature around
$1004\,\mathrm{nm}$ caused by the CRH in the original ``B" frame,
(Figure \ref{bd-crh-raw}), which creates an artificial feature in the
2D image. To correct this problem, we use the algorithm of \citet[as
  implemented in the IDL code ``\emph{la\_cosmic.pro}"]{vanDokkum01}
to clean the raw image before running the pipeline. The CRH-corrected
images are then run through the pipeline before extracting the 1D
spectrum (Figure~\ref{bd-crh-1d}). The CRH pre-cleaning is used
whenever strong CRHs are found to corrupt the 1D spectra.

\subsubsection{Bad columns}

\begin{figure}
  \subfloat[]{\label{bd-col-raw}\includegraphics[angle=0,scale=0.32]{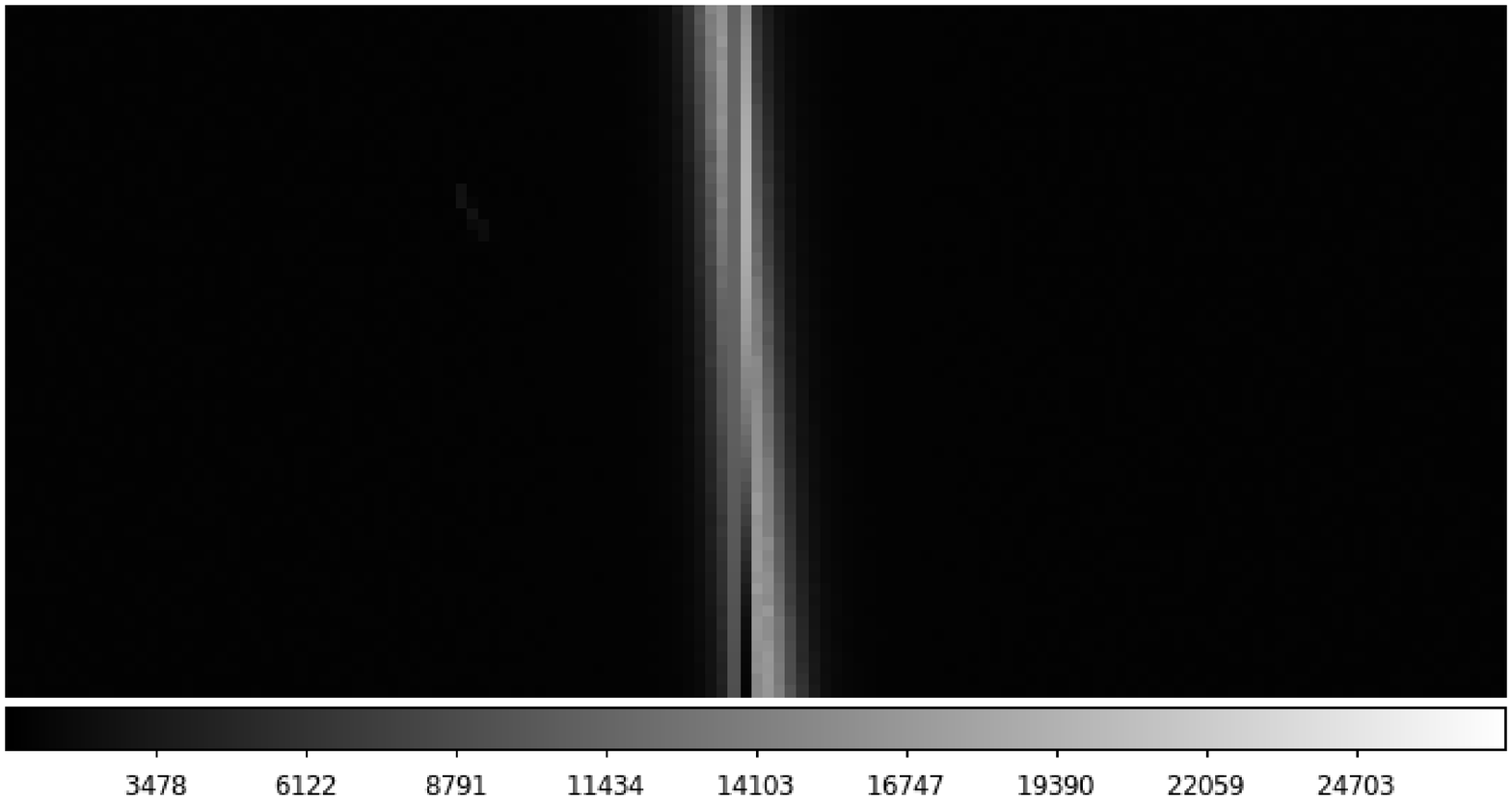}}\\
  \subfloat[]{\label{bd-col-2d}\includegraphics[angle=0,scale=0.32]{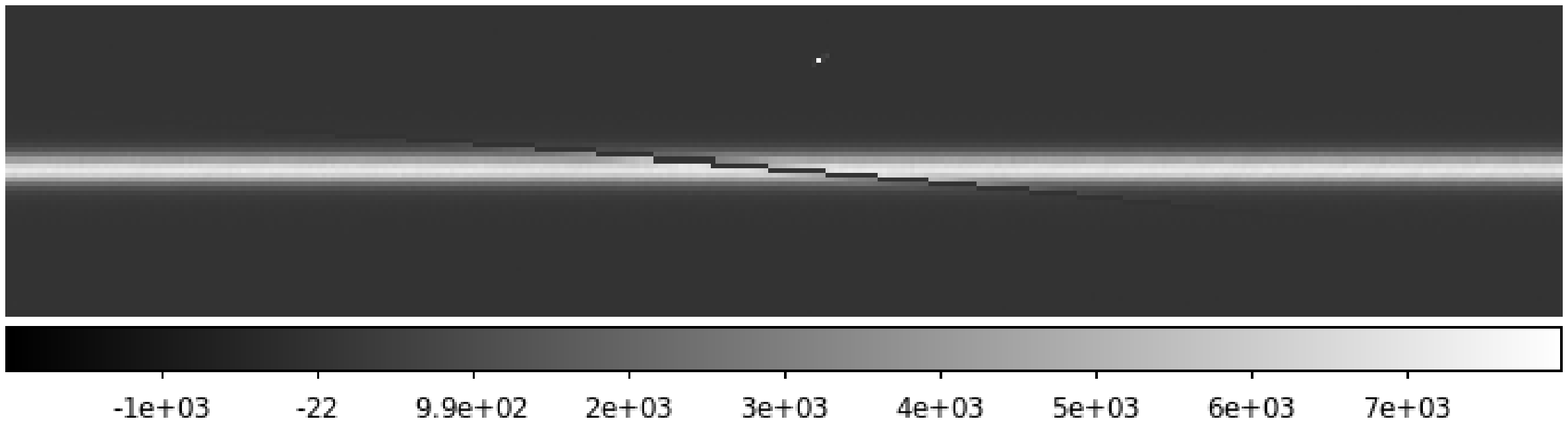}}
   \caption{Upper panel: Raw image of the telluric standard star
     Hip024505 zoomed into a region of bad columns in the VIS arm.
     Lower panel: Corresponding final, pipeline-corrected 2D spectrum of this
     star.}
 \label{bd-col}   
\end{figure}

We have found that the bad pixel maps produced by and used in the
pipeline reduction recipes do not always 
produce a complete list of bad pixels.
In particular, a few bad
columns in the VIS arm that are improperly mapped in the pipeline disrupt the
object spectra in order 26 at 635--638 nm.  Figure~\ref{bd-col} shows
these bad columns in the raw frame and the final 2D spectrum.  These
bad columns are difficult to correct effectively in the raw spectra
and make it difficult to reconstruct the profile of the star
in this region.  Moreover, due to instrumental flexure, these bad
columns can affect different exposures at slightly different
wavelengths.  
In the current version of the spectral library, 
we have set the fluxes to zero from
$635.1\, \mathrm{nm}$ to the end of order 26 in the final reduced 2D
order-by-order frames.




\subsubsection{Extraction of 1D spectra from pipeline-corrected 2D images}

After the pipeline reduction and our modifications, we extract
a one-dimensional spectrum from the pipeline-corrected, 
flat-fielded wavelength and geometrically-corrected, 
single-order 2D spectra using
our own weighted-extraction code in IDL, which is inspired by
the prescription of \cite{Horne86}. 
There are three extensions in total for each order in the
pipeline-produced 2D spectra: the first is the flux in counts, the
second is the error, and the third is the quality, which corresponds
to the bad pixel mask. We make use of the third extension as the first
guess for a bad pixel mask, and the second extension as the square
root of each pixel's variance. With 2 to 20 iterations, the bad pixel
mask is improved, and most of the CRHs or bad pixels are
masked. Spectra from each order are then extracted and merged using a
variance-weighted mean of each wavelength in the overlapping regions.

The extraction aperture was set to a fixed width of $4\farcs9$ (in
narrow-slit observations) or $10\farcs9$ (in wide-slit observations)
in the optimal-extraction code.  However, if significant CRHs or other
problems remain within the nominal extraction aperture of the final 2D
frame, the extraction aperture was modified to exclude these regions.

\begin{figure*}
   \centering
   \includegraphics[angle=90,scale=0.9]
   {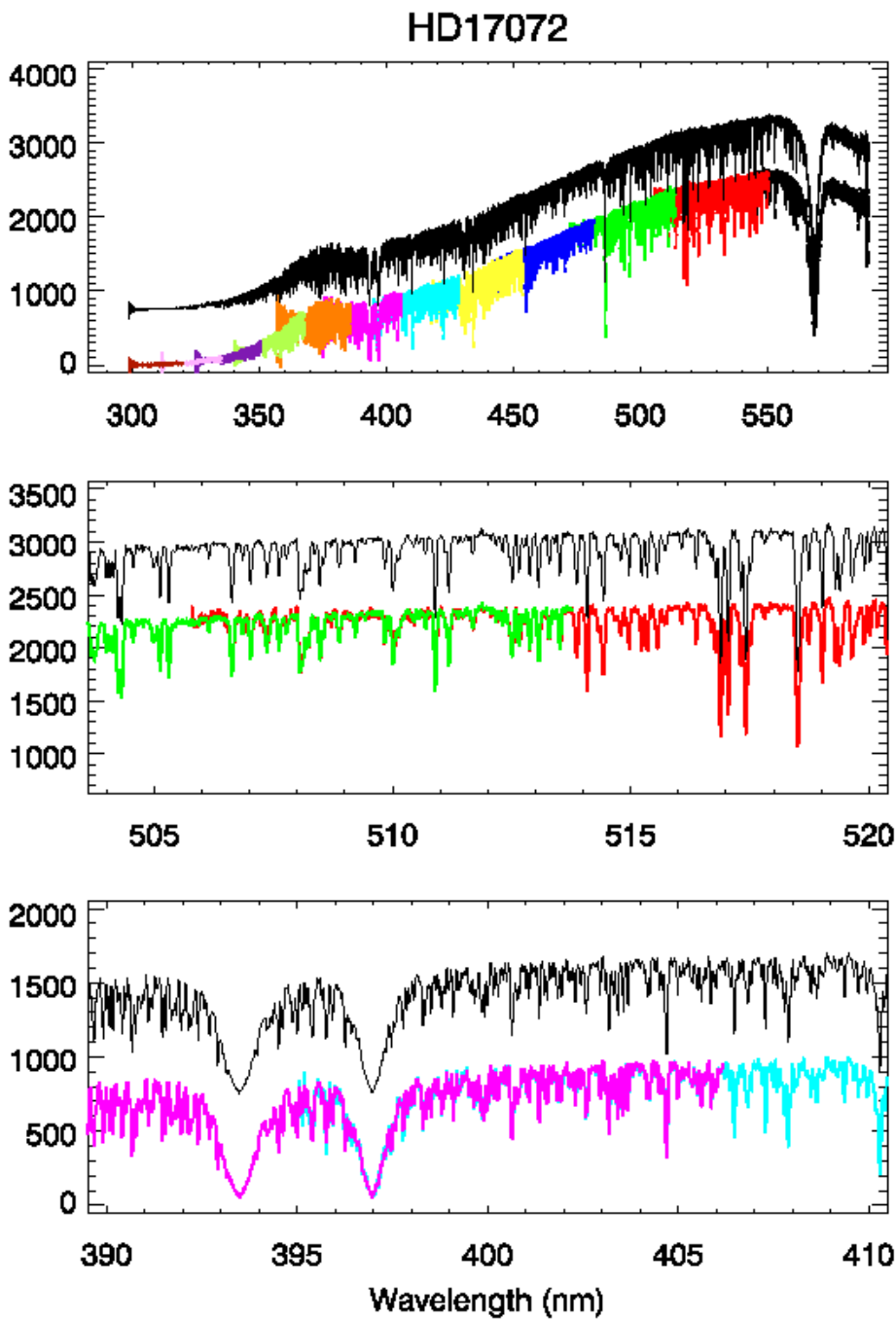}
   \caption{Order-by-order 1D spectrum extraction for the G2 star
     HD17072 in the UVB arm without flux calibration. Different
     colors correspond to different orders.  The continuous black
     spectrum is the combined spectrum, which is offset for clarity.  In the
     middle and bottom panels, we zoom in on two wavelength regions,
     504--520 nm and 390--410 nm, where orders overlap to see the
     extraction and combination process in detail.}
   \label{extra-order-1d}
\end{figure*}    

We show a G2 star, HD17072, observed in the UVB arm in
Figure~\ref{extra-order-1d} to illustrate this process.  We note that
the orange order between 355 and 370 $\mathrm{nm}$ shows noisy
features. However, the larger errors in this region yield lower
weights compared with the green order in the overlap region.  These
features therefore do not appear in the final merged spectrum (thin
black line).


\subsection{Saturation}

\begin{figure}
   \includegraphics[angle=90,scale=0.35]
   {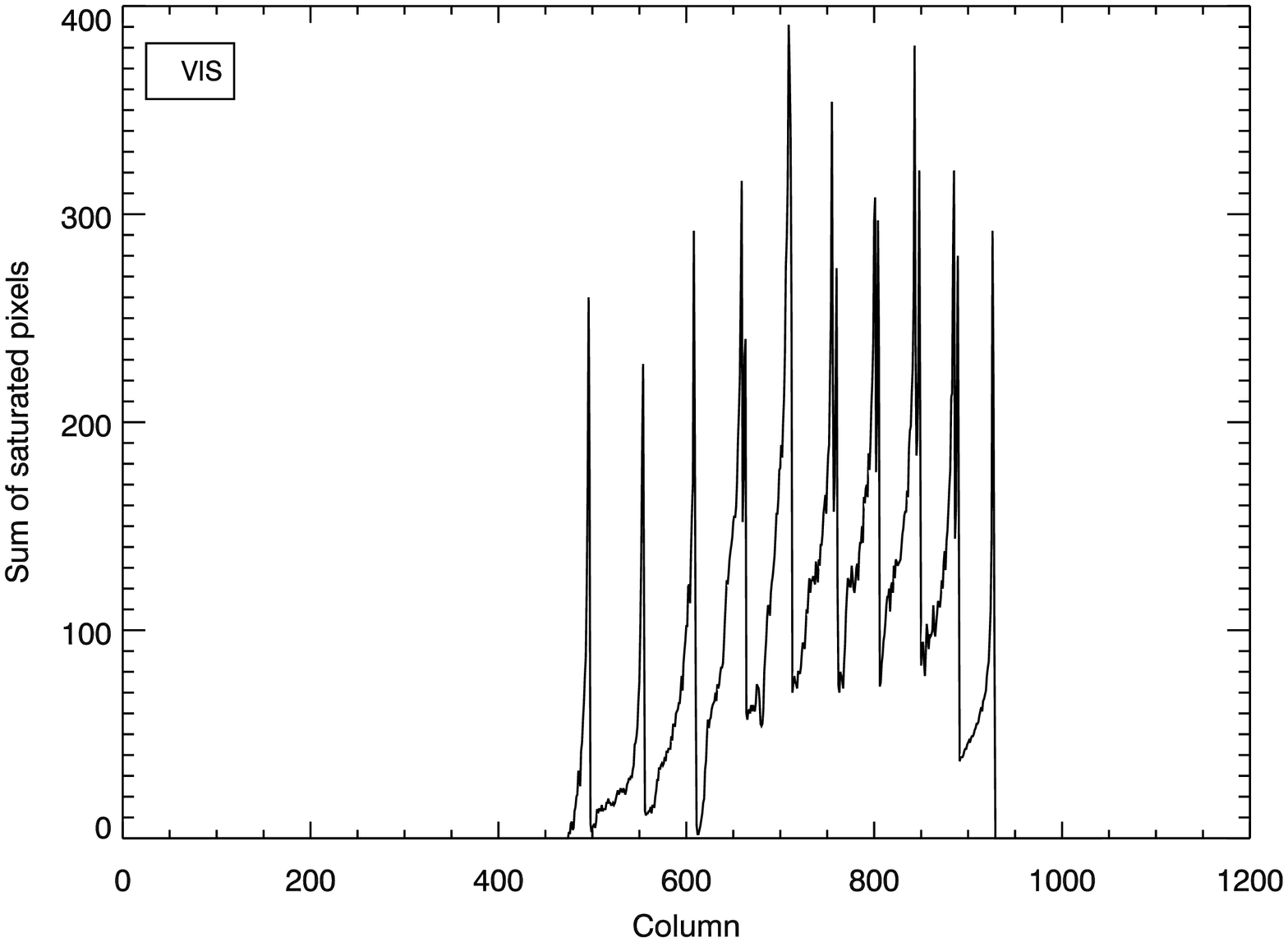}
   \caption{A saturated raw VIS arm frame after transformation of
     counts to a binary ``good pixel''/``saturated pixel'' basis, and
     summed along the wavelength direction. Note the sharp features
     with values in excess of 200, indicative of saturation due to
     overexposure.  See text for details.}
   \label{saturate}
\end{figure}    

In these first two periods, saturation was a common problem, as our
sample contained variable stars near X-shooter's bright target limit
(and our exposure times were, in the absence of accurate light curves,
determined for the \emph{mean} stellar brightnesses).  Given the large
total number of frames in these periods (1847 for the 258
observations, including NIR arm spectra), an automated
saturation-detection scheme was required.  Uncorrected CRHs
and bad pixels appear as nearly saturated pixels, 
so a scheme had to be developed to decide which
frames were strongly saturated and had to be discarded.
Every raw frame was transformed using
the following rules.  In the UVB and VIS arm images, pixels with
$\mathrm{counts}\geq65000$ are set to ``1", which means ``saturated'';
all others are set to ``0", which means ``good''\footnote{In the NIR
  arm, due to the different readout scheme of these detectors, pixels
  with $\mathrm{counts}\leq0$ or $\geq42000$ are set to ``1''.}.  Truly
saturated frames typically have saturated pixels clustered together in
individual orders.  We sum each transformed binary
``good''/``saturated'' image along the wavelength direction for 
better visualization of the saturation.  We show an example of this transformation for a
typical saturated raw frame in Figure~\ref{saturate}.  Here, saturated
pixels are accumulated along the orders, showing sharp features.  Raw
frames with such features are considered to be ``saturated" and
removed from the library. In the end, 173 ($9.3\%$) of 1847 extracted
spectra were found to be ``saturated" and removed.



\section{Telluric correction}

Ground-based spectroscopy is always subject to contamination from
the Earth's atmosphere. 
The sky subtraction  described
in Sec.~\ref{extra:calib} 
corrects for the additive component of this contamination, which
leaves the multiplicative effect of absorption.
In the visible and NIR portions of the
spectrum, water vapor, molecular oxygen, carbon-dioxide, and methane generate strong
absorption features.  Absorption features that originate in the
Earth's atmosphere are referred to as telluric features. Correction
for telluric contamination, therefore, is important for the XSL spectra
in the VIS and NIR arms. The correction for the 
continuous component of atmospheric extinction is discussed in
Sec.~\ref{extin:flx}.

In general, if we knew how much light was absorbed by the Earth's
atmosphere in a certain wavelength region, we could ``recover'' the
fluxes in that region by dividing the known absorbed fraction. This
requires a telluric template, either determined by modeling the
atmosphere by a radiative transfer code, such as ATRAN \citep{Lord92}
and LBLRTM\footnote{See
  \url{http://rtweb.aer.com/lblrtm_description.html}}, or by 
observations of ideally featureless stars
(in practice typically hot stars, whose only features are 
hydrogen and/or helium lines). Although radiative
transfer codes have been shown to be promising \citep{Seifahrt10} to
remove the atmospheric absorption, particularly at high spectral
resolution in the NIR, this method requires a molecular line
database and a model atmosphere based on meteorological data as input.  
Because we have thousands of spectra taken at different
airmasses, pressures, and humidities, this
method is currently computationally too expensive.


Instead, we use telluric standard star observations that are taken 
as part of the standard X-shooter calibration plan
directly after each of our science observations as a basis for telluric
correction of our data.  
The narrow-slit settings
are used for the telluric standard observations to match
the highest resolution of our science observations.


\begin{figure}
   \includegraphics[angle=90,scale=0.35]
   {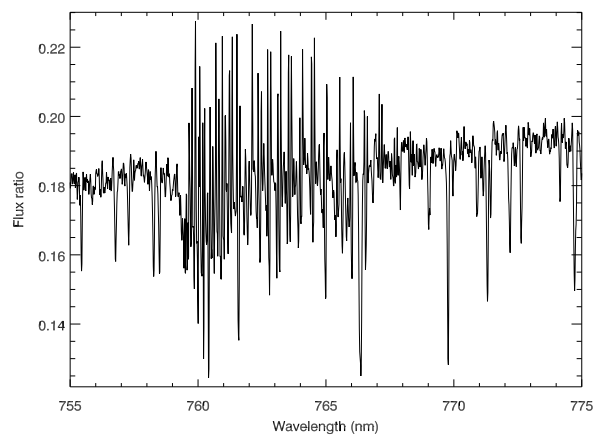}
   \caption{Ratio of counts between Cl* NGC 6522 ARP 4329 and the
     closest (in elapsed time) telluric standard star Hip094378,
     zoomed in the `A' band in the VIS arm.}
   \label{bad_tell_corr}
\end{figure}	
We find that the telluric absorption lines change strength on
timescales shorter than the ``long'' exposure time ($\ge 90$ seconds)
of faint XSL stars and the total overhead time of $\sim
900$ seconds, which results in an imperfect telluric
correction. In addition, small changes in spectral
resolution and wavelength zero-point occur even between successive 
observations. 
Figure~\ref{bad_tell_corr} shows the ratio of counts
between a science object (Cl* NGC 6522 ARP 4329, an M star in the
Galactic bulge) and the telluric standard star (Hip094378, a B2V star)
observed 37 minutes later.
The ratio shows that the closest telluric standard spectrum does not
give a satisfactory solution for the atmospheric transmission.  

To optimize the telluric correction, we have built a library of telluric
spectra, where the hot stars were first carefully wavelength-(re)calibrated.  
The final library is a large collection of different
atmospheric transmissions used to correct the telluric absorption
features in the XSL spectra, as described below.  We note that 
we only discuss the correction of the VIS arm spectra in this paper, as the
UVB spectra do not require telluric line correction, and we will discuss
the NIR spectra in another paper.

\subsection{The telluric library}

\begin{table*}
\centering
\caption{The input stars for the VIS-arm telluric library.}\label{tabtell}
\begin{tabular}{*{6}{l}}\hline
Name &RA (J2000.0)&DEC (J2000.0)& Exp. time (s)&airmass& Sp. type\\\hline 
Hip088947	&	18:09:22.50	&	$-$36:40:21.1	&	12.50	&	1.77	&	B1II	\\
Hip089086	&	18:10:55.35	&	$-$33:48:00.2	&	12.50	&	2.51	&	B1V	\\
Hip091038	&	18:34:15.85	&	$-$04:48:48.8	&	12.50	&	1.06	&	B1V	\\
$\cdots$&$\cdots$&$\cdots$&$\cdots$&$\cdots$&$\cdots$\\\hline 
\end{tabular}
\end{table*}

We have collected 178 VIS arm spectra of telluric standard stars with
spectral type A, B, and G in Periods 84 and 85. The data were reduced
identically to the program spectra.  As the G stars are much cooler
and have more lines than A and B stars, we have not used them to build
our telluric templates.  Moreover, 22 out of 175 spectra of A and B
had strong intrinsic emission lines and have not been used
either. Table~\ref{tabtell} lists the 152 telluric standard stars (and
the airmass at which they were observed) used in the telluric library.

\subsubsection{Identification of atmospheric transmission features}

The goal of building the telluric library is to produce a collection of
empirical 
atmospheric transmission spectra for use as a basis for correcting the
telluric absorption in the science spectra. To accomplish this, the
main telluric features in the observed telluric standard spectra
must be separated from the intrinsic features of these hot stars.


To extract the atmospheric transmission from the 1D spectrum
of a hot star, we need to identify the intrinsic spectral features of
the star. We use synthetic spectra from model atmospheres drawn from
the POLLUX
database\footnote{\url{http://pollux.graal.univ-montp2.fr/}}
\citep{Palacios10}
 for stars
with effective temperatures of 10000--15000 K and synthetic spectra from
\citet{Munari05} for stars with effective temperatures of 15000--27000 K.
We use the full-spectrum-fitting program pPXF \citep{Cappellari04} to
fit the hot star's intrinsic features with the synthetic spectra.  We
choose a subsample of templates from the collection of synthetic
spectra, according to 
the spectral type of each telluric standard star.

\begin{figure*}
   \includegraphics[angle=0,scale=0.7]
   {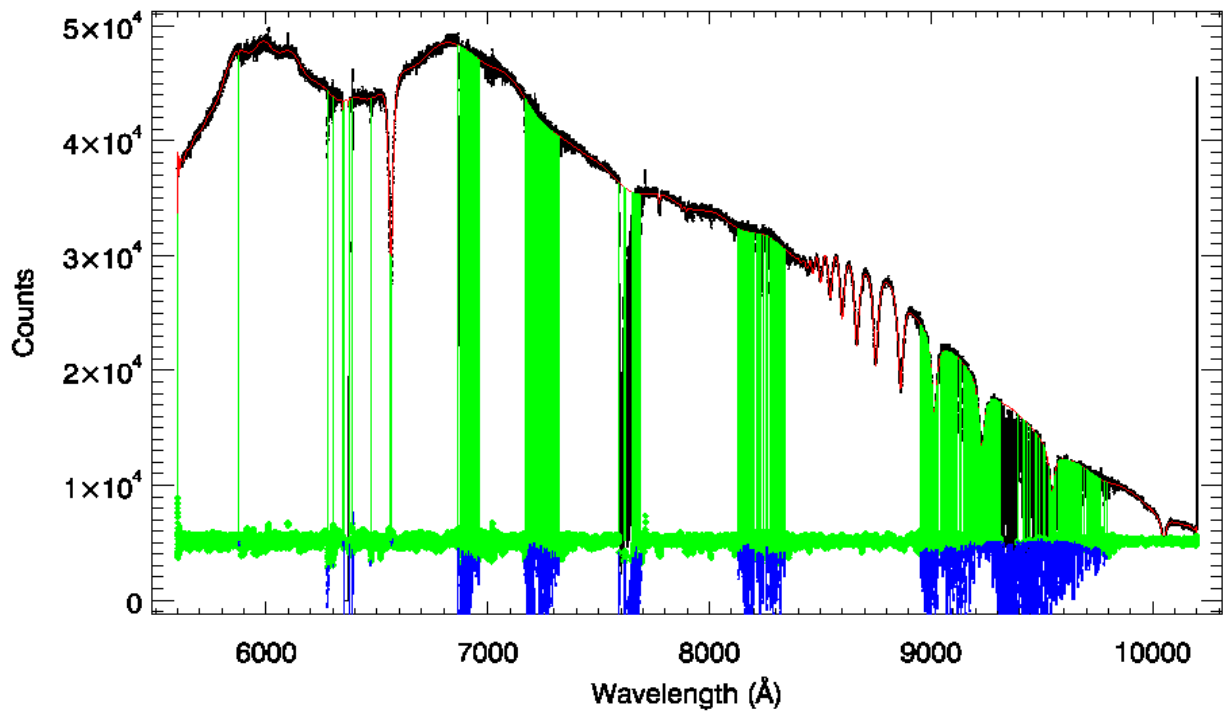}
   \caption{Extraction of the telluric absorption from the B8V star
     Hip012389. The black spectrum is the original 1D spectrum
     (without telluric or flux calibration).  This spectrum is fit by a
     synthetic LTE spectrum from a model atmosphere in the pixels
     outside of the vertical green lines. The red spectrum represents
     this fit. The lower part of the plot shows the residuals 
     from the best fit: green points represent regions of intrinsic stellar
     absorption, while blue points are (almost) the telluric absorption
     spectrum. We note that the residual between the best fit and the data 
     is scaled to the view level. It is, therefore, not surprising that some features are below 
     zero.}

   \label{tell_raw}
\end{figure*}    

Pixels that are fit well by the model template are marked as intrinsic
features of the hot stars, while any others are marked as contaminated by telluric
features. It is not always straightforward to recognize which pixel 
belongs to the star itself or is affected by the atmosphere
especially when the
atmospheric features are superimposed on the hydrogen Paschen
absorption lines of the hot stars, which is in the wavelength region
$\sim8900$--9800 \AA. Furthermore, some lines affected by non-local
thermodynamic equilibrium (NLTE) effects are not perfectly reproduced
by the local thermodynamic equilibrium (LTE) POLLUX and Munari models.  To separate the pixels
containing telluric features from those containing intrinsic stellar
features, at least two iterations are necessary.
In the first iteration, we define strong hydrogen
absorption regions uncontaminated by telluric absorption and fit
these with the model synthetic spectrum and a moderate-order
multiplicative polynomial.  Because some parts of the $\rm{H}\alpha$
line ($\lambda\ 6562.8$ \AA) are affected by NLTE
\citep{Thurl06} and the core is occasionally filled by weak
absorption, this line can be very hard to reproduce by the models
based on LTE. We therefore mask the core of the
$\rm{H}\alpha$ line, which is roughly $\pm4$ \AA\ around the line center,
  and replace this region with the same wavelength
  region of the fitted model after the fit.
Residuals from this first iteration are assumed to
represent the noise of each pixel. Pixels deviating by more than 3--5 standard
deviations from this first fit are 
mostly telluric absorption features and are
masked from the fit in the next step.

In the second iteration, 
the fit is performed with a high-order multiplicative polynomial, and
we are able to
match the continuum of the hot star in detail. A B8V star is shown
in Figure~\ref{tell_raw} to illustrate the fitting process in 
this iteration.  
Subsequent iterations were occasionally necessary to
improve the final fit.  The telluric features were then extracted by
dividing the original 1D telluric standard star spectrum by the best
fit of the final iteration.

We call the telluric features extracted by this process the ``raw
transmission spectra'', as there are several steps required before the
telluric library is ready to be used.  The quality of the raw transmission
spectra depends on the templates of the hot stars, the
signal-to-noise of the original X-shooter spectra, and the
multiplicative polynomial used in the fitting.

\subsubsection{Absolute wavelength calibration and spectral
  cleaning}\label{abs_tell_wave}

\begin{figure}
   \includegraphics[angle=90,scale=0.35]
   {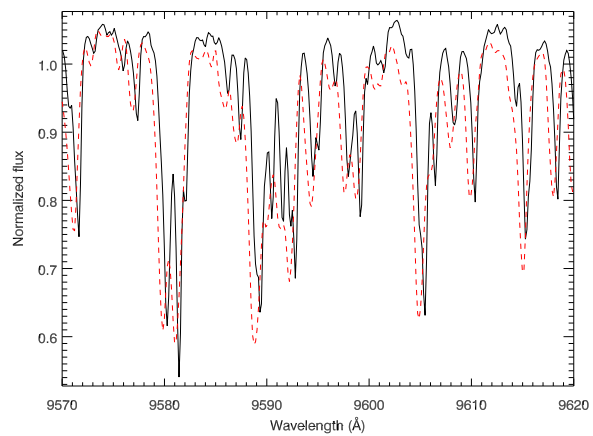}
   \caption{Two normalized telluric standard stars, Hip026545 (black
     line) and Hip012389 (red dashed line), zoomed into the wavelength
     region 9570--9620 \AA, showing the impact of the flexure of the
     X-shooter backbone on the wavelength calibration.  The wavelength
     shift is roughly 2 pixels between these two spectra, or
     $\approx10\,\kms$.}
   \label{tell_pre}
\end{figure}    

To build the telluric library, all the telluric features in different
standard stars must be at exactly the same wavelengths. We find that
the wavelengths for the same telluric feature are slightly shifted in 
different spectra, due
to the flexure and imperfect rotation of the X-shooter backbone.  In
Figure~\ref{tell_pre}, we show an example of this shift, which is
equivalent to $\approx10\,\kms$ in this case.  We therefore require an
extra wavelength calibration step for each raw transmission
spectrum.

Although we do not use theoretical atmospheric transmission spectra to
correct our spectra for telluric absorption, we use them as
references for the final wavelength calibration. We use a
high-resolution ($R \sim 60000$) model transmission spectrum computed
for us by J.~Vinther from the ESO Sky Model
Calculator\footnote{\url{http://www.eso.org/observing/etc/bin/gen/form?INS.MODE=swspectr+INS.NAME=SKYCALC}}
as a template to cross-correlate and correct the small residual
wavelength offsets in the raw transmission spectra. 
 
\begin{figure}
   \includegraphics[angle=90,scale=0.35]
   {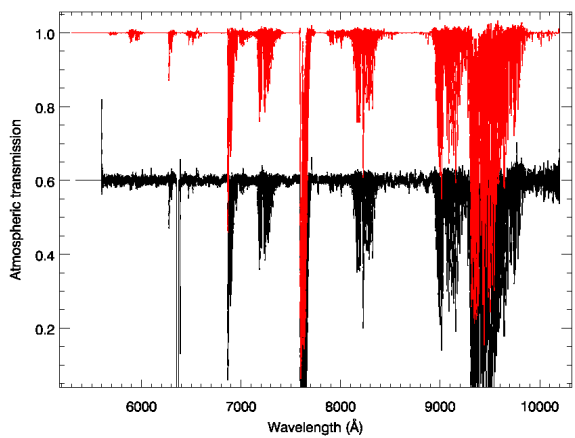}
   \caption{Extracted raw transmission spectrum from the B8V telluric
     standard star Hip012389 (black) and the shifted, cleaned
     ``final'' transmission spectrum from this star (red). The raw
     transmission spectrum has been offset for clarity.}
   \label{tell_norm}
\end{figure}

The extracted transmission spectra are inevitably influenced by the
signal-to-noise ratios of the original spectra (and any remaining bad
pixels).  To reduce the influence on the final telluric-corrected
spectra, especially those pixels which are not contaminated by
telluric features, we ``clean'' the shifted transmission spectra
based on the model transmission template. Pixels with values, which
deviate more than 10 sigma from the template, are set to unity, so as
not to introduce noise or artificial features into the
telluric-corrected science spectra.  Figure~\ref{tell_norm} shows the
raw transmission spectrum and the shifted, cleaned transmission
spectrum extracted from the telluric standard star Hip012389.

After carefully
checking the quality of the output transmission spectra, the telluric
library with 152 final transmission spectra is ready to be used.
 
\subsection{Final telluric correction}

Once the telluric library is available, we can perform the telluric
correction on the science spectra.  The simplest process would be to
divide a science spectrum by the temporally-closest transmission
spectrum. An improvement 
is obtained by exploiting the variety of telluric 
absorption properties present in our large telluric library.
 A combination of 
telluric templates can provide a better estimate of the 
telluric absorption in a given science exposure than the 
one telluric standard observed closest in time \citep{Chen11}.


The key to our telluric correction method, which is similar to producing the telluric
library itself, is the determination of the underlying stellar spectrum.
A quick glance at the extracted 1D spectra shows that it is easier to
define the continua of 
warm stars than cool stars, such as M-type giants and 
dwarfs, carbon stars, or LPVs.
We therefore classify the science spectra
according to their spectral types into two groups: those that have
easily-determined continua, O to K stars, and the cooler stars, which
do not.  For warm stars, we estimate the atmospheric transmission
using the optimization algorithm described below, 
which makes use of 
the original science spectra; for the cooler stars, we simply
use the temporally-closest transmission spectrum from the telluric
library.


\subsubsection{Reconstructing telluric absorption with principal
  component analysis}

Algorithms that estimate the telluric absorption 
by seeking the best linear combination of the 152 templates of the
telluric library seemingly succeed (i.e., they
produce a result that passes eye inspection), but they are dangerous: 
this linear problem is ill-conditioned. There are not 152 meaningful
degrees of freedom in this problem. In more physical terms, there are
not 152 fundamental Earth-atmospheric parameters that produce differences
between transmission curves larger than the amplitude of the noise in
the XSL data. The consequence is that the linear combination coefficients
derived from a standard linear-optimization algorithm are determined to 
a large extent by the noise in the spectra rather than by the 
telluric features of interest \citep[see][for further explanations in a conceptually similar context]{Ocvirk06}. 
Details of the science spectrum that are not telluric features may be erased. 
To avoid this issue, one needs to remove items from the telluric 
library until the linear problem is well behaved. We ask: but which ones? 
We use principal component analysis (PCA) to circumvent this problem.

Principal component analysis reveals the internal structure of a
data set in a way that best explains the variance in the data. It is
widely used in analyzing multidimensional data sets.  In data
reduction applications, PCA has been used for sky subtraction by
\cite{Wild05} and \cite{Sharp10}. Inspired by this idea, we have developed a
method to determine the telluric correction for our (warm) program
stars. The algorithm \citep{Wild05} is to determine the basis vectors
from those pixels, which vary most in the spectra of the telluric
library. The amplitudes given by the projection of these basis vectors
onto each science spectrum are used to reconstruct the telluric
features in the science spectrum.  Telluric correction is then a
straightforward division of science spectrum by the projected telluric
spectrum.

\paragraph{PCA components}

The telluric library contains 152 spectra, and each spectrum has a length of 43615
pixels.  Standard PCA analysis would require
diagonalizing a $43615\times43615$ matrix, which is computationally
expensive.  Instead, we assume that {\bf X} is the matrix that we
built based on the telluric library with the mean subtracted from
each individual spectrum.  The covariance matrix $C$ for computing the
eigenvectors then can be written as
\begin{equation}
 C = \frac{1}{N} {\bf X^{\dagger}}{\bf X}, 
\end{equation} 
where ${\bf X^{\dagger}}$ is the transpose of the matrix {\bf X} and
$N$ is the number of objects, which is 152 here.  The eigenvectors ${\bf
  u}_{i}$ then can be solved from the equation
\begin{equation}\label{eq solve0}
 C{\bf u}_{i} = \lambda_{i}{\bf u}_{i}, 
\end{equation} 
where $\lambda_{i}$ is the eigenvalue associated with ${\bf u}_{i}$.
If we expand Equation~\ref{eq solve0} and multiply by {\bf X} on
both sides, we have
\begin{equation}\label{eq solve1}
 \frac{1}{N}{\bf X}{\bf X^{\dagger}}{\bf X}{\bf u}_{i}=\lambda_{i}{\bf X}{\bf u}_{i}. 
\end{equation} 
Equation~\ref{eq solve1} has the same set of eigenvalues as
Equation~\ref{eq solve0} for the vector ${\bf v}_{i}={\bf X}{\bf  u}_{i}$.  
The dimension of the array ${\bf X}{\bf X^{\dagger}}$  at 152$\times$152 is much
smaller than in the standard formulation, dramatically 
reducing the computing time.  When the temporary eigenvector ${\bf v}_{i}$ is 
derived, we can derive the original eigenvector ${\bf u}_{i}$.  We wrote our 
own IDL PCA code following the above algorithm and performed the PCA on
our 152 telluric library spectra. The extracted 152 eigenvectors or
principal components are then ready to be used to reconstruct the
telluric features in the science objects.  

\begin{figure*}
   \includegraphics[angle=0,scale=1.]
   {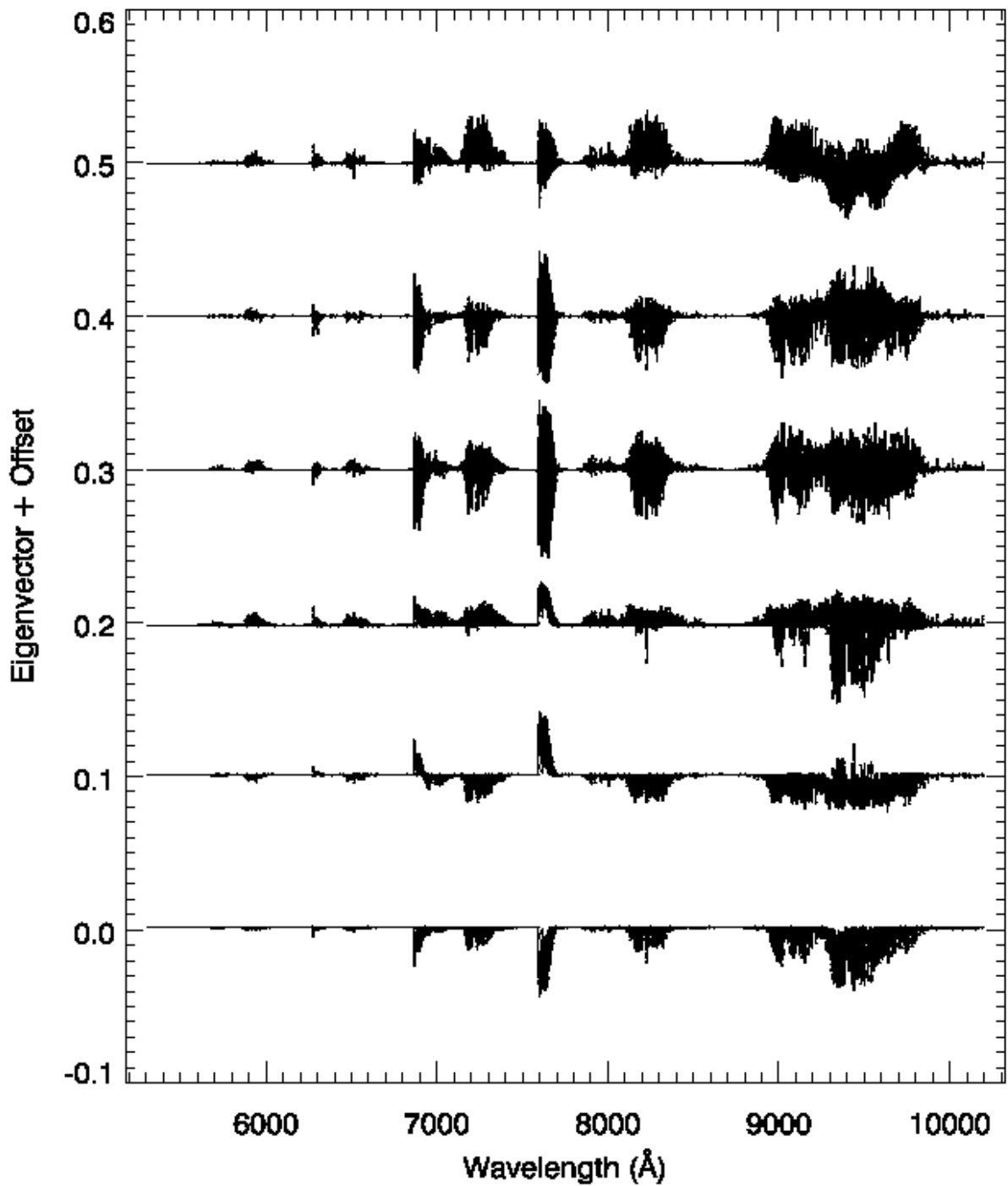}
   \caption{The first six principal components of the telluric library.
     The eigenvalues decrease from the bottom to the top of this
     figure (i.e., the first and largest eigenvector is the lowest).
     The eigenvectors have been offset for display purposes.}
   \label{pca6}
\end{figure*}

Figure~\ref{pca6} shows the first six principal components from the
bottom to top.
The first two components have a clear
physical meaning: the first component shows the mean spectrum of the
telluric library, while the second component appears to separate most
of the water vapor features (seen ``in absorption'')
from O$_2$ features (seen ``in emission'').

\paragraph{Reconstructing the telluric absorption spectra} 

Two steps remain before we can obtain the
final estimate of the telluric absorption in a science spectrum. 
First, we must normalize the science spectrum.
Second, we have to correct the wavelength offset of the telluric features
in the observed spectrum (due to instrument flexure) from the
wavelengths of the telluric library spectra that are more accurate.  
The resulting telluric absorption spectrum is used to correct the
science spectrum.  In this procedure, any intrinsic absorption or
emission features should not be removed from the science spectra. An
accurate estimation of the stellar continuum is therefore required.
We recall that we only perform the PCA reconstruction
for the spectra of warm stars.

To normalize a science spectrum, we need to determine
its continuum.
We have developed an IDL code that determines an approximate continuum
for a (warm) star. 
The code seeks the strong intrinsic features (e.g., 
$\mathrm{H}_\alpha$, $\mathrm{H}_\beta$, $\mathrm{Ca}\textsc{ii}$ triplet etc.) 
from a line list\footnote{\url{http://physics.nist.gov/PhysRefData/ASD/lines_form.html}},
according to the radial velocity given by ULYSS
and defines them as nodes. We flag pixels between those nodes
that have a second derivative
greater than zero and replace them by a linear interpolation
between the nearest unflagged pixels. By this process, weak and narrow lines
are rejected,
since we only care about a rough continuum.
Usually around three to five iterations are used 
to find the local continuum between two nodes.
Once the continuum of a
spectrum has been found, the science spectrum is easily normalized
by dividing the continuum before finding the PCA components.


We use pPXF to determine any residual wavelength shift in the telluric
lines 
of the normalized science spectrum, taking the first 
principal component as a template. The shift is applied to the PCA components, 
thus avoiding unnecessary rebinning of the science spectrum.

For wide-slit observations, pPXF also determines an
adequate broadening function for the lines of the telluric library. 
The principal components are then convolved with this function before they 
are used. We note 
that smoothed PCA components are not precisely the PCA components
of a smoothed telluric library. After a few tests, smoothing first was judged to be too
computationally expensive for our current purpose, considering that the wide-slit
observations are used only to correct the high resolution observations
for slit losses.

We finally project the mean-subtracted, normalized 
science spectrum onto the modified principal components and sum 
these projections.
The final estimate of the
transmission is derived by the combination of 40 principal  
components with the amplitudes that result from the projection.
However, stars with weak emission lines or with a high density of
lines present a challenge, and for these stars, we use only 20
components to avoid affecting their intrinsic features by
mistake.  Once the reconstructions have been constructed, the telluric
correction can be performed by dividing the normalized spectra by the reconstructed
transmission. The final telluric-corrected spectra are then 
derived by multiplying the resulting spectra by the continuum.

\begin{figure*}
   \centering
   \includegraphics[angle=90,scale=0.7]
   {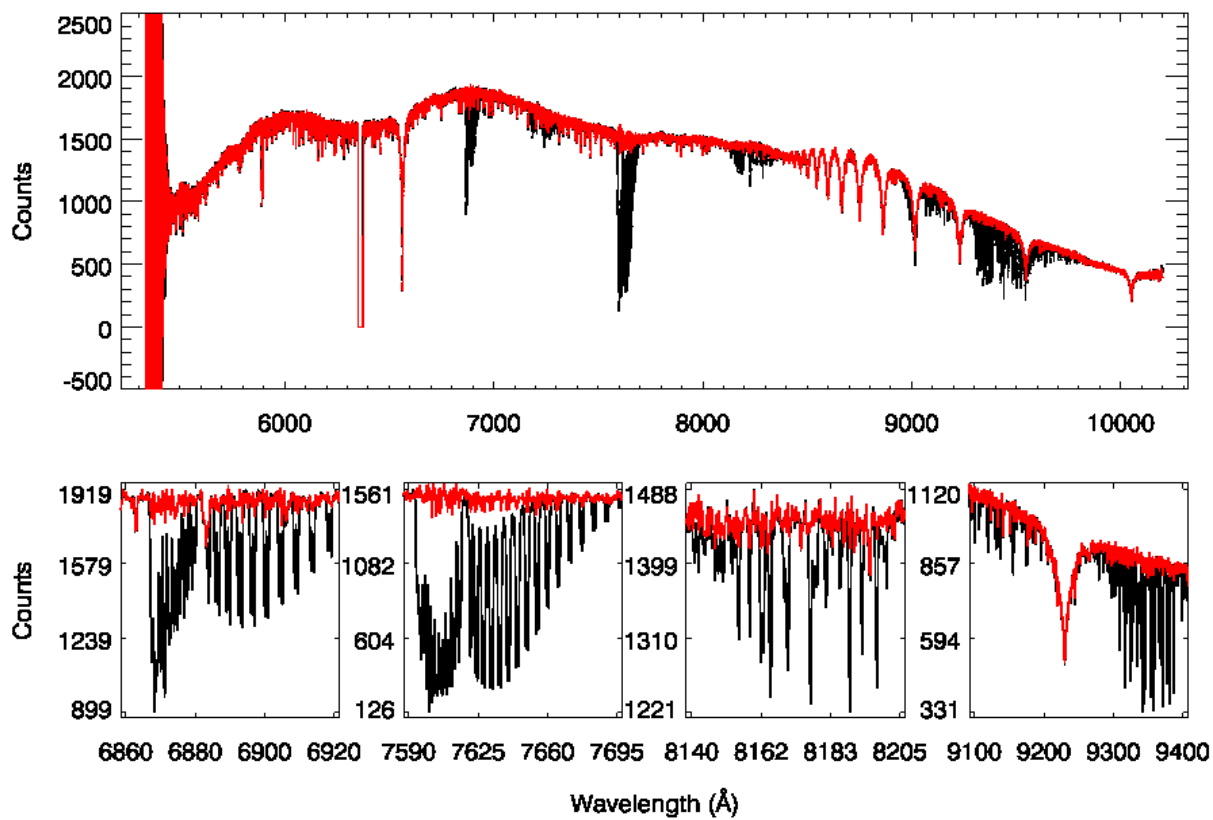}
   \caption{Spectrum of HD 164257 (A0) before (black) and after (red)
     telluric correction in the VIS arm, using the PCA reconstruction
     method described in the text.  The lower panel shows four
     zoomed-in regions to show the corrections in detail.}
   \label{telcr-1}
\end{figure*}    
\begin{figure*}
   \centering
   \includegraphics[angle=90,scale=0.7]
   {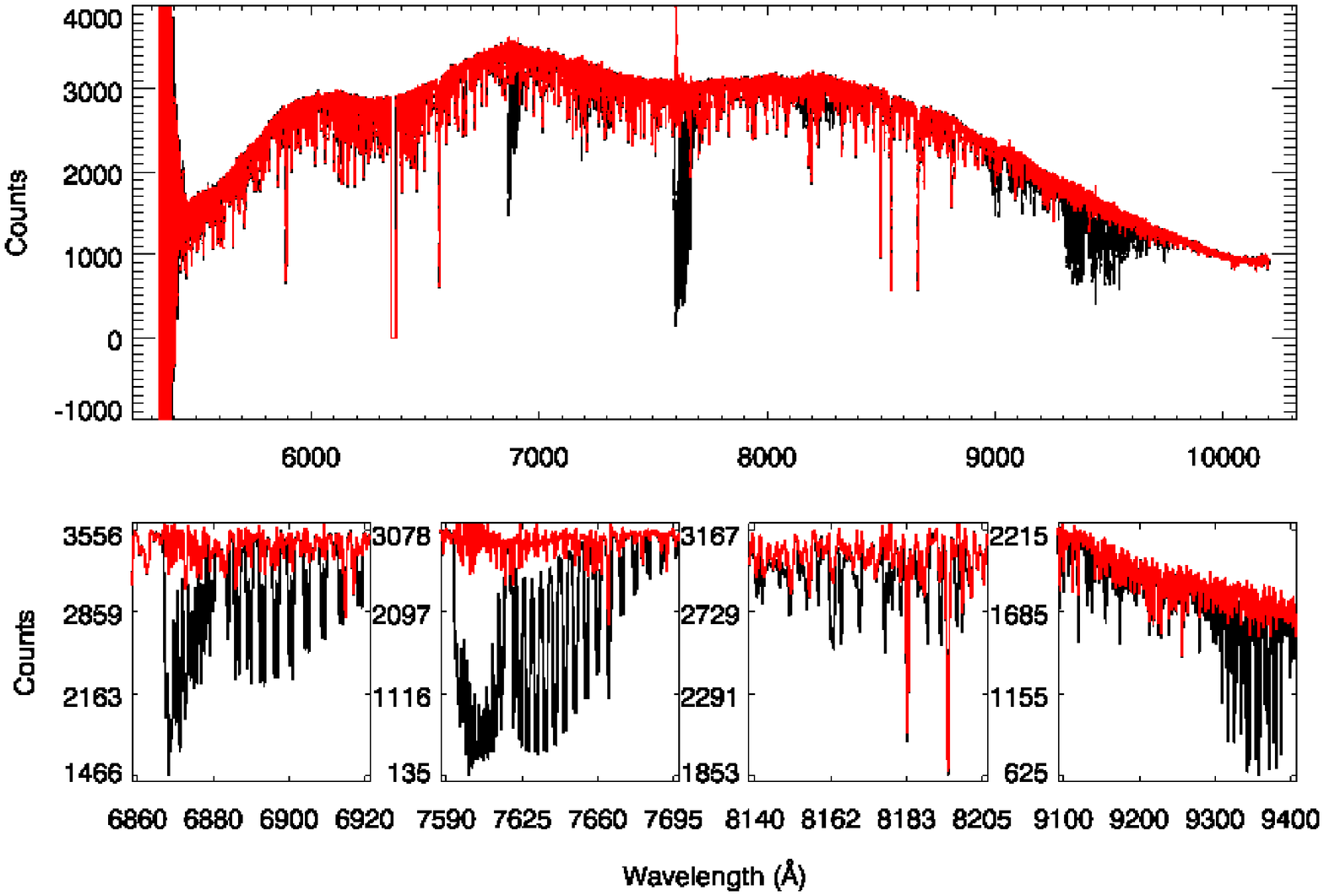}
   \caption{Spectrum of HD 193896 (G5IIIa) before (black) and after (red)
     telluric correction in the VIS arm, using the PCA reconstruction
     method described in the text.  The lower panel shows four
     zoomed-in regions to show the corrections in detail.
     The spikes around 7600 \AA\ are residuals after telluric correction.}
   \label{telcr-2}
\end{figure*}    
\begin{figure*}
   \centering
   \includegraphics[angle=90,scale=0.7]
   {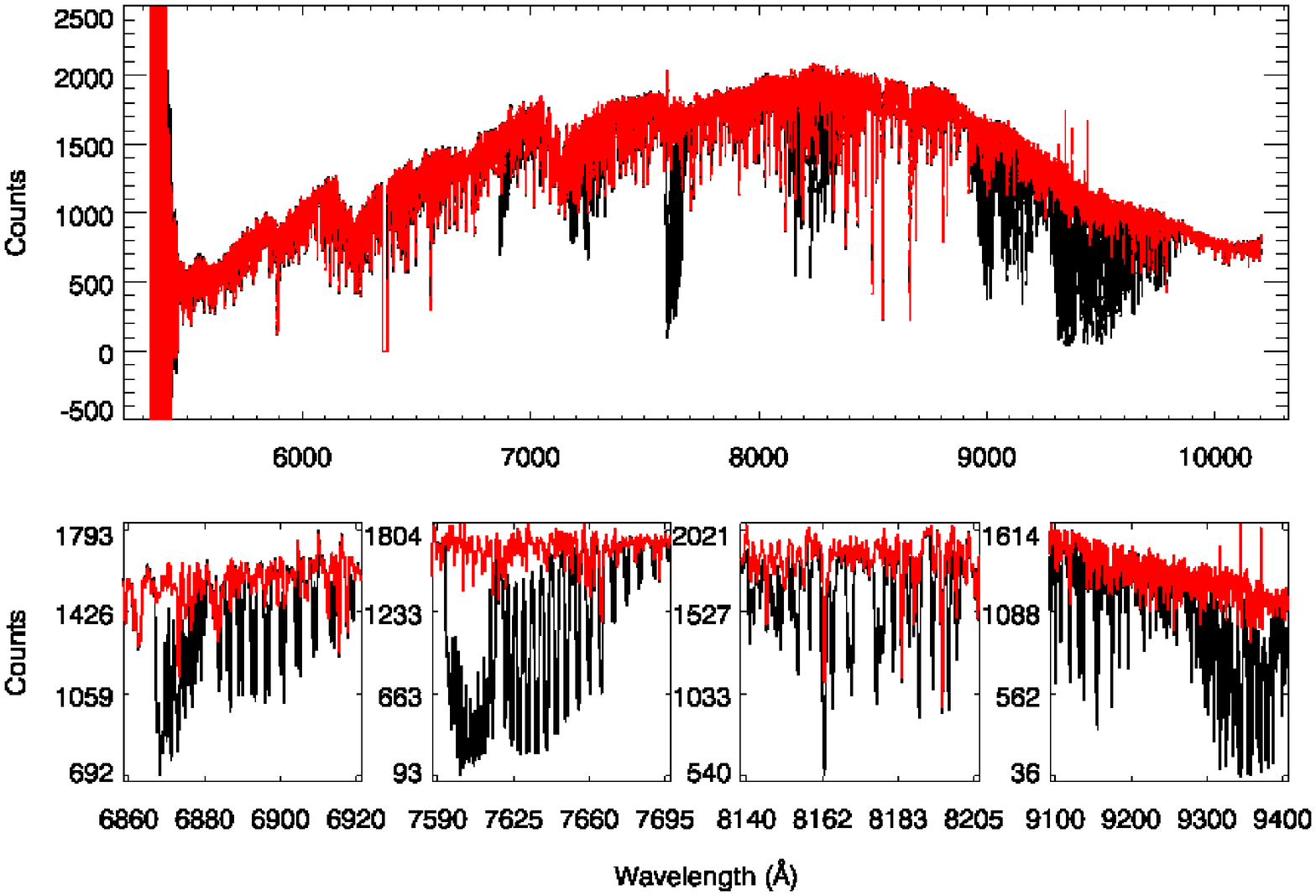}
   \caption{Spectrum of HD 79349 (K7IV) before (black) and after (red)
     telluric correction in the VIS arm, using the PCA reconstruction
     method described in the text.  The lower panel shows four
     zoomed-in regions to show the corrections in detail.
     The spikes around 9350 \AA\ are residuals after telluric correction.}
   \label{telcr-3}
\end{figure*}    

We show the original 1D spectra of HD 164257 (A0V), HD 193896 (G5IIIa),
and HD 79349 (K7IV) and their telluric-corrected versions in
Figures~\ref{telcr-1}, \ref{telcr-2}, and \ref{telcr-3},
respectively. We find that the telluric correction by the PCA
reconstruction has done a reasonable job for both the early-type stars
and some late-type stars.  In general, this method works better for
stars with simpler continua and high signal-to-noise ratios.
Occasionally, division by almost zero produces artificial spikes, 
such as those seen at 7600 \AA\ in Figure~\ref{telcr-2} or at 
9300 -- 9400 \AA\ in Figure~\ref{telcr-3}. When these occur, we flag the pixels 
that are incorrectly reconstructed and set them equal to zero in the final 
telluric-corrected spectra.

\subsubsection{Telluric correction for cool stars}

The PCA method requires an accurate representation of the stellar
continuum to work correctly. It is difficult to trace
the continua of cool stars. For instance, carbon stars have strong and
sharp molecular bands. One therefore has to trace each absorption
bandhead, which can be difficult because some molecular bands occur
exactly at the same wavelengths as the telluric absorption regions. In
this case, we use the temporally-closest transmission spectrum from
the telluric library.

As in the warm stars, 
we must correct 
residual wavelength calibration 
errors and match the line-broadening of the science and the 
temporally-closest transmission spectra before dividing. We calculate these 
corrections as described above, using the wavelength range of 
the atmospheric ``A'' band of O$_2$. The ``A''
band is 
chosen because
it is the strongest telluric absorption
feature in  the VIS arm spectra and is easily distinguished
from other molecular species. 
The telluric correction is then made by
directly dividing the science spectrum by this modified transmission
spectrum.

\begin{figure*}
   \includegraphics[angle=90,scale=0.7]
   {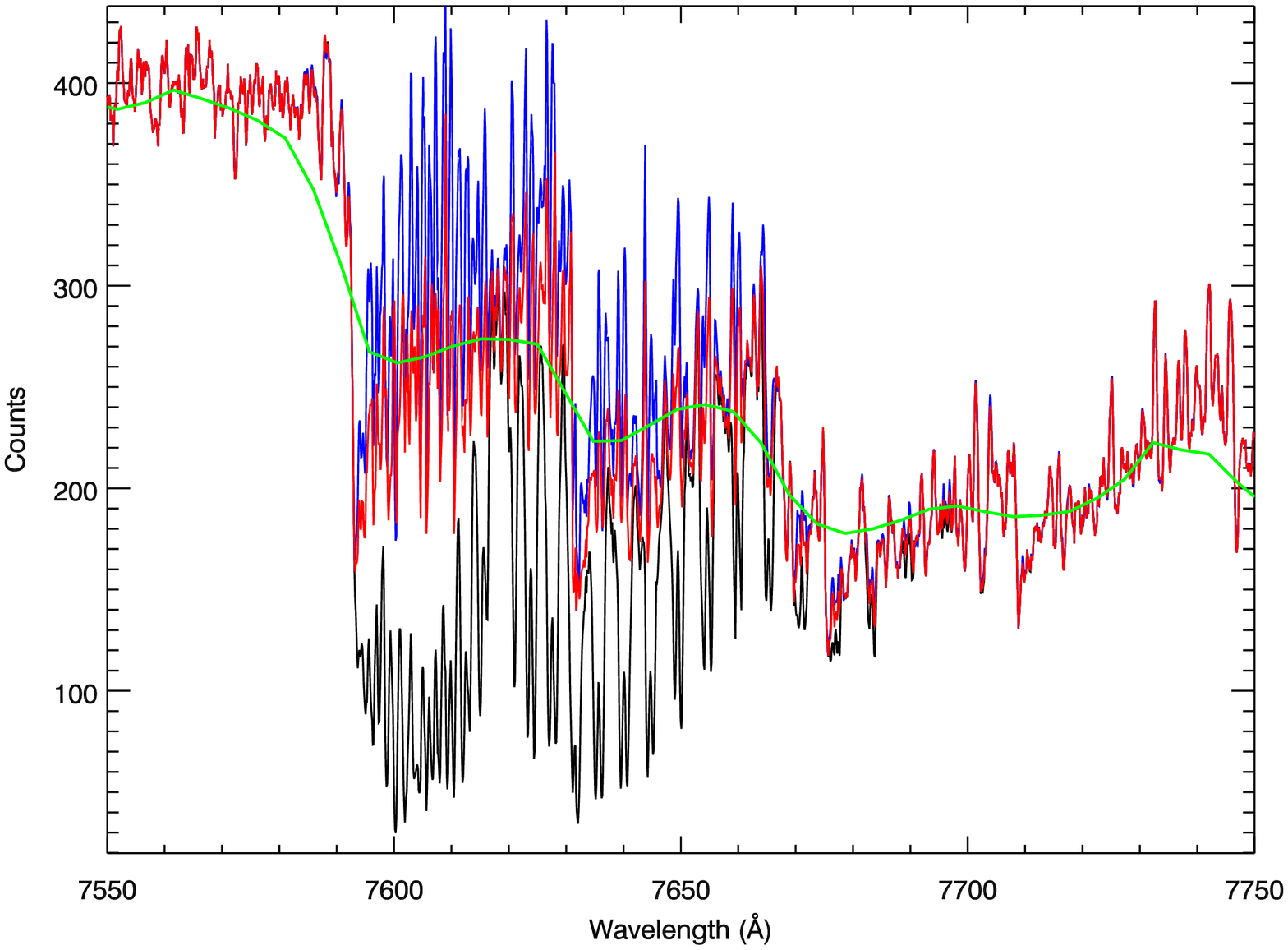}
   \caption{Telluric correction by two different methods for the cool
     bulge LPV star OGLEII\ DIA\ BUL-SC1\ 235. The black line is the
     original 1D extracted spectrum, and the blue and red lines are
     the spectrum after correction by the PCA reconstruction and the
     temporally-closest telluric absorption spectrum, respectively.
     The green spectrum is an M5III star with a
     similar temperature, HD175865, taken from NGSL as a comparison.}
   \label{telcr-4}
\end{figure*}    

By using the closest telluric absorption spectrum, there is the risk that 
one may over- or underestimate the real telluric spectrum of the science
object. In the case that the exposure time of the science object is
short and the accompanying telluric standard star spectrum is therefore very
close in time to the science object, this method does a better job to
preserve the intrinsic features of the stars than the PCA method 
for cool stars. We show an example of the telluric
correction made by both methods on an LPV star
OGLEII\ DIA\ BUL-SC1\ 235 in Figure~\ref{telcr-4}. 
To make sure where the molecular bands should be, we use an NGSL 
star HD175865 with similar temperature 
as a comparison, which has no telluric contamination. The PCA
reconstruction method, as shown in the blue spectrum, does not reproduce
the molecular bands properly. The correction by the closest
transmission spectrum, although not perfect, is closer to the
intrinsic features of the star, as we can distinguish the molecular
bands clearly. 

We use this method on all carbon stars, LPV stars, and most of the
cool M stars. We remind the reader that the method (using the closest
transmission spectrum) for cool stars may not be the final, best
solution for these stars, but we use it in the absence of accurate
spectral models of these (molecule-rich and typically variable) stars.

\subsubsection{Comparison of the two methods of telluric correction for warm stars}

\begin{figure*}
   \includegraphics[angle=90,scale=0.7]
   {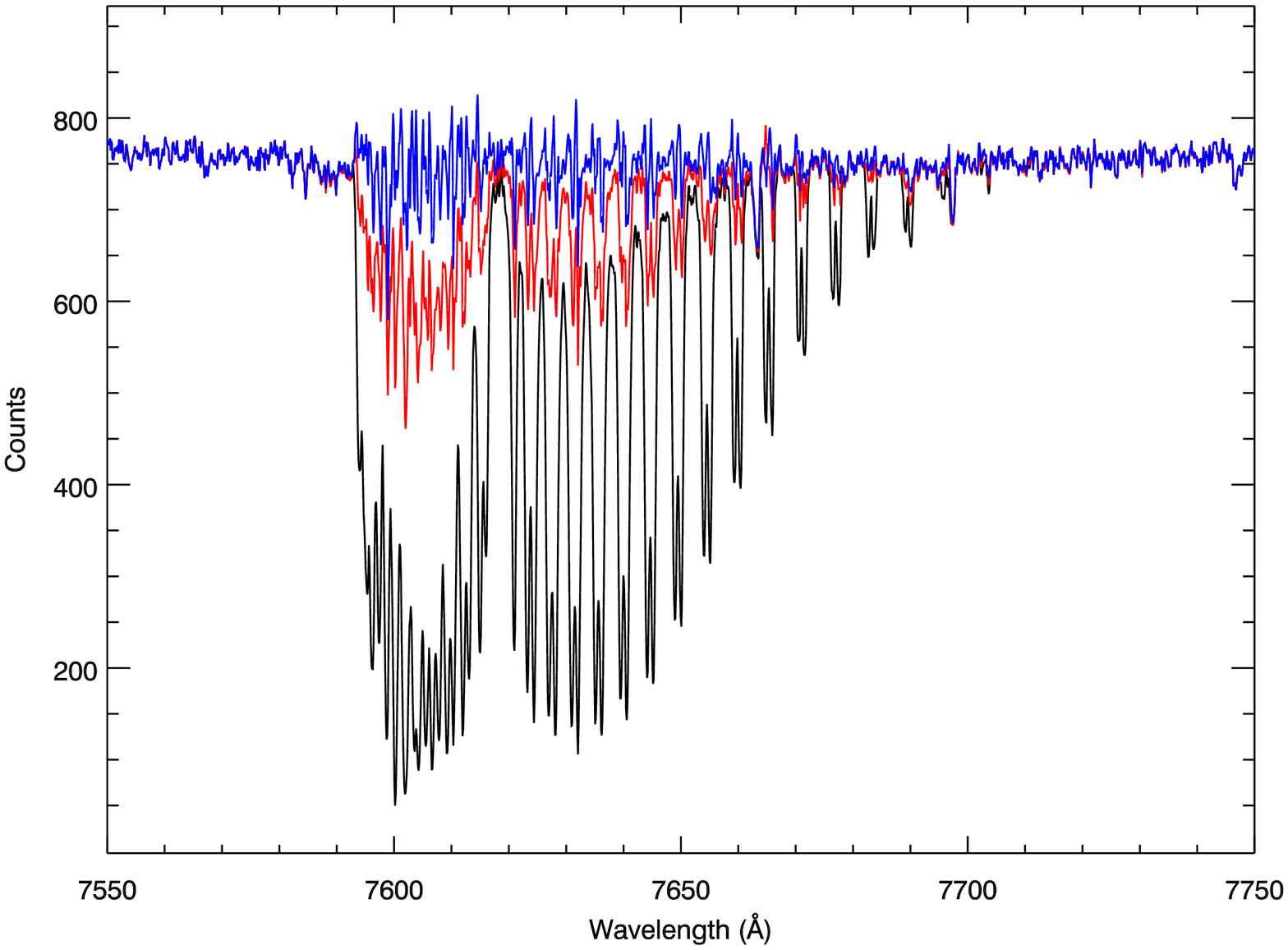}
   \caption{Telluric correction by the PCA reconstruction and
     temporally-closest telluric absorption methods on the G3V star
     G169-28. Symbols and wavelength ranges are the same as in
     Figure~\ref{telcr-4}.}
   \label{telcr-5}
\end{figure*}    


To test the PCA reconstruction method, we compare the results
with a simple correction using the closest telluric absorption spectra
on warm stars.
In Figure~\ref{telcr-5}, we show the telluric correction
achieved by our two methods for the G3V star G169-28. We zoom
into the atmospheric ``A'' band for a fair comparison. It is clear that
the PCA reconstruction does a better job than the
temporally-closest telluric absorption spectrum in this warm
star.  Again, whether we use the PCA reconstruction method to perform
the telluric correction depends on whether we can determine an
accurate representation of a star's continuum.

\section{Flux calibration}

To perform a reliable flux calibration, we observed several
spectrophotometric standards (BD+17\ 4708, GD\ 71, GD\ 153, EG\ 274,
 Feige\ 110, LTT\ 3218, and LTT\ 7987) with a wide slit 
 ($5\arcsec\times11\arcsec$) in ``stare'' mode at different
airmasses. All flux standard stars used here were observed with the readout
mode ``100k/1pt/hg''. The median signal-to-noise ratio of the flux standards 
(44 spectra in the UVB arm and 43 spectra in the VIS arm) spanned the range 
between 100 and 800 .

\subsection{Construction of the extinction curve}\label{extin:flx}
To reconstruct the intrinsic flux and spectral 
shape, it is important to correct the spectra for atmospheric
extinction. 
We start from the basic assumption
that the atmospheric (extinction curve) and instrumental (response curve) 
properties are uncorrelated. Determining the extinction curve then can be simply 
done by assuming 
the extinction curve does not change with time.

We reduced and extracted the spectra of flux standard stars with the same set of master 
bias and master flat-field frames in each arm. Telluric correction was performed in the VIS arm
for each flux standard. The spectra of the standards were compared with the
flux tables of the appropriate stars from the CALSPEC HST 
database\footnote{\url{http://www.stsci.edu/hst/observatory/cdbs/calspec.html}}
\citep{Bohlin07}.
By running \texttt{IRAF.standard} and using the Paranal extinction
curve \citep{Patat11} as a first guess, we derived the extinction
curve for our X-shooter observations in the UVB and VIS arms.

\begin{figure}
   \includegraphics[angle=90,scale=0.35]
   {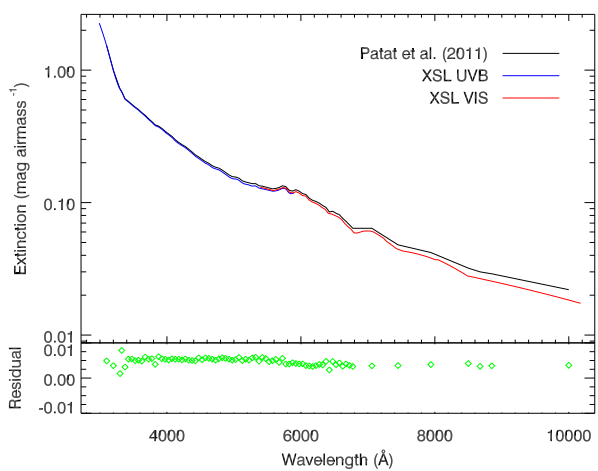}
   \caption{Extinction curve for XSL in the UVB arm (blue line) and
   the VIS arm (red line). The Paranal extinction curve of  \protect \cite{Patat11} 
   is shown as the black line for comparison. Lower panel shows the residual
   between XSL and \protect \cite{Patat11} calculated by the 
   \texttt{IRAF.standard} procedure and applied to derive the XSL extinction.}
   \label{xsl_extinction}
\end{figure}

The extinction curve of the XSL in Period 84 and 85 is shown
in Figure~\ref{xsl_extinction}, where the Paranal extinction
curve \citep{Patat11} is shown as a reference.
 We find that the inferred extinction curve is very similar
to that given by \cite{Patat11} with a slightly steeper extinction 
coefficient in the red. 

\subsection{Construction of the response curve}
We create our response curves as follows. First, for a given science spectrum,
we reduce its flux-standard star with the same set of master bias and master flat-field
frames as used for the science observation and extract the 1D spectrum. 
For the VIS arm spectra,
we perform the telluric correction on the 1D spectrum of the 
flux standard. Second, the airmass of the flux standard is used to derive 
the atmospheric extinction using the XSL extinction
curve. Third, the 1D spectrum of the flux standard is corrected by
$F^{\prime} = F_{ADU} / t_{exp} \times AtmExt$, where $F_{ADU}$ is the original 1D spectrum,
$t_{exp}$ is the exposure time of the flux standard, and $AtmExt$ is the derived
extinction term. We compare this corrected 1D spectrum with 
its flux table from the CALSPEC HST database. The final response curve is derived
by fitting a spline to the ratio between the reformed 1D flux standard spectrum and 
the corresponding flux table.

\begin{figure}
  \centering
    \subfloat{\includegraphics[angle=90,scale=0.36]{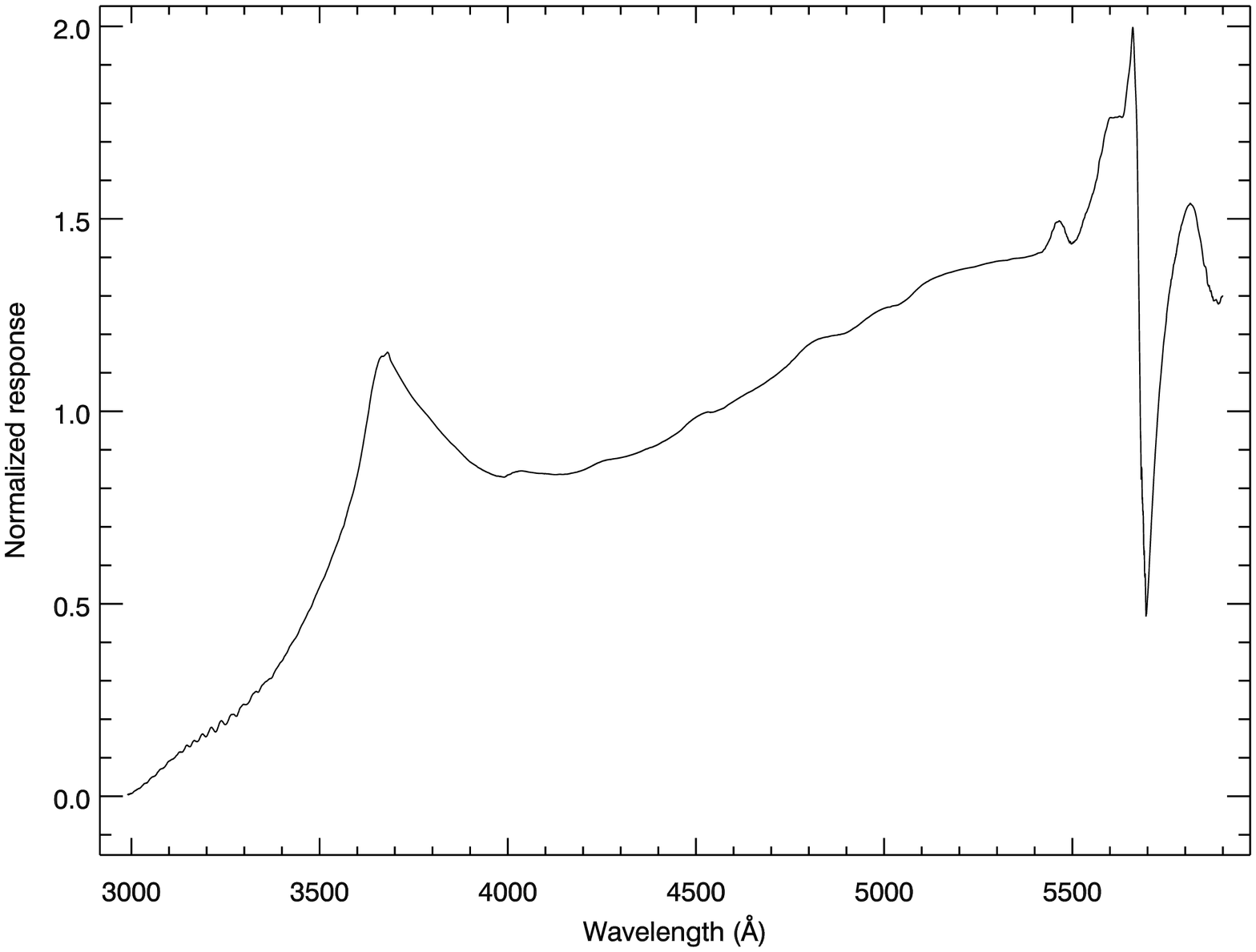}}\quad \\
    \subfloat{\includegraphics[angle=90,scale=0.36]{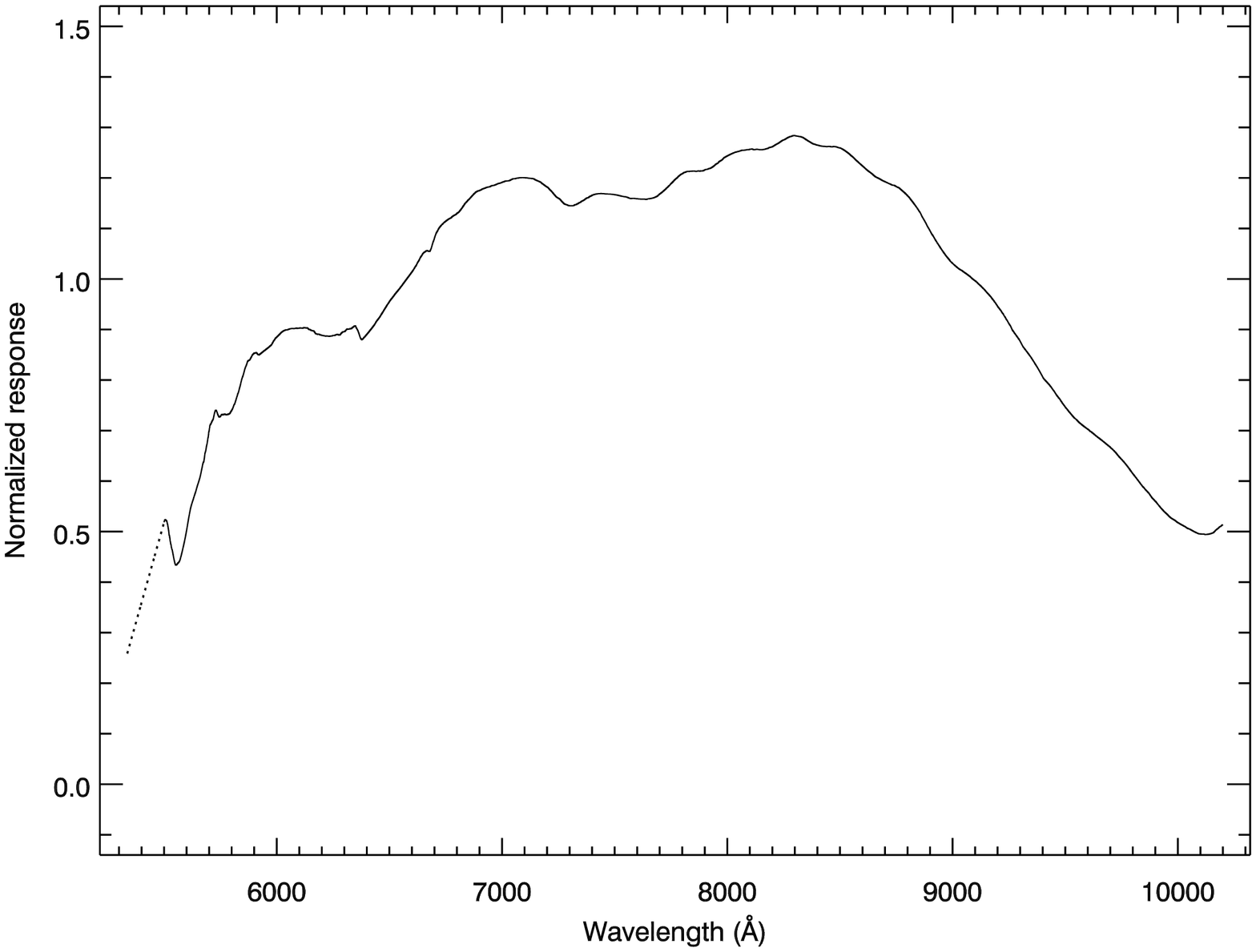}}
   \caption{Average response curves in the UVB (upper panel) and VIS arm (lower panel).
   The dotted line in the VIS arm indicates a low signal-to-noise region, interpolated
   for display purposes. Dichroic features are shown in both UVB and VIS arms around
   5600 -- 5800 \AA.
   }
   \label{response}
\end{figure}

Because the flat field and bias of X-shooter are not 
stable, at least in Periods 84 and 85, the response
curve derived from each flux standard can vary by around 5\%.
We show the averaged response curve in the UVB and VIS arm, respectively,
in Figure~\ref{response} to illustrate the overall response of the instrument.
The sharp feature around 3700 \AA\ in the top plot is due to the two flat field 
lamps used in the UVB arm. We note that there is a strong feature beyond
$\lambda\lambda$ 5500 \AA\ in the UVB arm (top panel). 
This is due to the dichroic used to split the 
beam between the UVB and VIS arms. This feature is also seen in the VIS 
arm before $\lambda\lambda$ 5800 \AA\ (lower panel in Figure~\ref{response}). 

\subsection{Flux calibration of individual frames}
We note that the dichroic features do not always appear in the same position
in the extracted 1D spectra, and therefore, it is difficult to completely remove those 
features in our final flux-calibrated spectra. To minimize the 
influence of the dichroic, we choose the observation closest in time 
for each science observation of a certain 
flux-standard star as the corresponding flux standard.
The chosen flux standard is reduced and extracted as described above. 
If the binning of the science observation is 
different from the selected flux standard, we rebin the selected flux standard 
first.

For each science spectrum, we generated its response curve following the procedure described above.
The flux-calibrated science spectrum is derived using 
the formula, 
\begin{equation}
 F_{cal, s}(\lambda) = \frac{F_{ADU,s}(\lambda)/ t_{exp}^{s} }{Respon \times AtmExt_{s}} , 
\end{equation} 
where $F_{ADU,s}$ is the 1D spectrum of the science object, $t_{exp}^{s}$ is the 
exposure time of the science object, $Respon$ is the derived response
curve, and $AtmExt_{s}$ is the extinction term calculated
from the airmass and the extinction curve of XSL.
We perform the flux calibration process on both narrow slit
and wide slit observations.

\subsection{Final flux calibration}\label{abs:flx}
A final flux calibration is carried out on the narrow-slit observations
using the shape of wide-slit observations to avoid flux losses.
We use the wide-slit exposure paired with each flux-calibrated, narrow-slit spectrum to do this.
If an unsaturated wide-slit observation
is available, we run pPXF to shift and broaden the narrow-slit spectrum. 
The flux correction is performed by multiplying the narrow-slit 
spectrum by the second-order polynomial determined by pPXF. For the narrow-slit spectra 
whose corresponding wide-slit spectrum are saturated, 
we leave their flux as in the previous step (individual flux
calibration). We note that stars without wide-slit flux correction
may be missing flux, especially in the UVB arm. 

\subsection{ADC issue in the UVB arm}
Some of our data, especially narrow-slit observations in the UVB arm that are observed 
from 18 July 2010 to 2 August 2010, are influenced by a failure of the atmospheric dispersion 
compensator (ADC). 
To correct for this ADC issue, we perform the following steps. First, we run pPXF to 
determine the possible shift and broadening between the narrow- and wide-slit observations.
Second, we convolve the narrow-slit observation to the same resolution and wavelength range as 
the wide-slit exposure according to the first step. Third, we smooth the wide-slit observation
and wide-slit-like narrow slit spectrum, respectively, with a boxcar size of 500 pixels
to avoid possible noise features. The final compensation curve is given by the ratio
of smoothed versions of the wide- and narrow-slit observations. Once the compensation
curve is derived, we multiply it to the original narrow-slit spectrum to perform the 
flux calibration.



\section{Quality checks}

\subsection{Spectral resolution}
We now determine the line-spread function (LSF) of our X-shooter spectra to 
accurately measure the spectral resolution and confirm the wavelength 
calibration of our observations. 
We fit the spectra of our F, G, and K stars (212 spectra) 
using the synthetic library of \cite{Coelho05} as templates. 
To determine the LSF, we use the function \texttt{ULY\_LSF} from ULySS,
which minimizes the difference between the observed spectra and a parametric 
model by full-spectrum fitting \citep{Koleva08, Koleva09}.
The fit is performed in wavelength intervals of 200 \AA\ spaced by 60 \AA. 

\begin{figure}
  \centering
    \subfloat{\includegraphics[angle=90,scale=0.34]{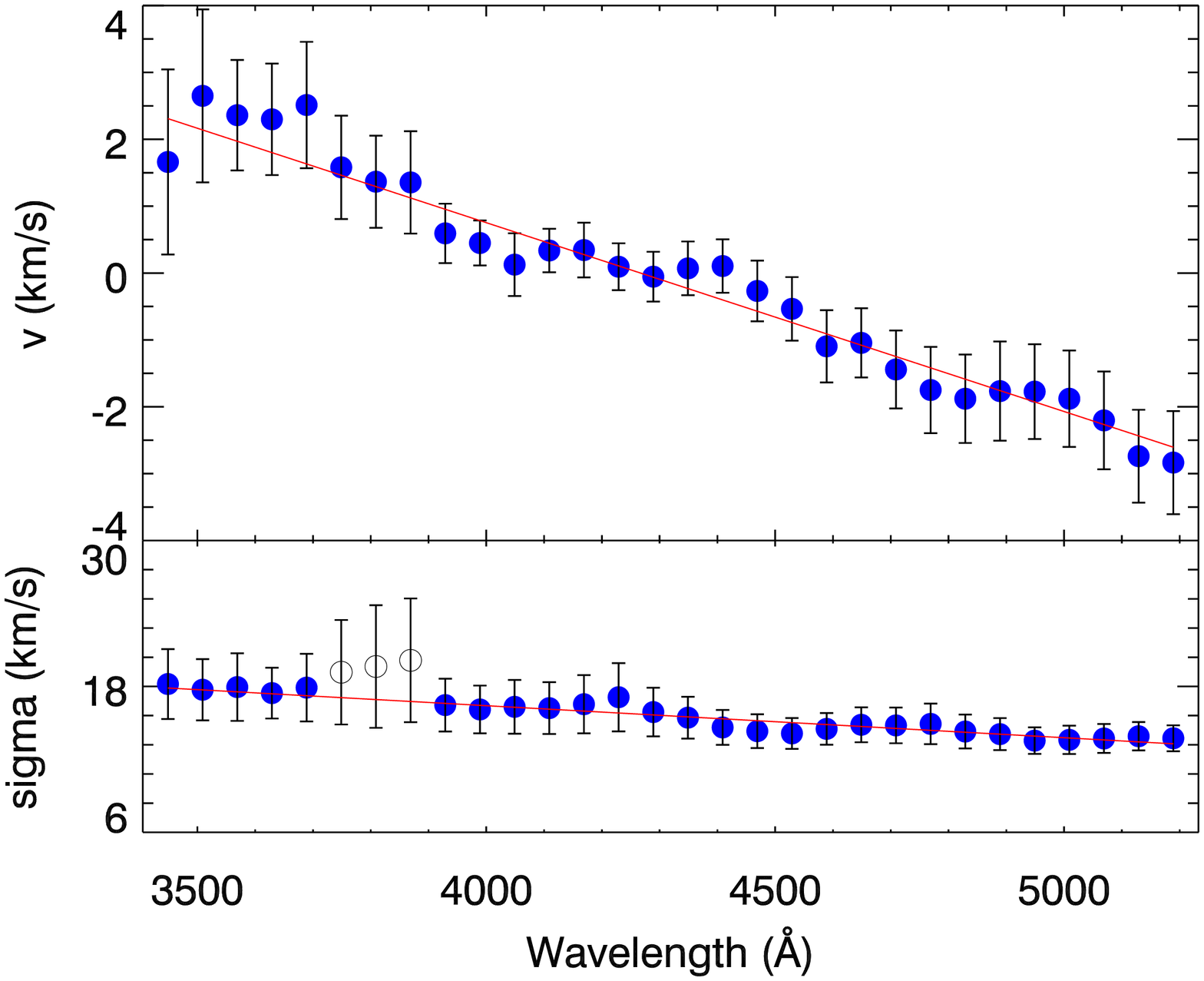}}\quad \\
    \subfloat{\includegraphics[angle=90,scale=0.34]{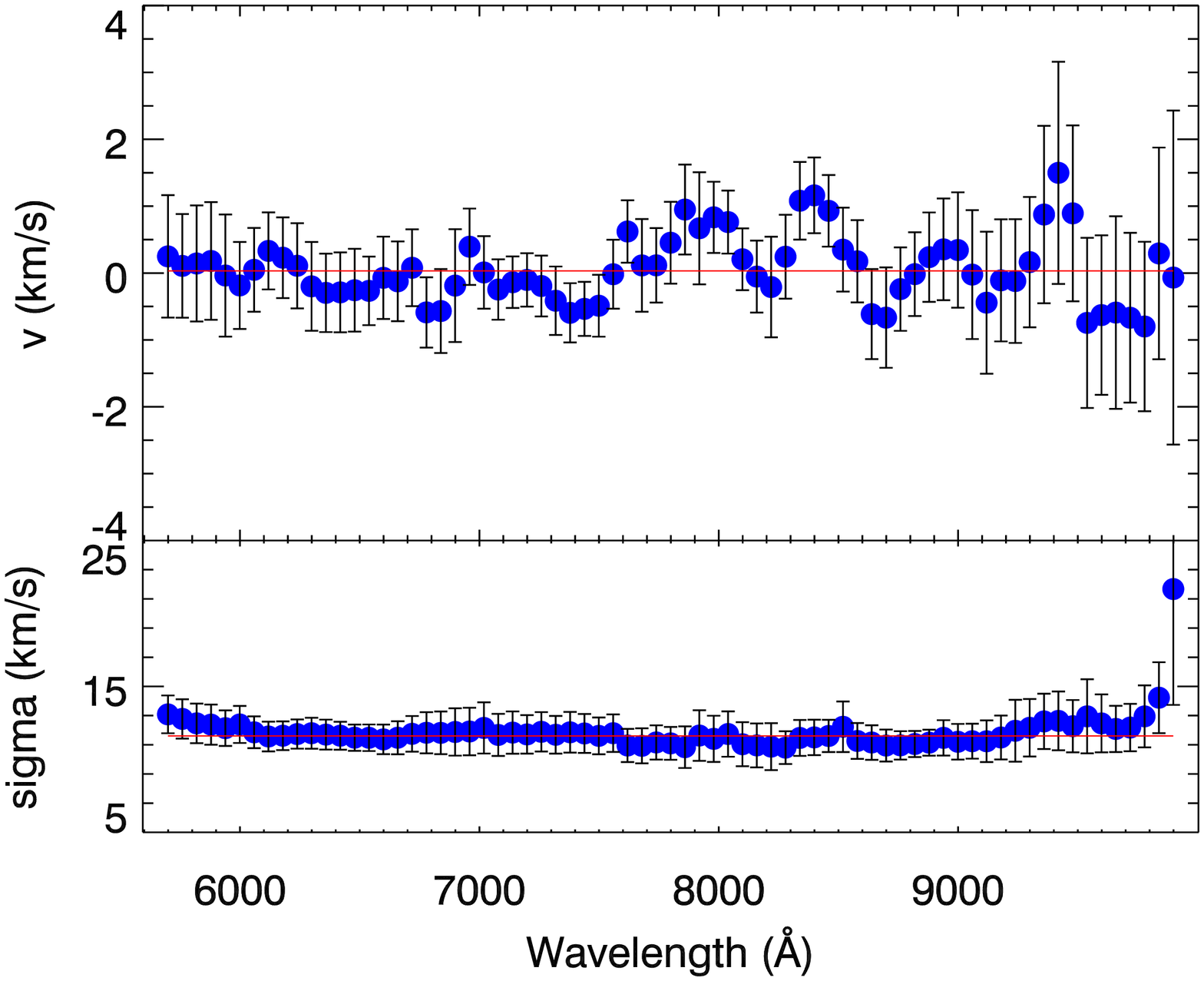}}\quad \\
   
   \caption{Average line-spread functions for the FGK stars (212 spectra) in the UVB
          (upper panel) and VIS (lower panel) arms. In each arm, the top panel shows the residual shift
	  of the spectra (blue dots). The bottom panel shows the detected instrumental velocity dispersion.
	  A simple first-order polynomial fit (red lines) of the line-spread function is marked 
	  in each panels. The error bars are the standard deviation in each wavelength bin.
          The three open circles show the region where two different flat field lamps are used in
	  the UVB arm, which may result in larger uncertainties.}
	  
   \label{lsf}
\end{figure}

Figure~\ref{lsf} shows the LSF for the F, G, and K
stars from XSL in the UVB (upper panels) and VIS (lower panels) arm, respectively.
In each arm, we estimate the mean instrumental velocity dispersion ($\sigma$) and
residual shift ($v$) using the IDL procedure \texttt{BIWEIGHT\_MEAN}. 
The mean difference of the residual velocity in the UVB arm spans the range between
$-2.6$ and $+2.3$ \kms. This may be due to the imperfect wavelength calibration. 
We find that the wavelength solution is very good in the VIS arm (upper panel in 
the lower plot). 

The instrumental velocity dispersion in the UVB arm ranges from $13.3$ to 
$18.1$ \kms, corresponding to a resolution $R = 9584 - 7033$. 
The fitted instrumental resolution in \kms\ is given by
\begin{equation}
\sigma_{\rm{UVB}}=15.625 - 0.0026\times(\lambda - 4300)\ (\kms)
\end{equation}
with $\lambda$ in \AA.
In the VIS arm, the instrumental variation is constant at $\sigma_{\rm{VIS}} = 11.62$ \kms
 (i.e., $R = 10986$), which is very close to the stated resolution $R = 11000$.

\subsection{Arm combination}
After carefully checking the resolution and flux calibration of our sample, we 
shift every spectra to zero velocity using the synthetic libraries of \cite{Coelho05, 
Palacios10}, and \cite{Allard11} as templates. 
The second wavelength calibration in the UVB arm is carefully performed 
to correct the small shifts shown in the upper panel of Figure~\ref{lsf}.
Multiple observations of the same non-variable star are combined
in the same arm. Multiple observations of the same variable star are kept
as different spectra. We merge multiple UVB and VIS arm spectra of each star 
into a single spectrum using the overlapped region, mostly from 5420 to 5650 \AA.
Some of the spectra have the dichroic features slightly shifted in wavelength, 
and in this case, we shift the overlapped regions accordingly.
The merged spectra of each star are carefully checked and evaluated to maintain
the appropriate spectral shape.

\subsection{Uncertainty of the PCA telluric correction}
\begin{figure}
   \includegraphics[angle=90,scale=0.42]
   {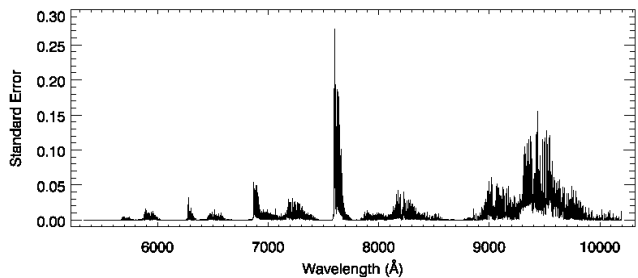}
   \caption{Root-mean-squared deviation of the ratio between the 
PCA-based and closest-in-time telluric corrections for 20-F stars from
XSL.}
   \label{pca-uncertainty}
\end{figure}

We have computed the root-mean-squared deviation of the ratio between the 
PCA-based and closest-in-time telluric corrections for 20 F stars from the sample.  
We plot this ratio in Figure~\ref{pca-uncertainty}.  We suggest that this is the maximum 
value of uncertainty that is likely
in the PCA-based correction for any given warm star, as the 
closest-in-time telluric correction could be very different than the  true telluric 
correction, if the closest-in-time calibrator in terms of the timescale of atmospheric changes in 
the molecular absorption lines was taken at a slightly different airmass 
at a time that is significantly after the program star.


\subsection{Photometric comparison}
We have calculated synthetic colors on the Johnson-Cousins $UVBRI$ and Sloan Digital
Sky Survey (SDSS) systems for our XSL stars and compared them with published values 
to check the reliability of our flux calibration.

\begin{figure}
   \includegraphics[angle=0,scale=0.5]
   {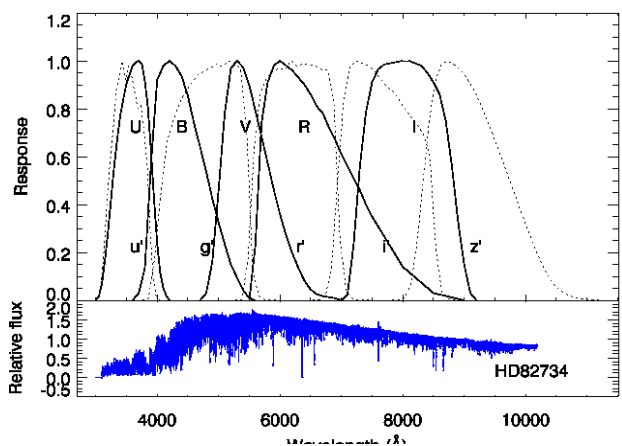}
   \caption{Upper panel: response functions of $UBVRI$ filters (solid lines) and the Monitor Telescope 
    $u^{\prime}$,$g^{\prime}$,$r^{\prime}$,$i^{\prime}$,$z^{\prime}$ filters (dotted lines).
    Both systems are normalized to 1. Bottom panel: a typical XSL spectrum of HD82734 (K0III)
    for comparison.}
   \label{filter}
\end{figure}

A number of studies have discussed the response functions required to reproduce the 
standard Johnson-Cousins $UVBRI$ photometric system \citep{Johnson66,Cousins71,Cousins73,
Landolt73, Landolt83, Bessell90}. Those which reproduce the observations most 
accurately are likely to be 
the work by \cite{Bessell90} and \cite{Fukugita95}. We therefore use the response functions
from \cite{Bessell90} for the $UVBRI$ system. The SDSS response 
functions \citep{Fukugita96} without the atmosphere
are adopted here to calculate the colors through the $u^{\prime}$ $g^{\prime}$ $r^{\prime}$ 
$i^{\prime}$ $z^{\prime}$ filters. We show the normalized $UBVRI$ (solid lines) and SDSS
(dotted lines) response functions in Figure~\ref{filter}. 

We have calculated the synthetic $B-V$, $U-B$, $R-I$, and $V-I$ colors of the XSL sample.
The synthetic magnitude of a filter $X$ in the Johnson-Cousins system is given by
\begin{equation}
 M_{X} = - 2.5\times  log_{10} \bigg[\frac{\int{\lambda f_{\lambda}(\lambda) R_{X}(\lambda) d \lambda}}
         {\int \lambda f^{\rm Vega}_{\lambda}(\lambda) R_{X}(\lambda) d \lambda}\bigg] + C_{X} , 
\end{equation} 
where $X$ can be any of the filters $UBVRI$, $R_{X}(\lambda)$ is the response function
of the filter $X$, $C_{X}$ is the $X$ magnitude of Vega, and 
$f_{\lambda}$ and $f^{\rm Vega}_{\lambda}(\lambda)$ are the flux densities 
of the object and Vega, respectively.

\begin{figure*}
\centering
   \subfloat{\label{ubngsl}\includegraphics[angle=0,scale=0.45]{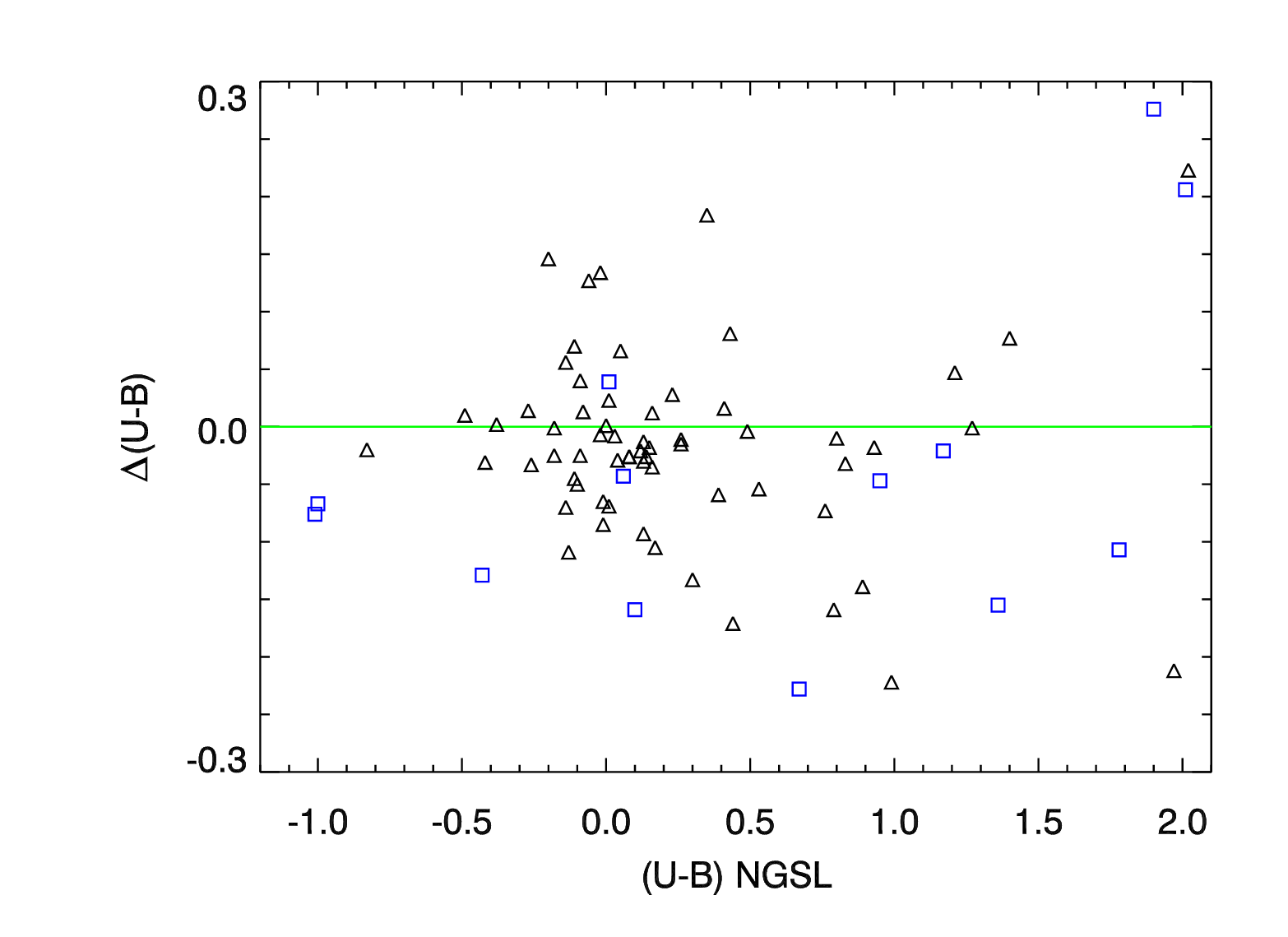}} \hspace*{-0.06\textwidth}               
    \subfloat{\label{bvngsl}\includegraphics[angle=0,scale=0.45]{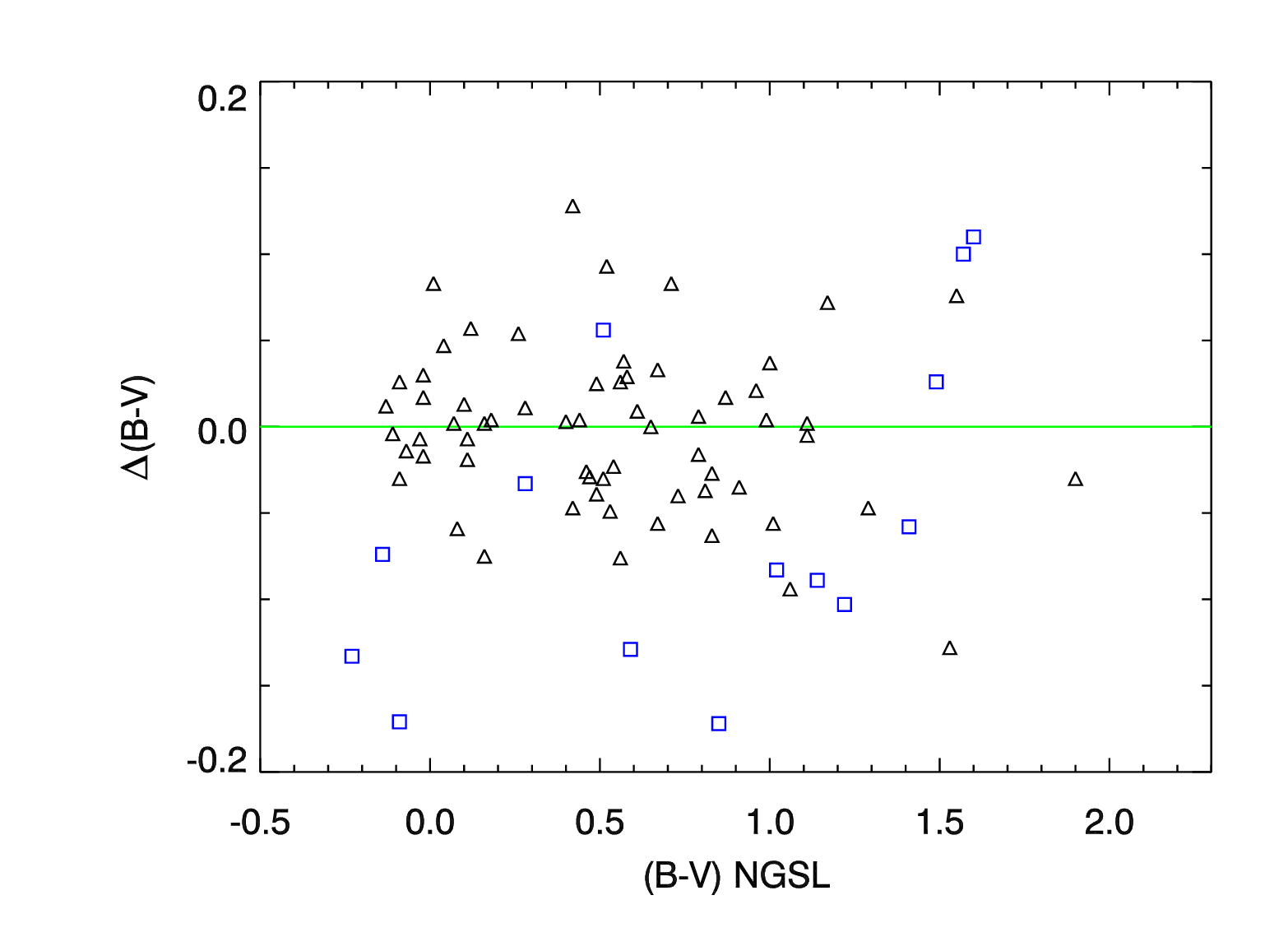}} \\                
   \subfloat{\label{ringsl}\includegraphics[angle=0,scale=0.45]{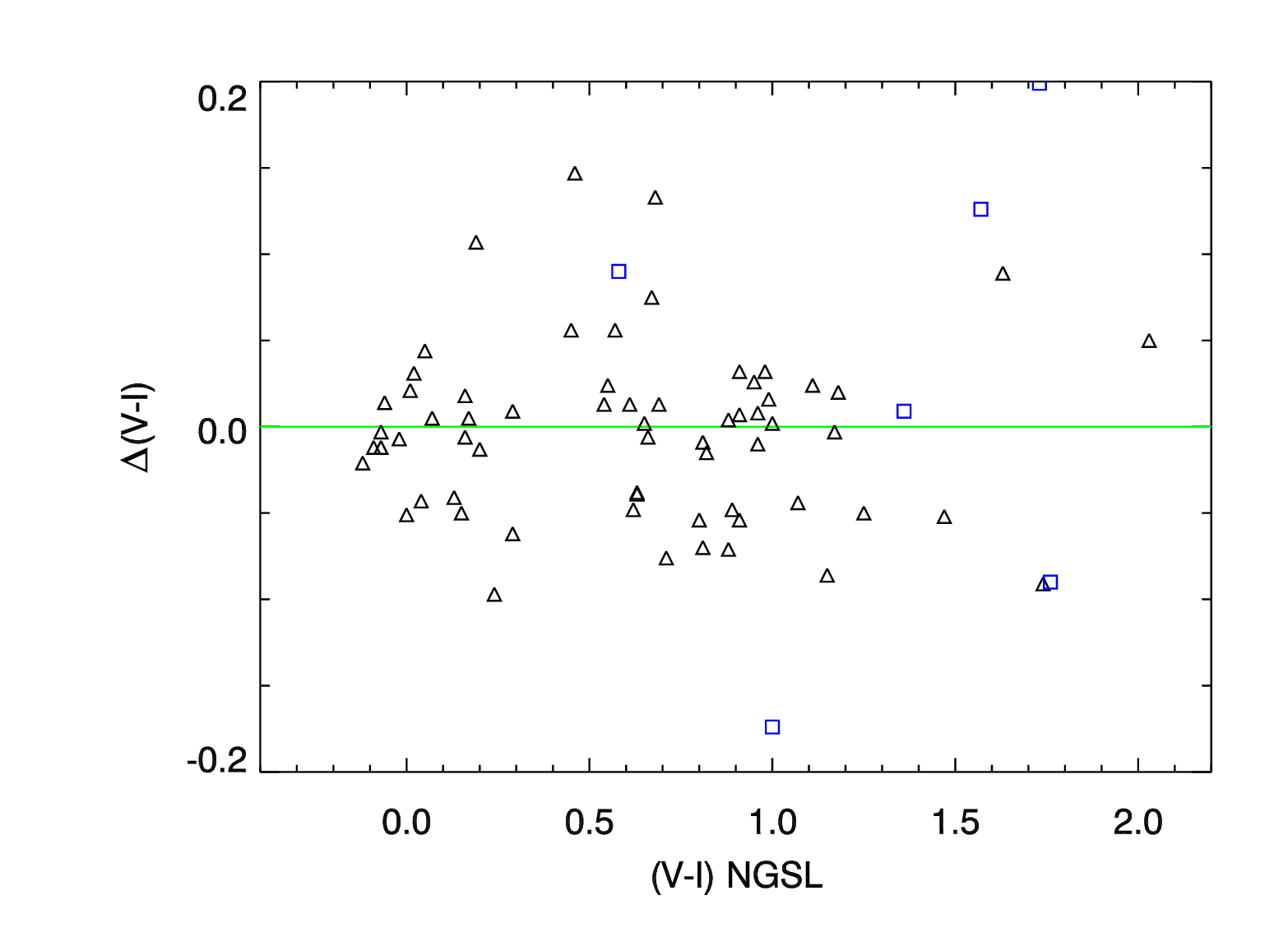}} \hspace*{-0.06\textwidth}              
   \subfloat{\label{vingsl}\includegraphics[angle=0,scale=0.45]{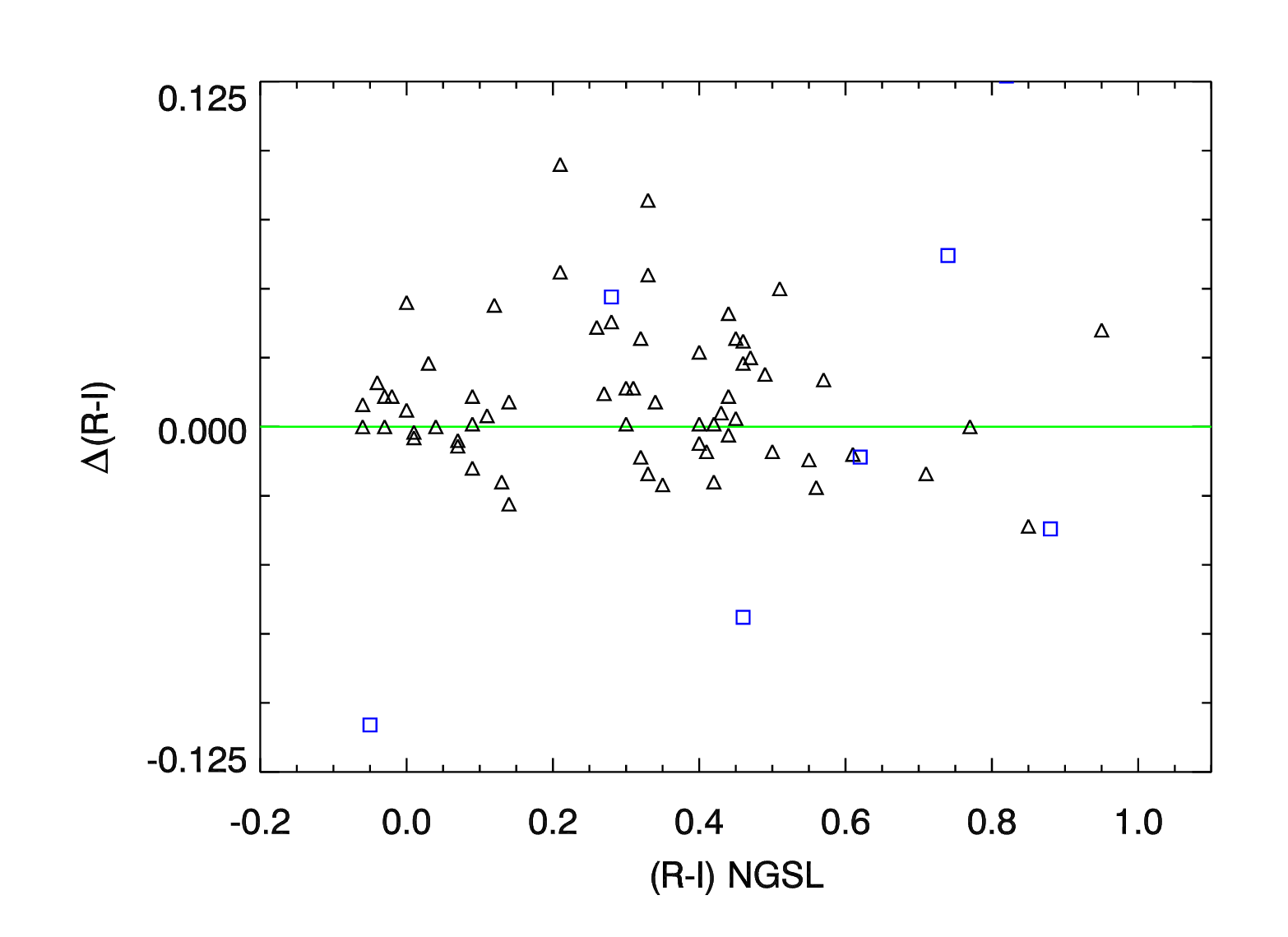}} \\                 
 \caption{Comparison of synthetic ($U-B$), ($B-V$), ($V-I$), and ($R-I$) colors in XSL and 
    NGSL.   The residuals in each panel 
    are calculated as $\rm color_{XSL} -color_{NGSL}$. Blue squares indicate those stars that likely 
    have flux losses, due to a lack of a useful wide-slit observation.}
  \label{ubvringsl}
\end{figure*}

\begin{table}
\caption{Mean color difference and rms scatter between XSL and NGSL.}
\centering
\label{sumngsl}
\begin{tabular}{*{5}{r}}\hline
     &  (U-B)& (B-V)& (R-I) & (V-I)\\\hline
XSL $-$ NGSL&$-$0.017  &  $-$0.002 &0.008&   $-$0.006\\  
\multicolumn{1}{c}{rms}    &0.071&  0.046 &  0.024&   0.049\\\hline   
\end{tabular}
\end{table}

To check the colors, we compare our synthetic colors with synthetic colors from
the NGSL library \citep{Gregg06} and observed colors from
the Bright Star Catalogue by \cite{Hoffleit83,Hoffleit91}. There are a total of 77 stars
in common with the NGSL library. 
Figure~\ref{ubvringsl} illustrates
the color comparison between NGSL and XSL. 
We calculate the mean offset and rms scatter for the
 ($U-B$), ($B-V$), ($V-I$), and ($R-I$) colors but avoiding stars
 which may have flux losses. The results are summarized in Table~\ref{sumngsl},
 where ``XSL - NGSL'' is the mean difference between colors of XSL and NGSL.
We find an agreement at level of 2.4--7.1\% between the XSL and NGSL synthetic colors.

\begin{figure}
  \centering
  \subfloat{\label{colorhoffi-1}\includegraphics[angle=0,scale=0.5]{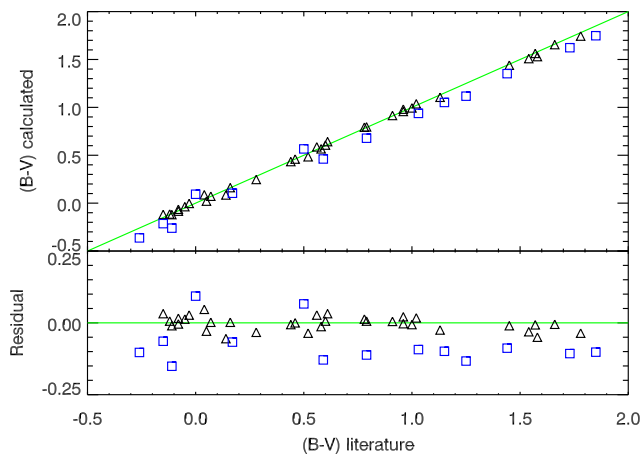}} \\               
  \subfloat{\label{colorhoffi-2}\includegraphics[angle=0,scale=0.5]{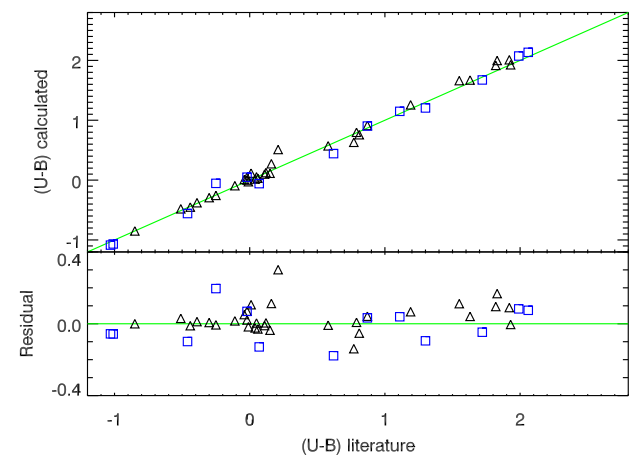}}\\
  \caption{Comparison of synthetic ($B-V$) and ($U-B$)  colors of XSL and 
    the Bright Star Catalogue. The residuals are calculated as in Figure~\ref{bvngsl}.
    Blue squares indicate those stars that likely have flux losses
    due to a lack of a useful wide-slit observation.}
  \label{colorhoffi}
\end{figure}

\begin{table}
\caption{Mean color difference and rms scatter between XSL 
and the Bright Star Catalogue (BSC).}
\centering
\label{sumbright}
\begin{tabular}{*{5}{r}}\hline
     & & (B-V)& & (U-B)\\\hline
XSL$-$BSC& & -0.024 &&   0.016\\  
\multicolumn{1}{c}{rms}   & &  0.058 & &  0.080\\\hline   
\end{tabular}
\end{table}

To further check the flux calibration, 
we also compare our colors 
with  $U-B$ and $B-V$ colors from the Bright Star Catalogue. There are 
in total 54 stars in common. We show the comparison in Figure~\ref{colorhoffi}.
The XSL stars with flux losses are marked as blue squares. The star, which has a
0.3 magnitude difference between literature and measured $U-B$, is the variable star
HD170756, which has a significant $U-B$ scatter in the literature values 
($\sim 0.5\ \mathrm{mag}$) as well. The mean offsets and rms scatter
measured from each color are summarized in Table~\ref{sumbright}.
The large scatter in the $U-B$ residual is partly due to variable stars 
and partly due to the low 
signal-to-noise region in the $U$ band. The outliers in the $B-V$ residual panel are mainly from the 
flux losses in the $B$ band in the UVB arm, as discussed in Sec.~\ref{abs:flx}.


\begin{figure}
   \includegraphics[angle=90,scale=0.6]
   {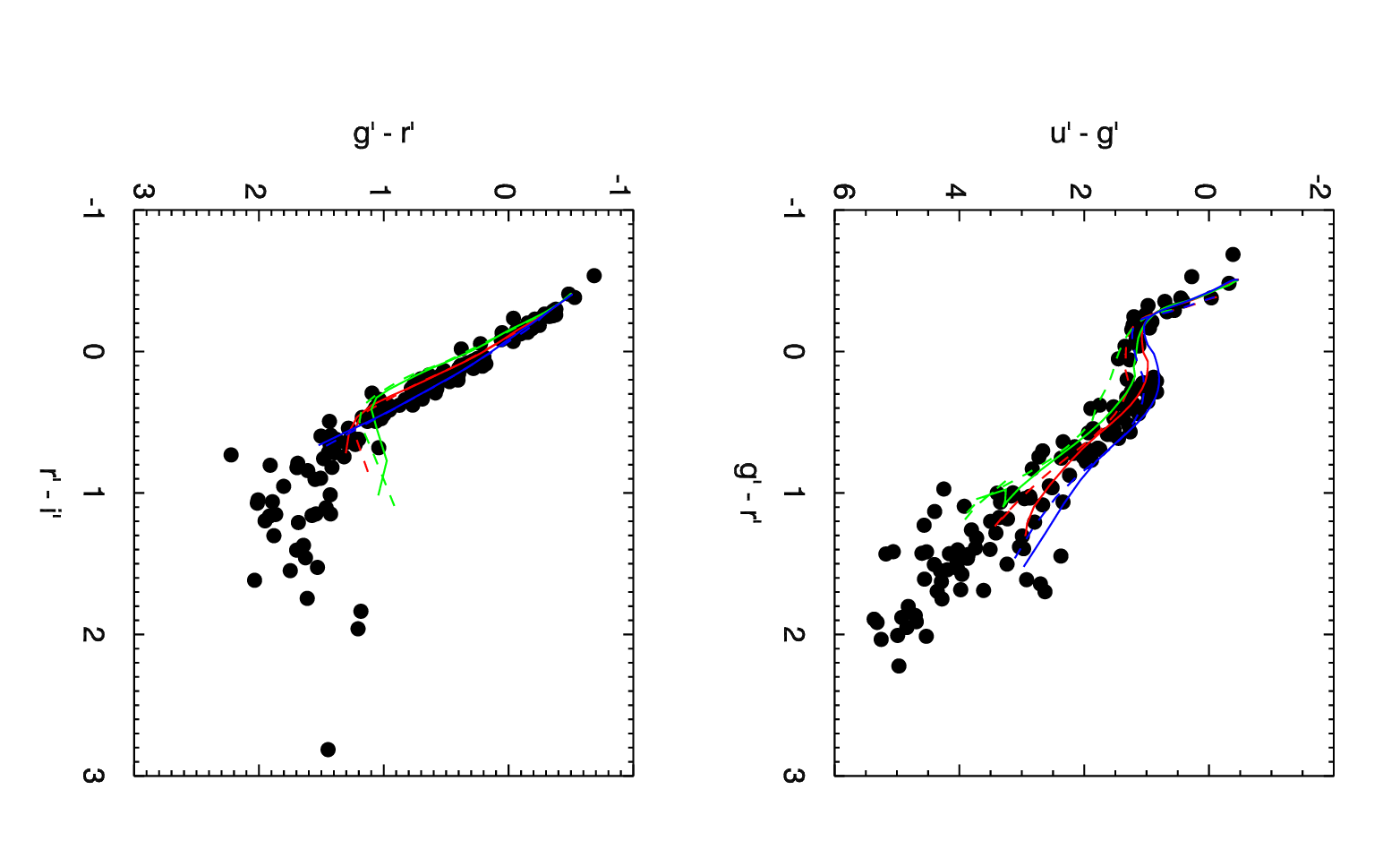}
   \caption{Synthetic SDSS colors of the XSL sample (black dots) compared with 
   synthetic model colors by \protect \cite{Lenz98}. Only three metallicities
   are shown in for the model colors: $\rm [M/H] = -2.0$ (blue lines), 
   0.0 (red lines), and  +1.0 (green lines). Solid lines indicate models with 
   $\log g = 4.5$ and dashed lines indicate models with $\log g = 2.5$. }
   \label{sdsscolor}
\end{figure}

The SDSS $u^{\prime} - g^{\prime}$, $g^{\prime} - r^{\prime}$, and 
$r^{\prime} - i^{\prime}$ colors 
are calculated following the definition of AB magnitudes: 
\begin{equation}
 M_{\rm AB} = - 2.5\times \log_{10} [ f_{\nu}^{\rm eff} ]- 48.6,
\end{equation} 
where $f_{\nu}^{\rm eff} = \frac{\int d{\nu} f_{\nu} R_{\nu}}{ \int d{\nu}R_{\nu} }$
\citep[see][for details]{Fukugita95}.
Since there are few literature stars, which have the 
SDSS colors in common with our sample, we use the model colors computed
by \cite{Lenz98} as a rough check. Models with $\log g = 4.5$ and $\log g = 2.5$
are chosen to represent the main sequence and giants, respectively. Three
 metallicities $\rm [M/H] = -2.0, 0$, and 1.0 are used to cover the metallicity
 range of the XSL sample. 
 Figure~\ref{sdsscolor} shows the  
$u^{\prime} - g^{\prime}$ vs $g^{\prime} - r^{\prime}$  colors and 
$r^{\prime} - i^{\prime}$ vs $g^{\prime} - r^{\prime}$ colors for the XSL 
sample  compared with the model colors. In general, the agreement between the 
models and data is very good. 
The coolest model from Lenz et al. has $T_\mathrm{eff}=3500\,\mathrm{K}$, 
which is significantly warmer than the coolest XSL stars, and so the reddest model 
colors are not be as red as the reddest XSL stars. 

\subsection{Comparison with literature spectra}
Since our sample is selected from different literature sources, it is 
interesting to compare the XSL spectra with spectra of the same stars
in other libraries. The first year 
of XSL has 77 stars in common with NGSL, 40 stars in common with MILES, 
34 stars in common with ELODIE, 26 stars in common with IRTF, and 25 
stars in common with CaT \citep{Cenarro01}.

\begin{sidewaysfigure*}
   \includegraphics[angle=0,scale=1.3]
   {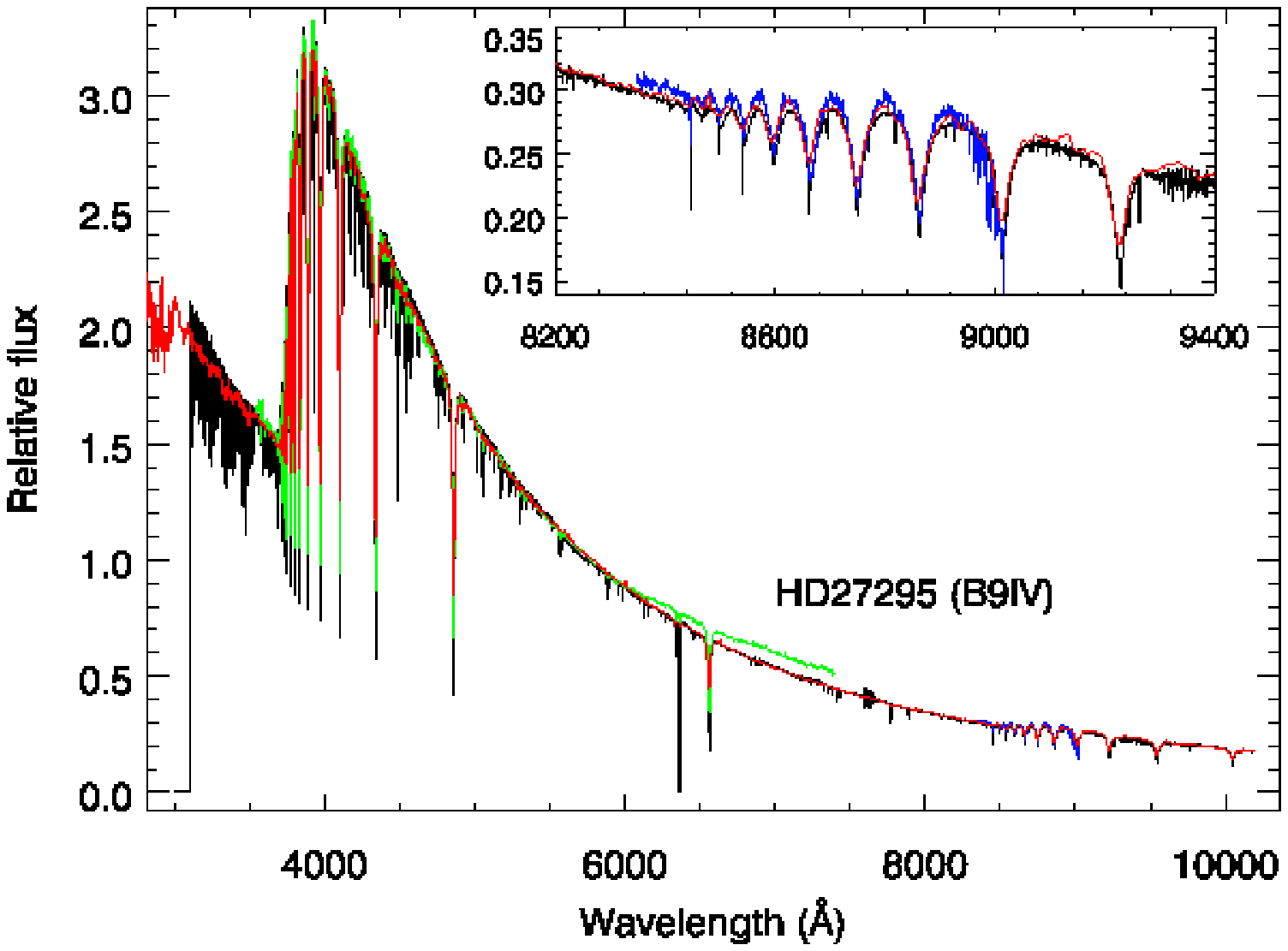}
   \caption{Spectra of HD27295 from XSL (black), MILES (green), NGSL (red), 
   and CaT (blue). The inset shows
   a zoomed-in region for a detailed comparison of the telluric corrections.
    }
   \label{hd27295}
\end{sidewaysfigure*}

\begin{sidewaysfigure*}
   \includegraphics[angle=0,scale=1.3]
   {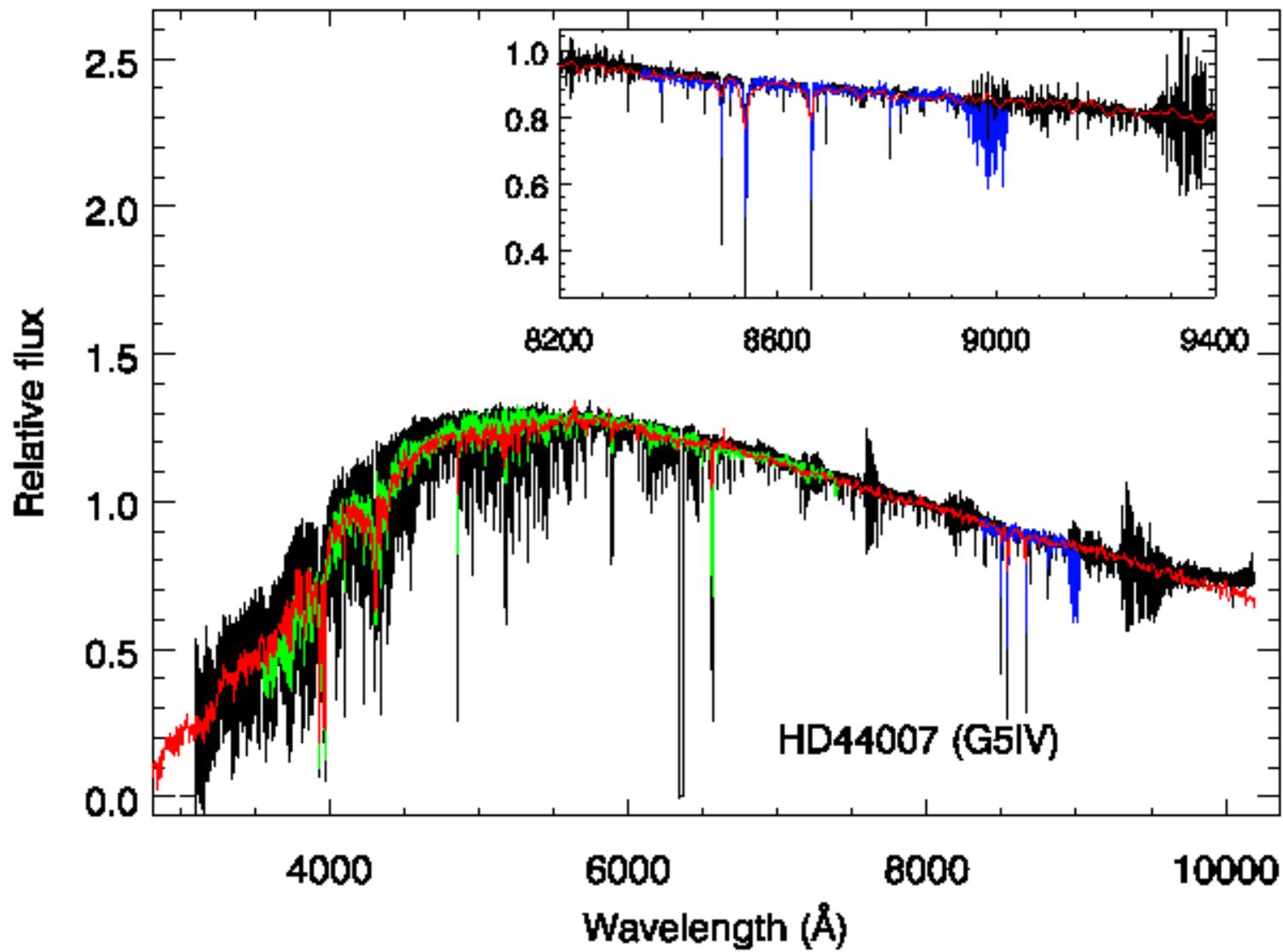}
   \caption{Spectra of HD44007 from different spectral libraries. Colors are 
   as in Figure~\ref{hd27295}. }
   \label{hd44007}
\end{sidewaysfigure*}

\begin{sidewaysfigure*}
   \includegraphics[angle=0,scale=1.3]
   {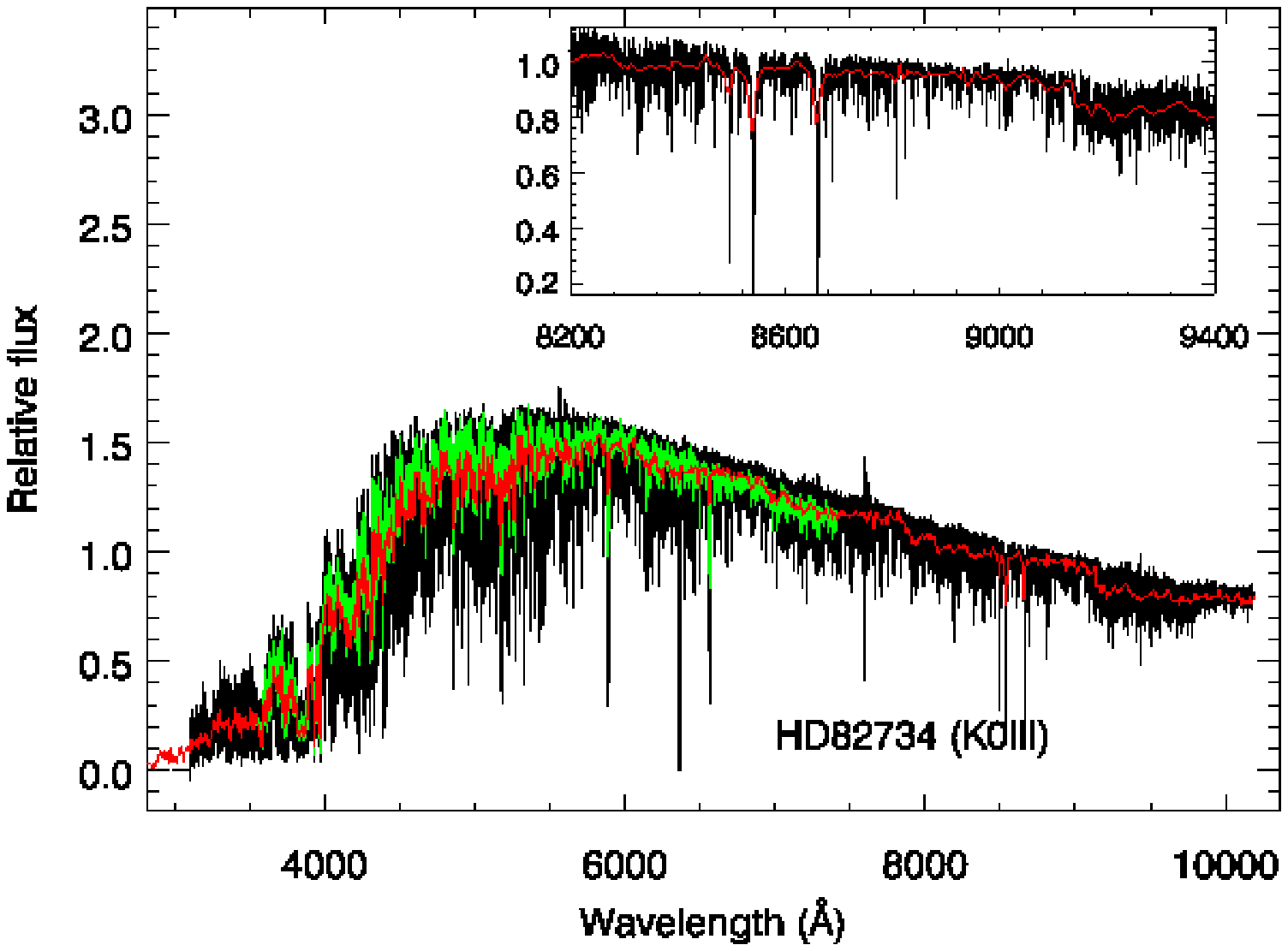}
   \caption{Spectra of HD82734 from different spectral libraries. Colors are 
   as in Figure~\ref{hd27295}.}
   \label{hd82734}
\end{sidewaysfigure*}

We show several examples of the spectral slope comparison between XSL, NGSL,
MILES and CaT in Figures~\ref{hd27295}, \ref{hd44007},  and \ref{hd82734}.
With the higher resolution data of XSL, we can resolve spectral features
in detail. We see that the flux calibration of 
XSL agrees well with MILES and NGSL in general, 
with the occasional exception of the very red part of the 
MILES spectra for HD27295 due to the second-order problem 
of some MILES stars \citep[see Figure~\ref{hd27295} and the discussion in Sec. 4.3 of ][]
{miles06}. 
Further, the comparison with the CaT library shows that library
is not telluric corrected \citep{Cenarro01},
as seen from the features around 9000 \AA\ in Figures~\ref{hd27295} and
\ref{hd44007}.

\begin{figure*}
\centering
   \includegraphics[angle=0,scale=1.0]
   {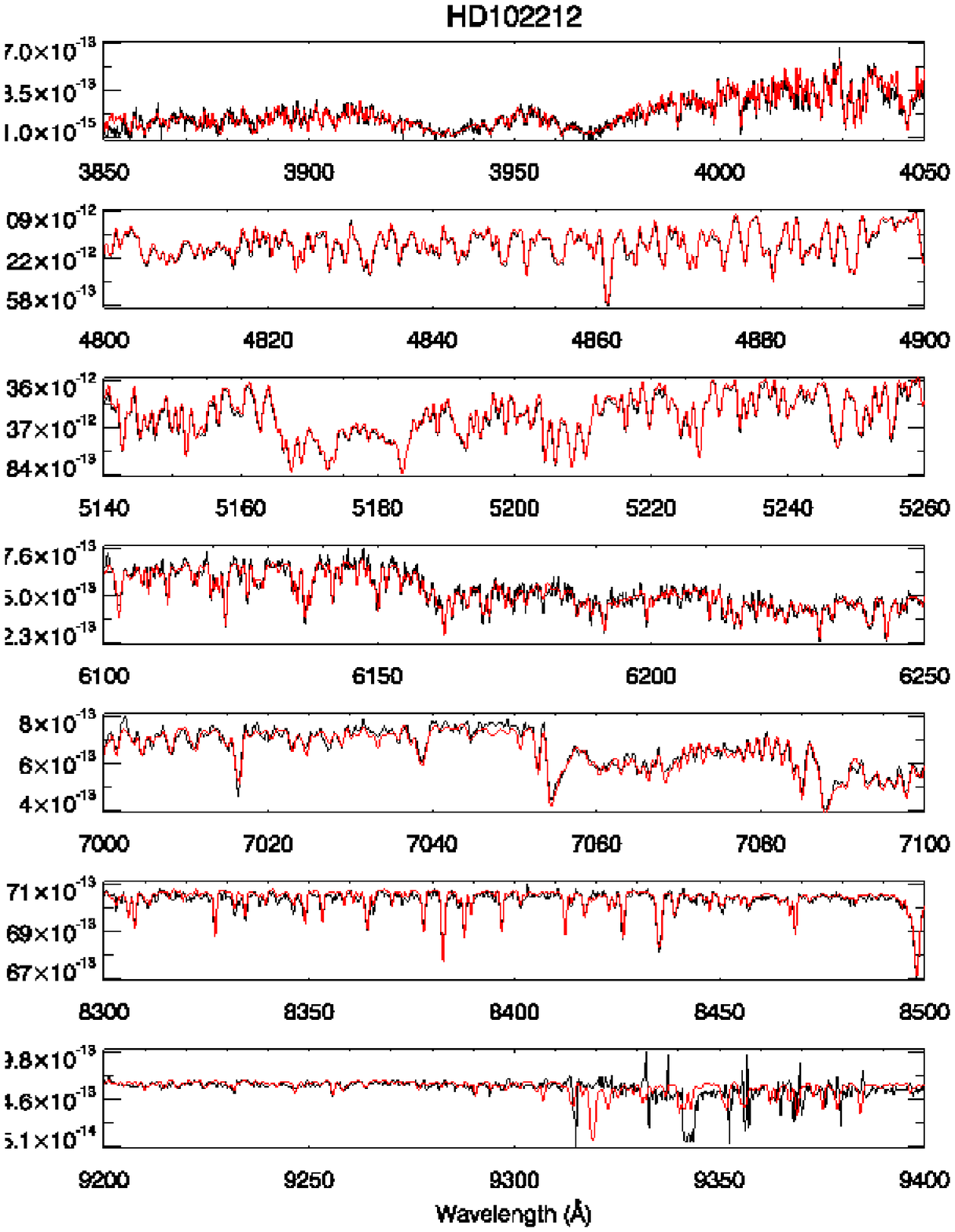}
   \caption{Detailed spectral line comparison between XSL (black) and UVES-POP (red)
       of HD102212 (M1III), where the UVES-POP spectrum is smoothed to the resolution
       of XSL. The XSL spectrum around 9350 \AA\ is heavily contaminated by telluric 
       features, and therefore the stellar features are difficult to recover.}
   \label{hd102212}
\end{figure*} 

\begin{figure*}
\centering
   \includegraphics[angle=0,scale=1.0]
   {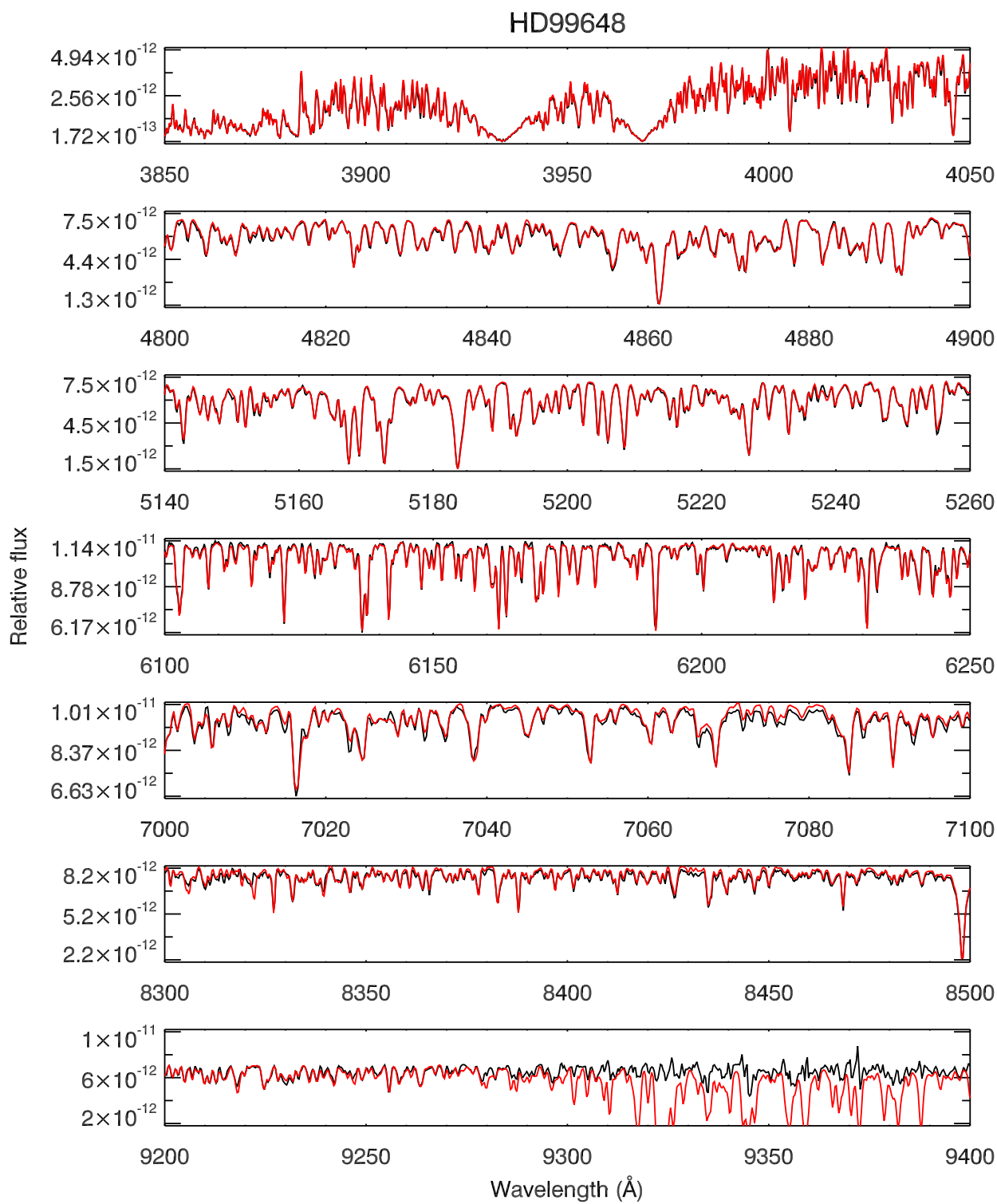}
   \caption{As for Figure~\ref{hd102212} but for 
       HD99648 (G8Iab).Telluric features are clearly seen in the UVES-POP 
       spectrum, especially around 9300--9400 \AA.}
   \label{hd99648}
\end{figure*} 

\begin{figure*}
\centering
   \includegraphics[angle=0,scale=1.0]
   {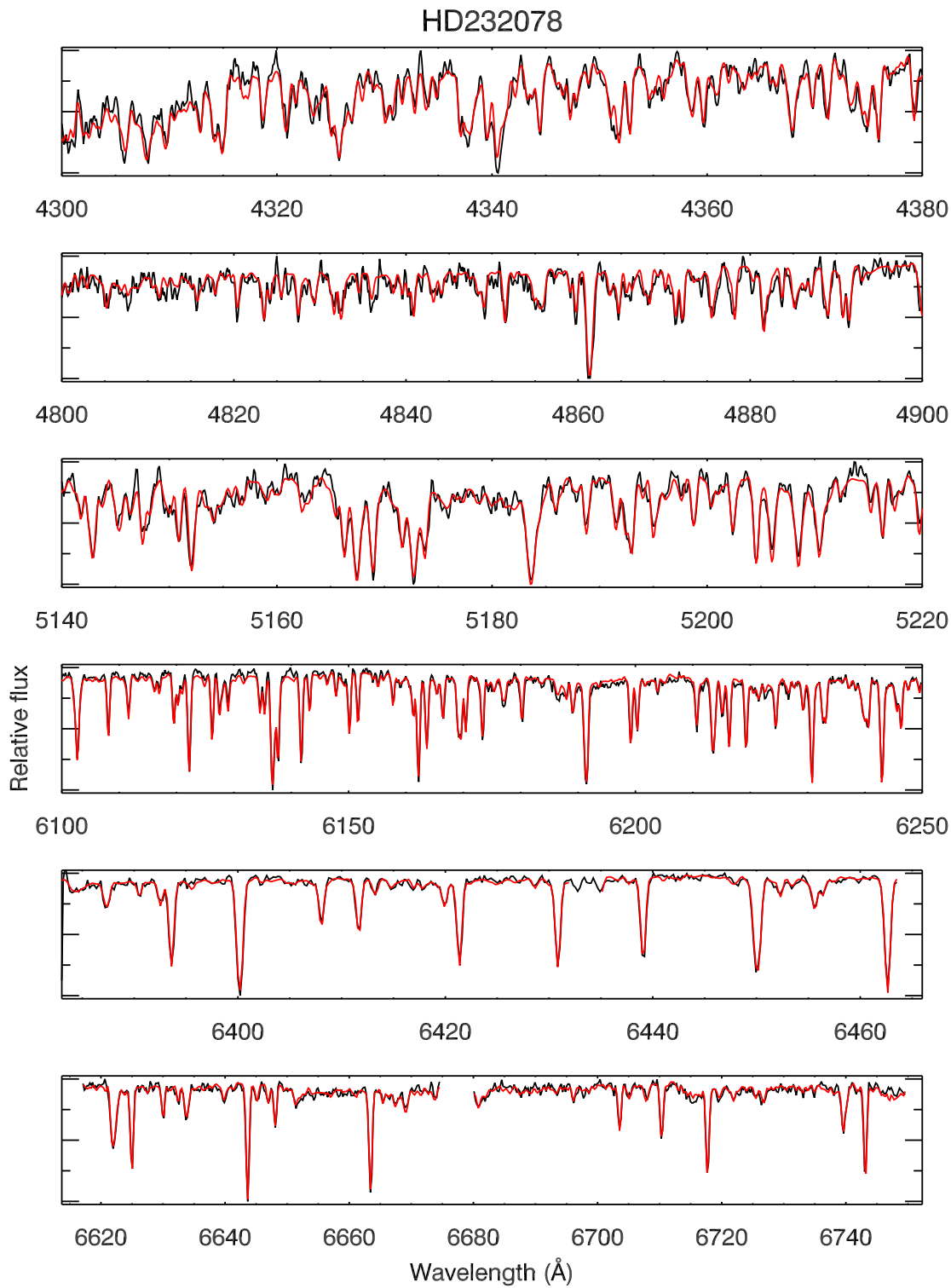}
   \caption{Detailed spectral line comparison between XSL (black) and ELODIE (red)
       of HD232078 (K3IIp), where the ELODIE spectrum is smoothed to the resolution
       of XSL. The gap in the two bottom panels represents bad pixels in ELODIE. }
   \label{hd232078}
\end{figure*} 

\begin{figure*}
\centering
   \includegraphics[angle=0,scale=1.0]
   {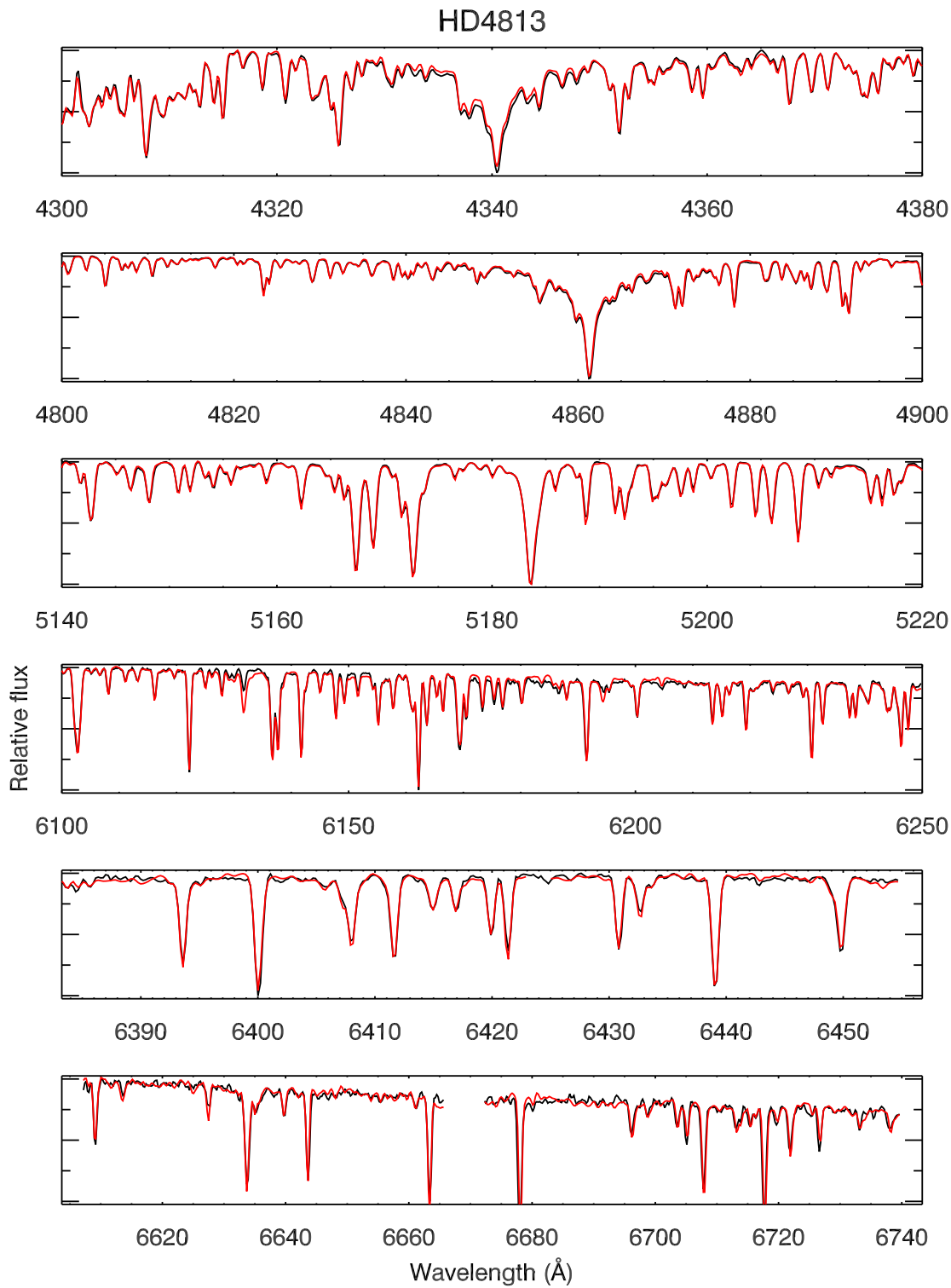}
   \caption{As for Figure~\ref{hd232078} but for HD4813 (F7IV-V). }
   \label{hd4813}
\end{figure*}

As a further check on the quality of XSL, we compare our spectra with
two higher-resolution spectral libraries: UVES-POP \citep{Bagnulo03}
and ELODIE\footnote{Note that we use the high-resolution version of
  the ELODIE library with $R\sim42000$ for this comparison.}
\citep{Prugniel01,Prugniel04,Prugniel07}. To make this comparison, we
first smooth the UVES-POP and ELODIE spectra to the resolution of the
XSL spectra and then use pPXF to match the continua and velocity zero
points, masking bad pixels when necessary.
Figures~\ref{hd102212}--\ref{hd4813} show the results of these comparisons.  
The typical residual between XSL and UVES-POP is
2--4\%; the typical residual between XSL and ELODIE is 2--6\%.  
We find very good agreement in the line shapes and
depths between XSL and the two higher-resolution libraries for both
warm and cool stars.  This gives us confidence that the XSL spectra
will be a useful basis for moderate-resolution studies of both stars
and composite stellar populations.

\begin{figure*}
\centering
   \includegraphics[angle=0,scale=1.0]
   {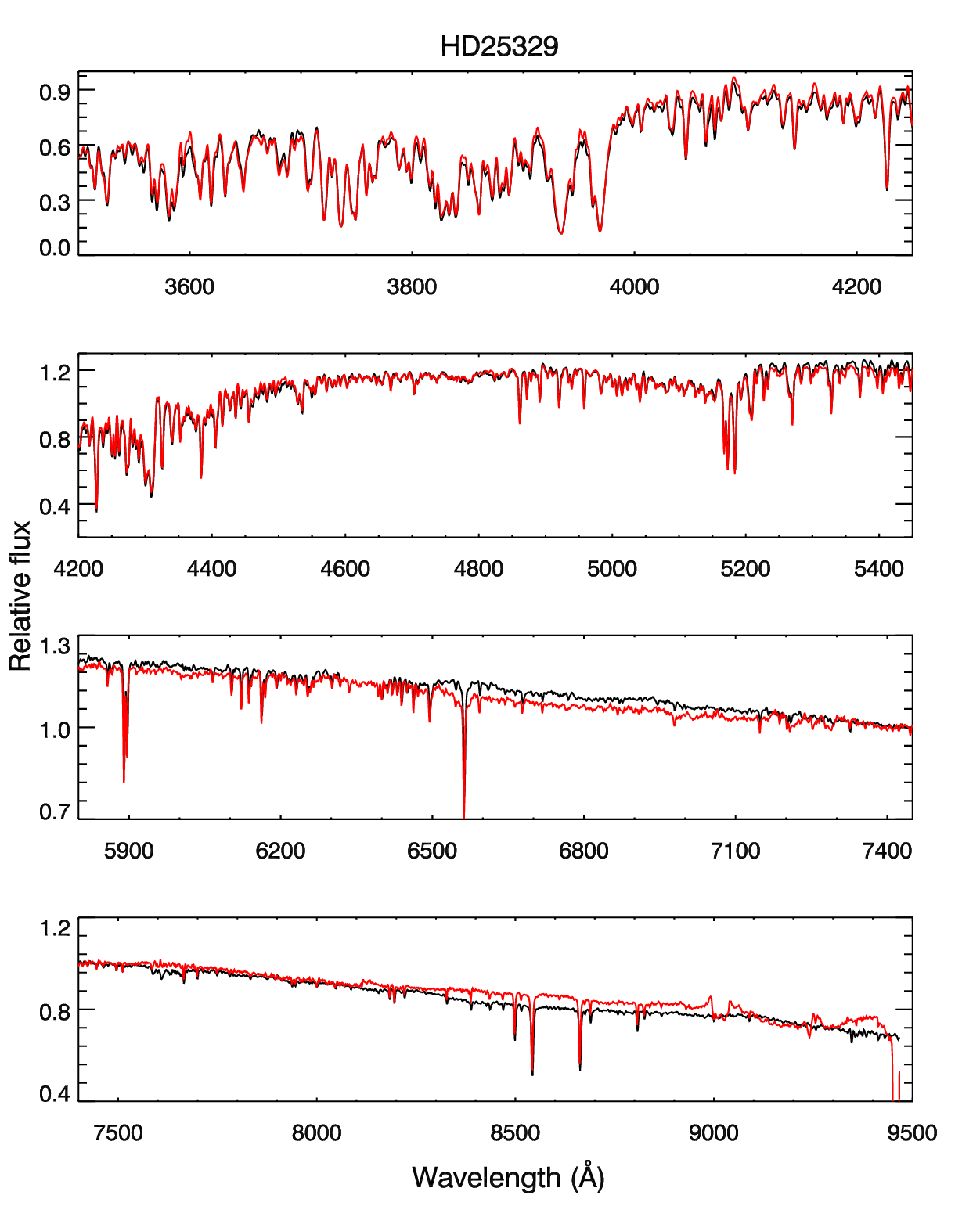}
   \caption{Detailed spectral line comparison between XSL (black) and MIUSCAT (red)
       of HD25329 (K1V), where the XSL spectrum is smoothed to the resolution
       of MIUSCAT. The gap in the third panel represents bad pixels in XSL. }
   \label{hd25329-zoom-mius}
\end{figure*} 

\begin{figure*}
\centering
   \includegraphics[angle=0,scale=1.0]
   {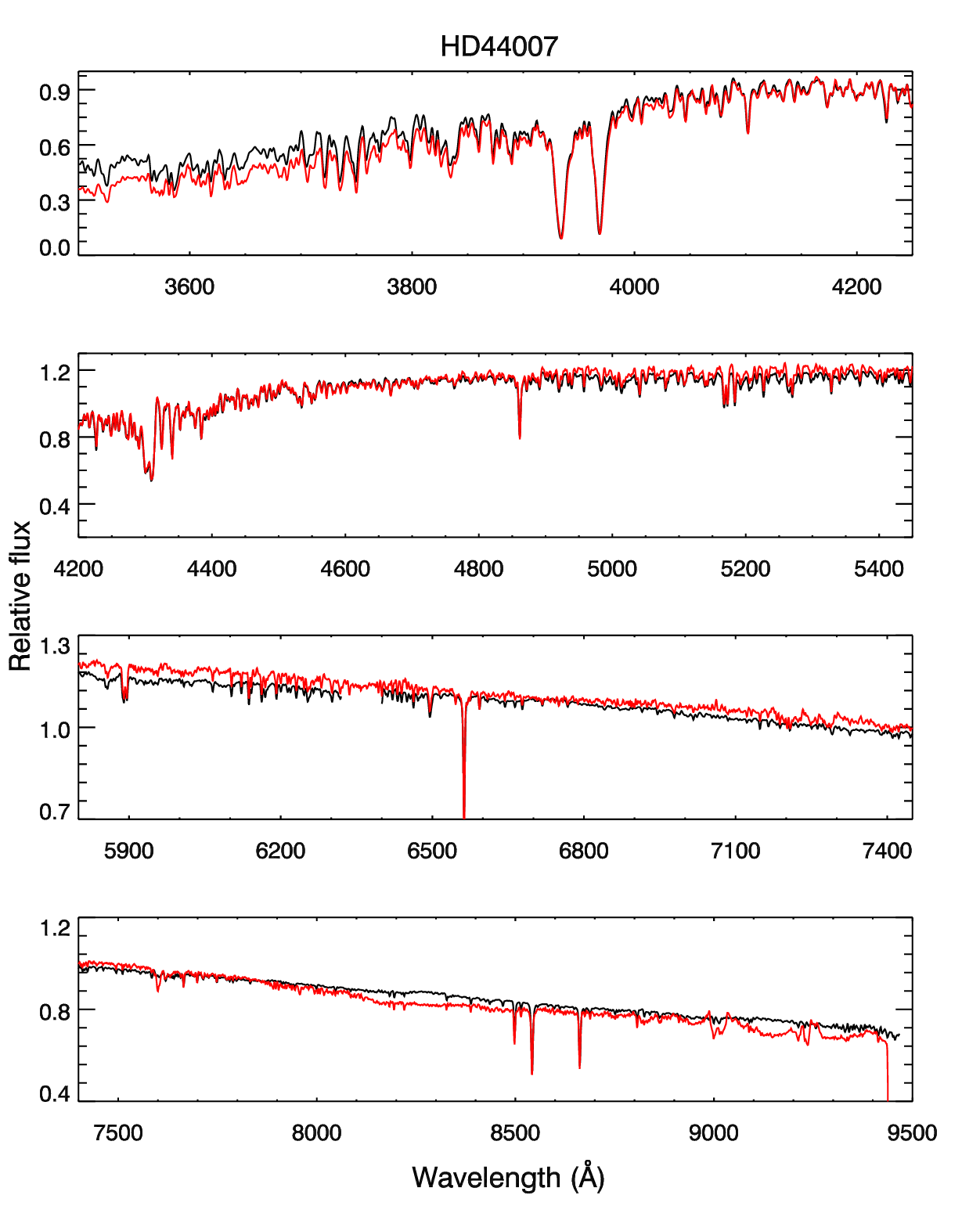}
   \caption{As for Figure~\ref{hd25329-zoom-mius} but for HD44007 (G5IV:w...).}
   \label{hd44007-zoom-mius}
\end{figure*}

\begin{figure*}
\centering
   \includegraphics[angle=0,scale=1.0]
   {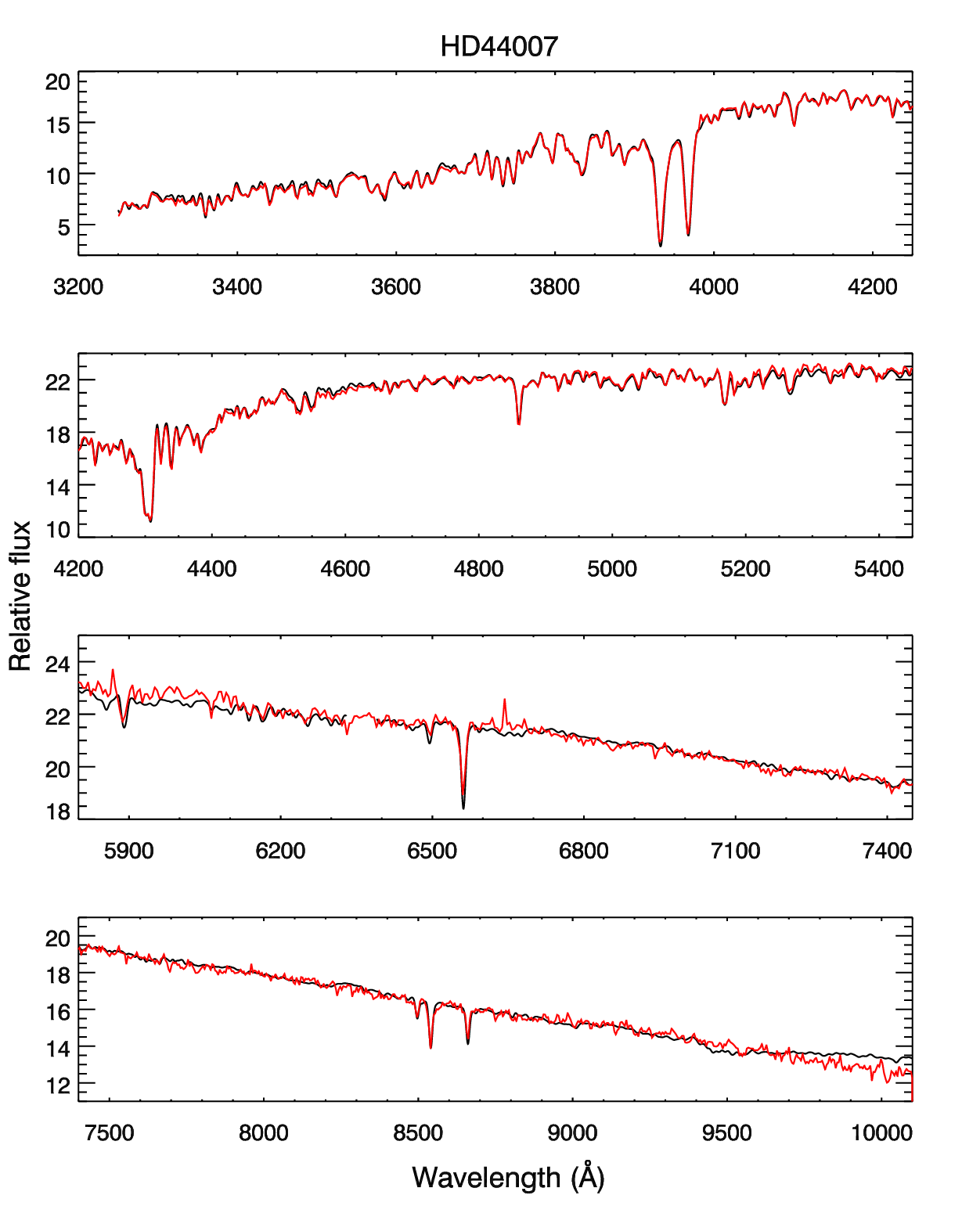}
   \caption{Detailed spectral line comparison between XSL (black) and NGSL (red)
       of HD44007 (G5IV:w...), where the XSL spectrum is smoothed to the resolution
       of NGSL. The gap in the third panel represents bad pixels in XSL. }
   \label{hd44007-zoom}
\end{figure*}

Comparison of XSL with the intermediate resolution spectral library MIUSCAT \citep{miuscat1} shows 
a good agreement for lines unaffected by telluric contamination. 
Figures~\ref{hd25329-zoom-mius} and \ref{hd44007-zoom-mius} 
show the smoothed version of XSL and MIUSCAT for two  stars in common, HD25329 
and HD44007, over the wavelength range 3500 \AA\ to 10000 \AA. 
The typical rms flux residual between XSL and MIUSCAT is $2\%$.
We note that the flux differences exist: for instance, a flux difference of $\sim7 \%$ over
the wavelength range 3600--4200 \AA\ is found in HD44007. This may be due to the extinction used
in MIUSCAT star, since comparison of XSL with NGSL for the same star shows excellent 
agreement (see Figure~\ref{hd44007-zoom}). Flux differences of $2.3\%$ are found in both stars
over the wavelength range 8000--9400 \AA, which likely arises from the joined libraries, which are CaT
and INDO-US in MIUSCAT. 
To join the MILES with CaT spectral ranges, \cite{miuscat1} used the $(V-I)$ color--temperature--metallicity relations of 
\cite{Alonso96,Alonso99}. This matching is performed on the basis of these generic relations 
\citep[see][]{miuscat1} according to the parameters of the stars. This might explain some of 
the difference in flux between the XSL and MIUSCAT in that specific spectral range,
where the matching of the different libraries in MIUSCAT may cause slight flux calibration issues.
Regions beyond $\sim$ 9000 \AA\
in MIUSCAT are contaminated by telluric absorption.

\begin{figure*}
\centering
   \includegraphics[angle=0,scale=1.0]
   {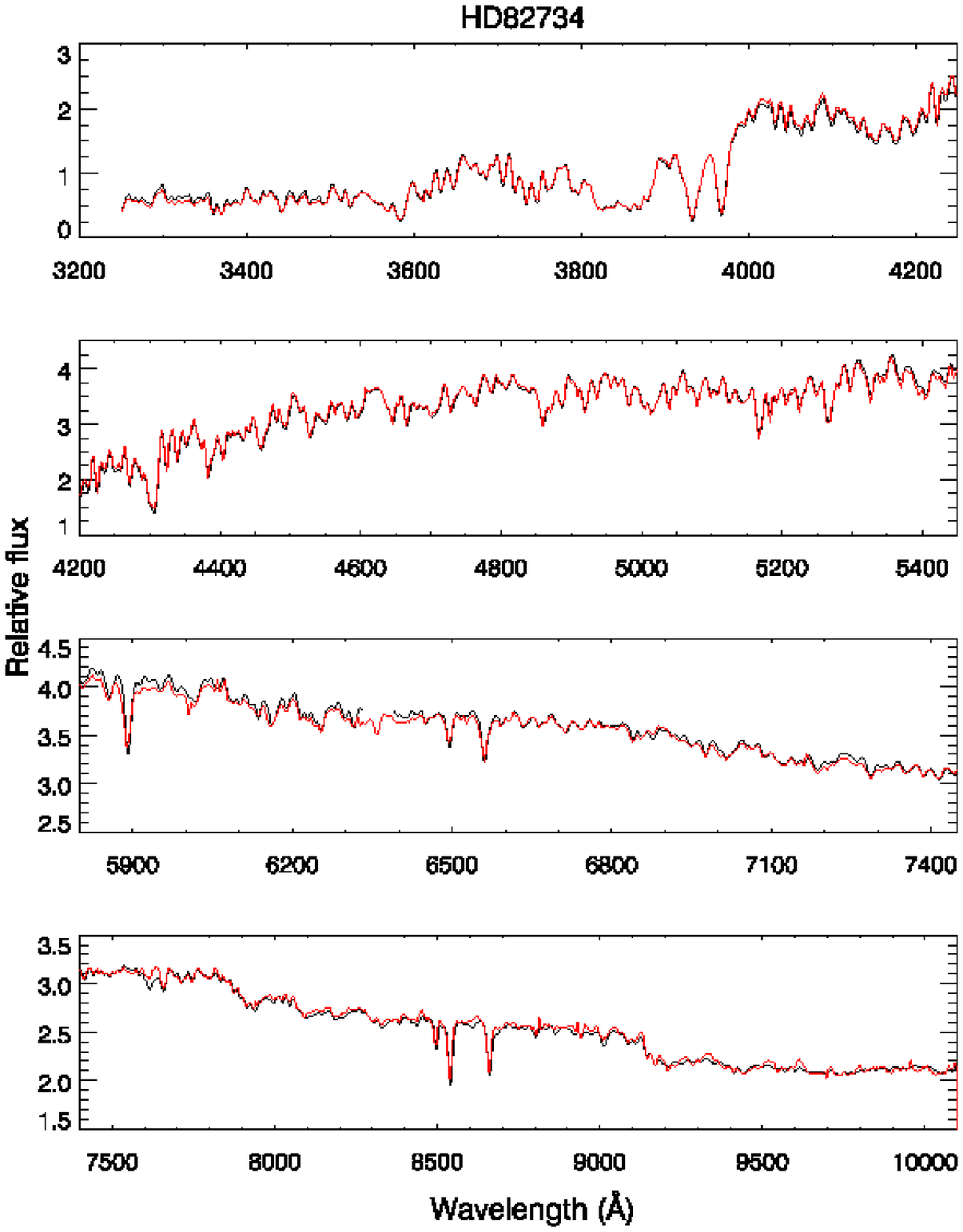}
   \caption{As for Figure~\ref{hd44007-zoom} but for HD82734 (K0III). }
   \label{hd82734-zoom}
\end{figure*} 

Comparison of XSL with the lower resolution spectral library NGSL also shows 
good agreement. Figures~\ref{hd44007-zoom} and \ref{hd82734-zoom} 
show the smoothed version of XSL and NGSL for two  stars in common HD44007 
and HD82734 at wavelength range from 3200 \AA\ to 10000 \AA. 
The typical rms flux residual between XSL and NGSL is $1\%$. 

\section{Summary}
We are building a new, moderate-resolution stellar spectral library, the X-shooter
Spectral Library (XSL). The pilot survey\footnote{\url{http://xsl.u-strasbg.fr/}}
(this work) contains 237 unique stars covering the spectral range
$\lambda\lambda$ 3100 -- 10185 \AA\ at a resolution $R \sim 10000$.
We have identified a number of issues with the X-shooter pipeline and 
presented our solutions. A telluric library is built for telluric 
correction of the XSL data using a PCA-based method. Flux and wavelength 
calibrations are carefully performed and are shown to be consistent 
with published spectra.
the X-shooter Spectral Library is still under construction, and 
the final database will contain more than 700 stars.
This library will provide a vital tool for extragalactic astronomers to
extract even more information from their observations than previously possible
and will provide stellar astronomers with a unique empirical panchromatic spectral
library for further studies of a wide range of stellar types.


\section*{Acknowledgments} 
We thank the referee, G. Worthey, for the careful review and 
helpful comments that improved the final manuscript.
We thank our collaborators on XSL, 
S. Meneses-Goytia, A.J. Cenarro, J. {Falc\'on-Barroso}, 
E. {M\'armol-Queralt\'o}, {P. S\'anchez-Bl\'azquez}, 
C.J. Walcher, P. Hauschildt and M. Koleva. We thank also B. Davies and G. Zhang for useful discussions.  We
would also like to extend our great thanks to V. Manieri,
A. Modigliani, J. Vernet, and the ESO staff for their help during the
XSL observations and reduction process.
This research used the POLLUX database
( \url{http://pollux.graal.univ-montp2.fr} )
operated at LUPM  (Universit\'e Montpellier II - CNRS, France
with the support of the PNPS and INSU.
This  work has been supported by the Programa  Nacional de Astronom{\'{\i}}a y
Astrof{\'{\i}}sica of MINECO,  under grant AYA2010-21322-C03-02.

\bibliographystyle{aa}    
\bibliography{biblio_yanping}    

\appendix

\section{Log of telluric standard stars}\label{tabeltel-log}
We present the observation log of telluric standard stars in the
VIS arm used to create the telluric library in Table~\ref{tabtell-a}. 

\clearpage
\onecolumn

\begin{center}
\begin{longtable}{l l r r l l }
\caption{The input stars for the VIS-arm telluric library.}\label{tabtell-a}\\

\hline
\hline

Name &RA (J2000.0)&DEC (J2000.0)& Exp. time (s)&Airmass& Sp. type \\

\hline                                                  
\endfirsthead
{{\bfseries \tablename\ \thetable{} -- Continued.}} \\
\hline
\hline

Name &RA (J2000.0)&DEC (J2000.0)& Exp. time (s)&Airmass& Sp. type \\

\hline                                                  
\endhead

Hip088947	&	18:09:22.50	&	$-$36:40:21.1	&	12.50	&	1.77	&	B1II	\\	
Hip089086	&	18:10:55.35	&	$-$33:48:00.2	&	12.50	&	2.51	&	B1V	\\	
Hip091038	&	18:34:15.85	&	$-$04:48:48.8	&	12.50	&	1.06	&	B1V	\\	
Hip026766	&	05:41:08.13	&	$-$03:37:57.2	&	12.50	&	1.14	&	B2	\\	
Hip047963	&	09:46:30.37	&	$-$44:45:18.2	&	1.25	&	1.39	&	B2.5IV	\\	
Hip030660	&	06:26:34.44	&	$-$04:35:50.6	&	1.25	&	1.37	&	B2.5V	\\	
Hip087314	&	17:50:28.39	&	$-$53:36:44.7	&	12.50	&	2.06	&	B2/B3Vnn	\\	
Hip087505	&	17:52:43.76	&	$-$38:38:33.5	&	12.50	&	1.45	&	B2II	\\	
Hip054082	&	11:04:00.23	&	$-$57:57:19.2	&	12.50	&	1.20	&	B2III	\\	
Hip039691	&	08:06:41.61	&	$-$48:29:50.6	&	1.25	&	1.26	&	B2IV	\\	
Hip092470	&	18:50:41.37	&	$-$47:46:47.1	&	1.25	&	1.63	&	B2IV	\\	
Hip055667	&	11:24:22.05	&	$-$42:40:08.8	&	12.50	&	1.06	&	B2IV-V	\\	
Hip007873	&	01:41:17.90	&	$-$75:39:49.1	&	1.25	&	1.81	&	B2V	\\	
Hip025751	&	05:29:54.77	&	01:47:21.3	&	1.25	&	1.51	&	B2V	\\	
Hip025869	&	05:31:20.89	&	$-$06:42:30.2	&	1.25	&	1.79	&	B2V	\\	
Hip072518	&	14:49:37.04	&	$-$68:56:10.3	&	12.50	&	1.80	&	B2V	\\	
Hip045044	&	09:10:28.57	&	$-$26:17:12.8	&	12.50	&	1.48	&	B3/B4V	\\	
Hip100556	&	20:23:26.26	&	$-$14:15:23.2	&	12.50	&	1.53	&	B3II/III	\\	
Hip088702	&	18:06:29.54	&	$-$58:34:15.4	&	12.50	&	2.34	&	B3III	\\	
Hip049076	&	10:00:58.59	&	$-$50:28:39.8	&	12.50	&	2.43	&	B3IV	\\	
Hip044509	&	09:04:05.78	&	$-$47:26:29.2	&	1.25	&	1.32	&	B3IV/V	\\	
Hip014898	&	03:12:15.38	&	$-$20:37:10.5	&	1.25	&	1.00	&	B3V	\\	
Hip021575	&	04:37:54.34	&	07:19:03.5	&	1.25	&	1.19	&	B3V	\\	
Hip026581	&	05:39:02.40	&	$-$05:11:40.1	&	5.00	&	1.07	&	B3V	\\	
Hip027937	&	05:54:41.12	&	$-$49:37:37.2	&	12.50	&	2.11	&	B3V	\\	
Hip041823	&	08:31:36.69	&	$-$45:47:05.2	&	1.25	&	1.66	&	B3V	\\	
Hip052977	&	10:50:13.00	&	$-$62:38:08.6	&	12.50	&	1.37	&	B3V	\\	
Hip067969	&	13:55:10.04	&	$-$68:52:55.4	&	1.25	&	1.99	&	B3V	\\	
Hip074110	&	15:08:45.23	&	$-$61:38:01.6	&	1.25	&	1.32	&	B3V	\\	
Hip108022	&	21:53:03.77	&	25:55:30.5	&	12.50	&	1.58	&	B3Ve	\\	
Hip108022	&	21:53:03.77	&	25:55:30.5	&	12.50	&	1.58	&	B3Ve	\\	
Hip087928	&	17:57:42.74	&	$-$56:53:46.5	&	12.50	&	2.16	&	B4III	\\	
Hip098960	&	20:05:32.03	&	$-$32:59:59.4	&	12.50	&	1.01	&	B4III	\\	
Hip093225	&	18:59:23.80	&	$-$12:50:25.9	&	12.50	&	1.21	&	B4V	\\	
Hip093225	&	18:59:23.80	&	$-$12:50:25.9	&	6.00	&	1.22	&	B4V	\\	
Hip028064	&	05:56:00.54	&	12:57:46.0	&	12.50	&	1.39	&	B5	\\	
Hip036235	&	07:27:42.84	&	$-$03:04:17.2	&	1.25	&	1.45	&	B5	\\	
Hip080371	&	16:24:21.32	&	$-$25:01:31.4	&	12.50	&	1.62	&	B5III	\\	
Hip089224	&	18:12:29.30	&	$-$39:20:32.3	&	12.50	&	1.58	&	B5III	\\	
Hip022840	&	04:54:50.71	&	00:28:01.8	&	1.25	&	1.38	&	B5V	\\	
Hip026405	&	05:37:14.52	&	$-$01:40:03.8	&	1.25	&	1.12	&	B5V	\\	
Hip026405	&	05:37:14.52	&	$-$01:40:03.8	&	1.25	&	1.44	&	B5V	\\	
Hip027303	&	05:47:04.58	&	$-$28:38:20.1	&	4.00	&	1.01	&	B5V	\\	
Hip045145	&	09:11:52.91	&	$-$45:14:05.2	&	5.00	&	1.72	&	B5V	\\	
Hip047495	&	09:40:56.05	&	$-$54:59:00.3	&	12.50	&	1.25	&	B5V	\\	
Hip051934	&	10:36:32.78	&	$-$59:11:25.2	&	12.50	&	1.31	&	B5V	\\	
Hip054006	&	11:02:54.76	&	$-$72:14:32.0	&	12.50	&	1.98	&	B5V	\\	
Hip085885	&	17:33:06.02	&	$-$33:03:28.0	&	15.00	&	1.04	&	B5V	\\	
Hip085885	&	17:33:06.02	&	$-$33:03:28.0	&	30.00	&	1.04	&	B5V	\\	
Hip093836	&	19:06:35.12	&	$-$01:20:46.1	&	6.00	&	1.38	&	B5V	\\	
Hip094378	&	19:12:34.54	&	$-$37:31:26.5	&	12.50	&	1.03	&	B5V	\\	
Hip094378	&	19:12:34.54	&	$-$37:31:26.5	&	12.50	&	1.20	&	B5V	\\	
Hip094378	&	19:12:34.54	&	$-$37:31:26.5	&	1.25	&	1.21	&	B5V	\\	
Hip096840	&	19:41:05.53	&	13:48:56.5	&	12.50	&	1.69	&	B5V	\\	
Hip096840	&	19:41:05.53	&	13:48:56.5	&	12.50	&	1.39	&	B5V	\\	
Hip096840	&	19:41:05.53	&	13:48:56.5	&	6.00	&	1.56	&	B5V	\\	
Hip105164	&	21:18:11.07	&	$-$04:31:10.1	&	1.25	&	1.13	&	B5V	\\	
Hip105164	&	21:18:11.07	&	$-$04:31:10.1	&	12.50	&	1.14	&	B5V	\\	
Hip105164	&	21:18:11.07	&	$-$04:31:10.1	&	12.50	&	1.07	&	B5V	\\	
Hip029464	&	06:12:29.16	&	$-$36:33:49.5	&	1.25	&	1.04	&	B5Vn	\\	
Hip009534	&	02:02:35.26	&	$-$21:57:56.8	&	1.25	&	1.01	&	B6V	\\	
Hip025903	&	05:31:43.84	&	$-$49:01:06.5	&	12.50	&	1.68	&	B6V	\\	
Hip027534	&	05:49:53.52	&	$-$66:54:04.3	&	5.00	&	1.54	&	B6V	\\	
Hip039640	&	08:06:03.23	&	$-$34:45:18.2	&	25.00	&	1.45	&	B6V	\\	
Hip089960	&	18:21:16.45	&	$-$30:56:29.3	&	12.50	&	1.01	&	B6V	\\	
Hip089960	&	18:21:16.45	&	$-$30:56:29.3	&	12.50	&	1.12	&	B6V	\\	
Hip088309	&	18:02:05.22	&	$-$35:40:09.9	&	12.50	&	2.31	&	B7II	\\	
Hip013327	&	02:51:29.59	&	15:04:55.5	&	12.50	&	2.28	&	B7V	\\	
Hip013327	&	02:51:29.59	&	15:04:55.5	&	1.25	&	1.30	&	B7V	\\	
Hip013327	&	02:51:29.59	&	15:04:55.5	&	1.25	&	1.34	&	B7V	\\	
Hip020789	&	04:27:17.45	&	22:59:46.8	&	0.70	&	1.63	&	B7V	\\	
Hip046848	&	09:32:48.00	&	$-$50:00:15.7	&	45.00	&	1.41	&	B7V	\\	
Hip063465	&	13:00:17.13	&	$-$66:50:51.9	&	1.25	&	1.40	&	B7V	\\	
Hip003541	&	00:45:12.83	&	23:35:26.8	&	1.25	&	1.50	&	B8	\\	
Hip015378	&	03:18:19.78	&	$-$09:43:53.6	&	1.25	&	1.39	&	B8	\\	
Hip024632	&	05:17:05.30	&	09:55:38.2	&	1.25	&	2.01	&	B8	\\	
Hip028697	&	06:03:33.93	&	26:31:44.9	&	1.25	&	1.98	&	B8	\\	
Hip086954	&	17:46:07.33	&	05:31:48.7	&	12.50	&	1.17	&	B8	\\	
Hip087290	&	17:50:09.30	&	00:40:44.3	&	20.00	&	1.56	&	B8	\\	
Hip092409	&	18:49:54.80	&	08:31:10.9	&	1.25	&	1.42	&	B8	\\	
Hip094526	&	19:14:19.24	&	25:00:47.6	&	1.25	&	1.80	&	B8	\\	
Hip025098	&	05:22:22.15	&	$-$56:08:03.8	&	1.25	&	1.62	&	B8.5V	\\	
Hip025098	&	05:22:22.15	&	$-$56:08:03.8	&	12.50	&	1.43	&	B8.5V	\\	
Hip038184	&	07:49:28.81	&	$-$13:21:10.6	&	12.50	&	1.22	&	B8/B9II	\\	
Hip102376	&	20:44:36.37	&	$-$71:14:01.9	&	1.25	&	1.66	&	B8/B9IV	\\	
Hip009022	&	01:56:09.41	&	$-$49:50:11.2	&	1.25	&	1.11	&	B8/B9V	\\	
Hip021599	&	04:38:13.21	&	$-$68:48:42.0	&	1.25	&	1.41	&	B8/B9V	\\	
Hip021599	&	04:38:13.21	&	$-$68:48:42.0	&	1.25	&	1.45	&	B8/B9V	\\	
Hip021599	&	04:38:13.21	&	$-$68:48:42.0	&	1.25	&	1.43	&	B8/B9V	\\	
Hip021599	&	04:38:13.21	&	$-$68:48:42.0	&	1.25	&	1.40	&	B8/B9V	\\	
Hip021599	&	04:38:13.21	&	$-$68:48:42.0	&	1.25	&	1.52	&	B8/B9V	\\	
Hip021599	&	04:38:13.21	&	$-$68:48:42.0	&	1.25	&	1.56	&	B8/B9V	\\	
Hip038584	&	07:54:02.21	&	$-$41:42:10.5	&	12.50	&	1.26	&	B8II	\\	
Hip060771	&	12:27:24.83	&	$-$63:47:20.4	&	12.50	&	1.29	&	B8IV	\\	
Hip092989	&	18:56:40.50	&	$-$37:20:35.7	&	7.00	&	1.03	&	B8IVs...	\\	
Hip092989	&	18:56:40.50	&	$-$37:20:35.7	&	7.00	&	1.03	&	B8IVs...	\\	
Hip092989	&	18:56:40.50	&	$-$37:20:35.7	&	12.50	&	1.03	&	B8IVs...	\\	
Hip092989	&	18:56:40.50	&	$-$37:20:35.7	&	7.00	&	1.05	&	B8IVs...	\\	
Hip008485	&	01:49:22.66	&	$-$72:24:42.7	&	1.25	&	1.57	&	B8V	\\	
Hip008485	&	01:49:22.66	&	$-$72:24:42.7	&	1.25	&	1.57	&	B8V	\\	
Hip012389	&	02:39:31.73	&	$-$64:16:54.2	&	1.25	&	1.66	&	B8V	\\	
Hip026545	&	05:38:43.53	&	$-$40:42:26.3	&	12.50	&	1.26	&	B8V	\\	
Hip032348	&	06:45:07.23	&	$-$31:39:15.5	&	1.25	&	1.06	&	B8V	\\	
Hip052160	&	10:39:22.83	&	$-$64:06:42.4	&	1.25	&	2.14	&	B8V	\\	
Hip057451	&	11:46:36.40	&	$-$64:45:57.5	&	12.50	&	2.05	&	B8V	\\	
Hip062448	&	12:47:53.64	&	$-$24:51:06.0	&	12.50	&	1.06	&	B8V	\\	
Hip066454	&	13:37:23.47	&	$-$46:25:40.4	&	1.25	&	1.27	&	B8V	\\	
Hip070243	&	14:22:19.72	&	$-$34:47:12.5	&	12.50	&	1.42	&	B8V	\\	
Hip101552	&	20:34:47.39	&	$-$30:28:24.5	&	12.50	&	1.01	&	B8V	\\	
Hip092159	&	18:46:59.99	&	$-$50:49:20.1	&	1.25	&	1.66	&	B8Vn...	\\	
Hip003082	&	00:39:09.57	&	$-$03:59:33.6	&	25.00	&	1.11	&	B9	\\	
Hip005387	&	01:08:55.86	&	09:43:49.8	&	1.25	&	1.53	&	B9	\\	
Hip012392	&	02:39:34.75	&	01:22:07.2	&	25.00	&	1.27	&	B9	\\	
Hip022914	&	04:55:52.17	&	13:37:51.3	&	1.25	&	1.31	&	B9	\\	
Hip045252	&	09:13:22.87	&	05:48:55.9	&	12.50	&	1.20	&	B9	\\	
Hip045252	&	09:13:22.87	&	05:48:55.9	&	25.00	&	1.20	&	B9	\\	
Hip086960	&	17:46:10.92	&	05:39:29.6	&	12.50	&	1.16	&	B9	\\	
Hip087807	&	17:56:15.03	&	$-$04:34:35.0	&	12.50	&	1.07	&	B9	\\	
Hip101607	&	20:35:29.64	&	06:56:49.5	&	12.50	&	1.79	&	B9	\\	
Hip108127	&	21:54:22.03	&	12:45:03.8	&	12.50	&	1.90	&	B9	\\	
Hip112376	&	22:45:38.00	&	03:37:52.2	&	25.00	&	1.16	&	B9	\\	
Hip003356	&	00:42:42.89	&	$-$38:27:48.5	&	12.50	&	1.09	&	B9.5V	\\	
Hip003356	&	00:42:42.89	&	$-$38:27:48.5	&	1.25	&	1.17	&	B9.5V	\\	
Hip031589	&	06:36:40.20	&	$-$58:24:32.0	&	1.25	&	1.28	&	B9.5V	\\	
Hip031589	&	06:36:40.20	&	$-$58:24:32.0	&	1.25	&	1.41	&	B9.5V	\\	
Hip072154	&	14:45:30.20	&	00:43:02.2	&	6.00	&	1.30	&	B9.5V	\\	
Hip104297	&	21:07:44.69	&	$-$17:27:21.3	&	12.50	&	1.07	&	B9.5V	\\	
Hip106243	&	21:31:09.63	&	12:08:14.9	&	12.50	&	1.25	&	B9.5V	\\	
Hip108875	&	22:03:19.03	&	11:23:11.6	&	12.50	&	1.54	&	B9.5V	\\	
Hip025555	&	05:27:45.61	&	15:52:26.6	&	12.50	&	1.90	&	B9.5Vn	\\	
Hip092963	&	18:56:22.66	&	$-$01:47:59.5	&	12.50	&	2.58	&	B9III	\\	
Hip050200	&	10:14:53.68	&	$-$42:22:59.0	&	12.50	&	1.07	&	B9IV	\\	
Hip001191	&	00:14:54.52	&	$-$09:34:10.5	&	12.50	&	1.08	&	B9V	\\	
Hip024505	&	05:15:24.37	&	$-$26:56:36.6	&	1.00	&	1.02	&	B9V	\\	
Hip030591	&	06:25:43.65	&	$-$48:10:36.8	&	12.50	&	1.81	&	B9V	\\	
Hip038280	&	07:50:34.25	&	$-$01:36:35.2	&	28.00	&	2.32	&	B9V	\\	
Hip041399	&	08:26:43.89	&	$-$18:50:12.2	&	32.00	&	1.78	&	B9V	\\	
Hip058859	&	12:04:16.70	&	$-$51:28:21.1	&	12.50	&	1.12	&	B9V	\\	
Hip062483	&	12:48:16.47	&	$-$44:43:07.6	&	12.50	&	1.08	&	B9V	\\	
Hip087054	&	17:47:18.94	&	$-$42:01:44.7	&	12.50	&	2.17	&	B9V	\\	
Hip087054	&	17:47:18.94	&	$-$42:01:44.7	&	50.00	&	2.19	&	B9V	\\	
Hip087054	&	17:47:18.94	&	$-$42:01:44.7	&	40.00	&	1.75	&	B9V	\\	
Hip091286	&	18:37:12.62	&	11:25:17.4	&	12.50	&	1.48	&	B9V	\\	
Hip092188	&	18:47:19.04	&	$-$37:32:55.2	&	1.25	&	1.47	&	B9V	\\	
Hip108058	&	21:53:36.04	&	$-$10:18:41.8	&	1.25	&	1.78	&	B9V	\\	
Hip110573	&	22:24:00.49	&	15:16:53.3	&	12.50	&	1.32	&	B9V	\\	
Hip112542	&	22:47:42.77	&	$-$14:03:23.1	&	12.50	&	1.02	&	B9V	\\	
Hip113821	&	23:02:56.79	&	$-$47:50:54.0	&	12.50	&	1.69	&	B9V	\\	
Hip113821	&	23:02:56.79	&	$-$47:50:54.0	&	12.50	&	1.69	&	B9V	\\	
Hip006751	&	01:26:53.55	&	03:32:08.3	&	1.25	&	1.15	&	B9V+...	\\	
Hip026796	&	05:41:26.93	&	$-$33:24:02.7	&	12.50	&	1.73	&	A0V	\\	
Hip017543	&	03:45:23.73	&	$-$71:39:29.3	&	1.25	&	1.48	&	Ap...	\\\hline

\end{longtable}
\end{center}

\section{The first-year XSL sample}

Our telluric-corrected, flux-calibrated, UVB- and VIS-arm-merged spectral library 
contains 246 spectra of 237 individual stars, which cover wavelengths 
from 3100--10185 \AA. Table~\ref{slib} lists the object name of XSL sample, 
the observation Modified Julian Date (MJD), the airmass of the observation, and 
the star's spectral type. An asterisk (``*'') in column ``Flux note'' means that the star
may have flux losses due to a missing wide-slit observation (see Section \ref{abs:flx}).
When the merged spectrum of a certain star is combined from different observations,
the MJD and airmass of the first observation are taken for this star.

\begin{figure*}
   \includegraphics[angle=0,scale=1.2]
   {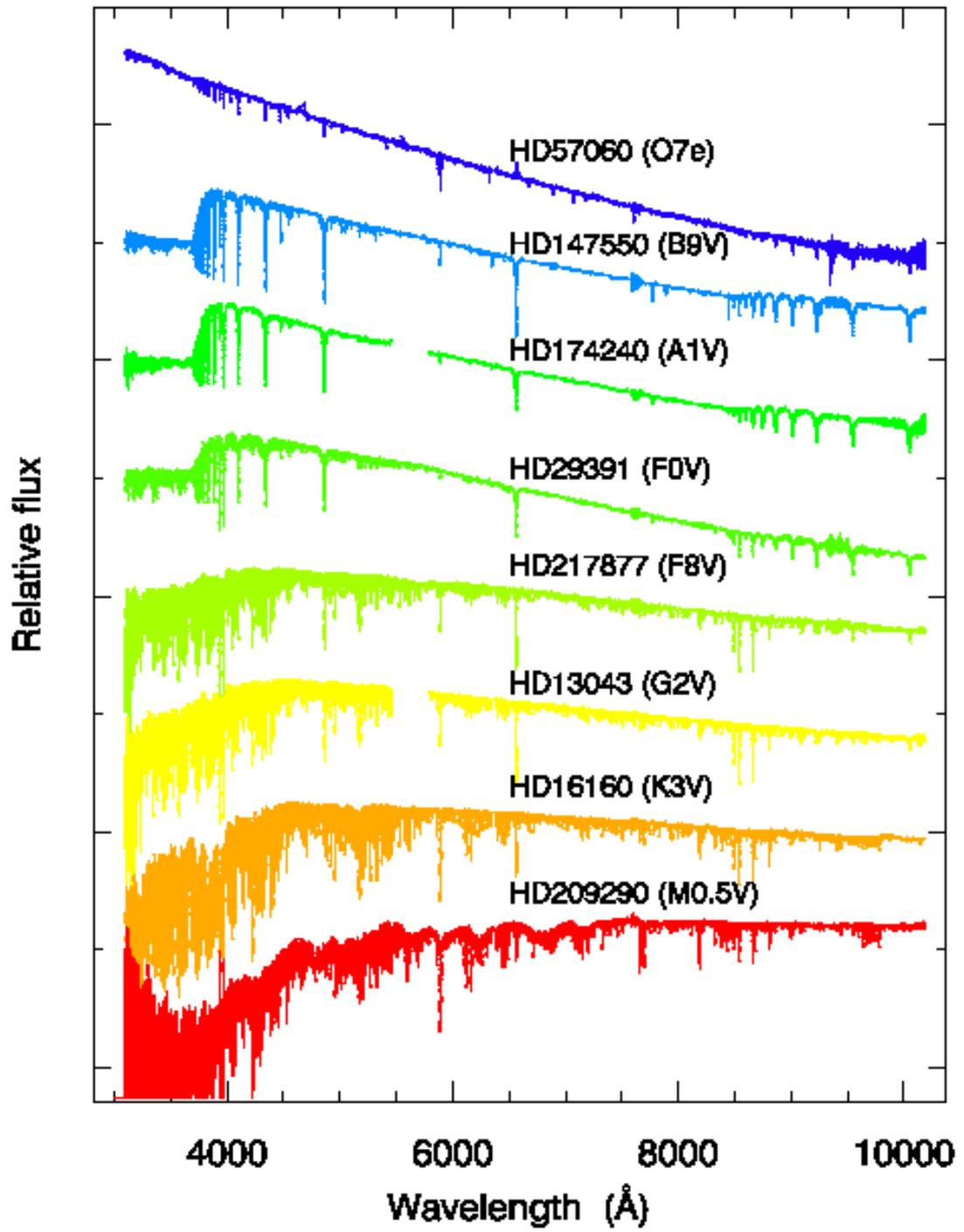}
   \caption{OBAFGKM sequence of the XSL sample in $\log (F_\lambda)$. Gaps 
   around 5700 \AA\ indicate strong dichroic features between the UVB and
   VIS arms. }
   \label{warmsample}
\end{figure*}

\begin{figure*}
   \includegraphics[angle=0,scale=1.2]
   {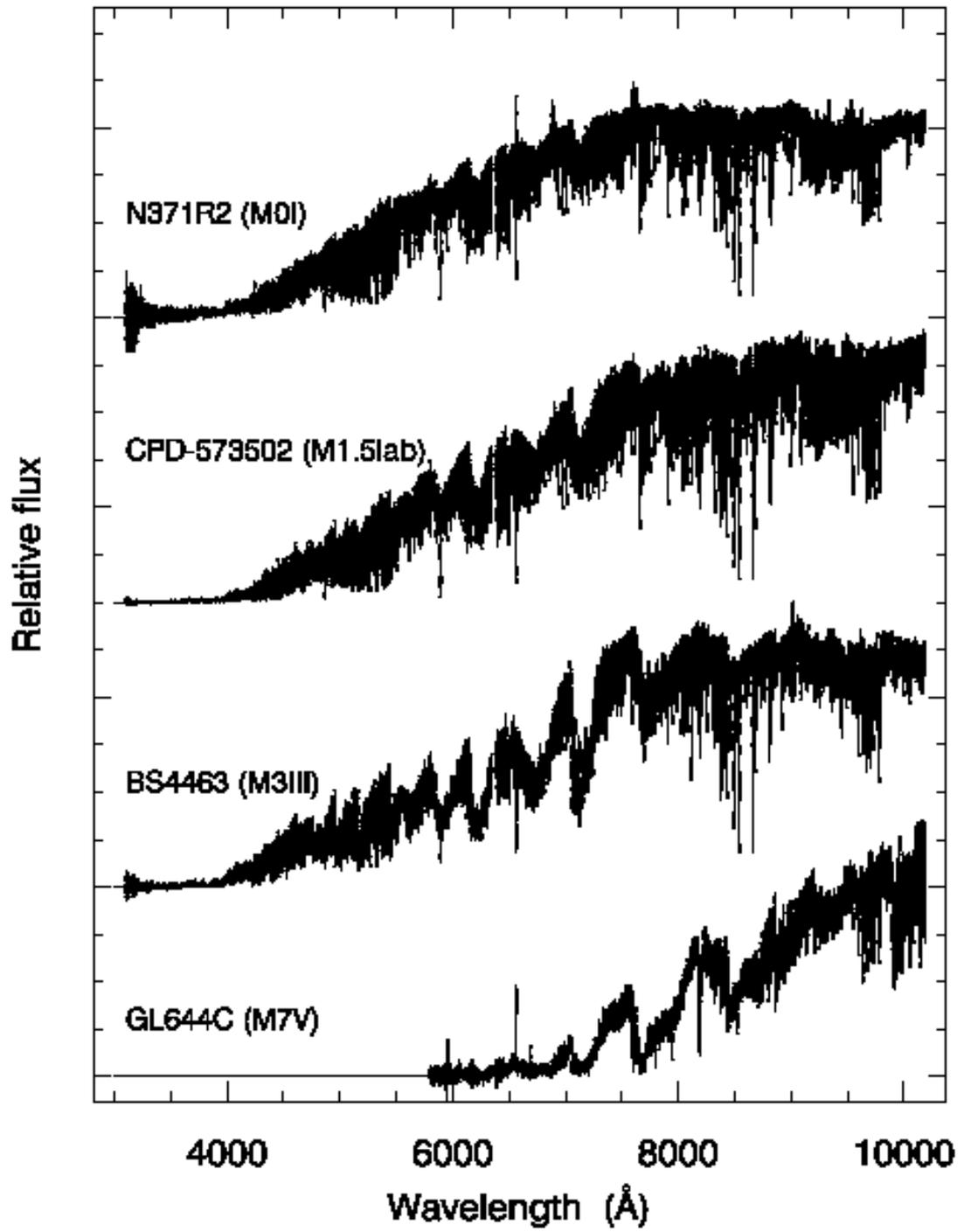}
   \caption{Sequence of M stars sorted by spectral type. Flat regions in 
   the star GL644C indicate spectrum in very low signal-to-noise ratio regions.}
   \label{m_sample}
\end{figure*}

\begin{figure*}
   \includegraphics[angle=0,scale=1.2]
   {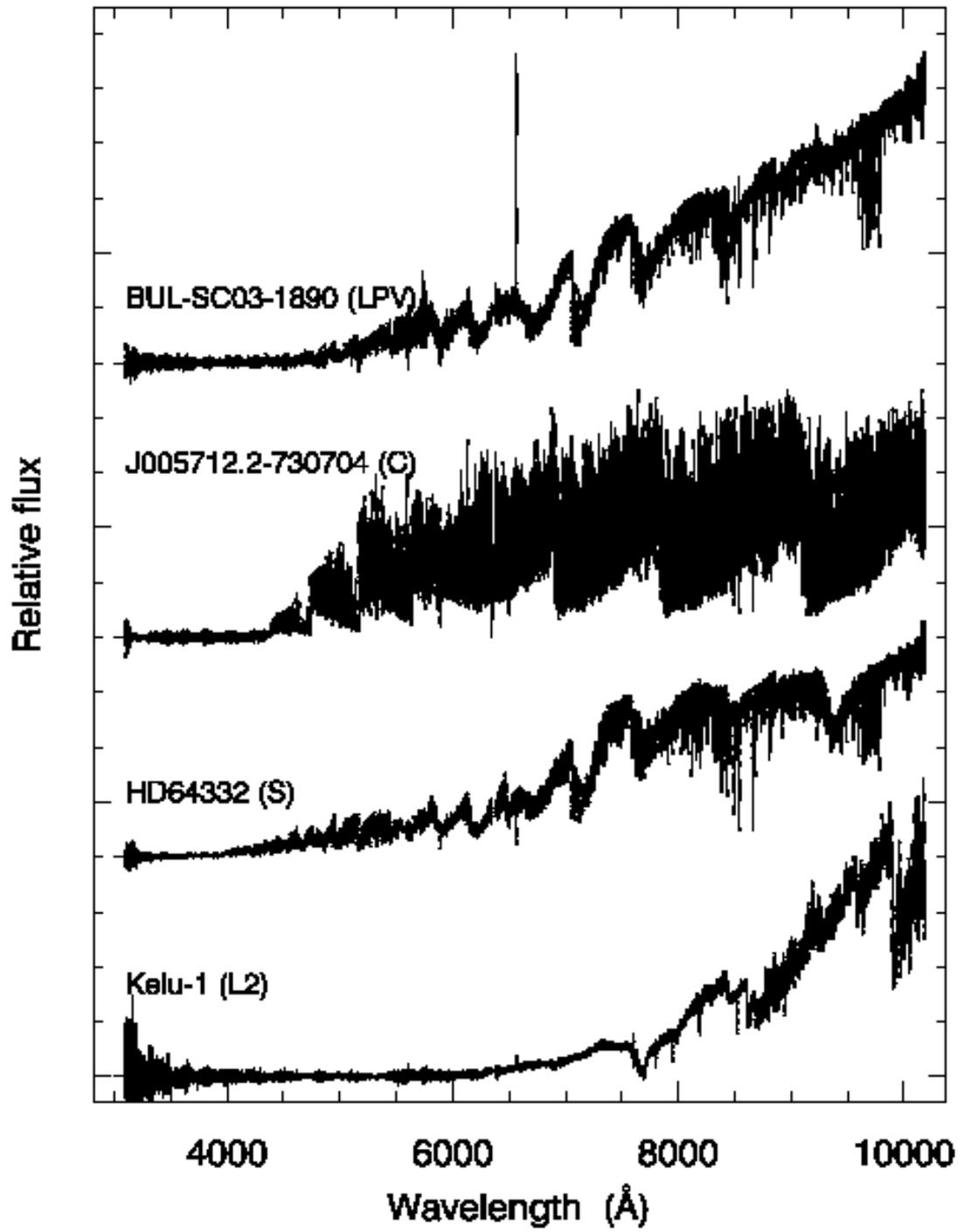}
   \caption{A sample of LPV, C, S, and L stars in the XSL.}
   \label{coolsample}
\end{figure*}

Typical XSL spectra in the OBAFGKM sequence are shown in 
Figure~\ref{warmsample}. We show an M-star sequence in Figure~\ref{m_sample} to 
illustrate our collection of these cool stars. Other cool stars, such as 
LPV, C, S, and L stars, are shown in Figure~\ref{coolsample}.

\clearpage
\onecolumn
\footnotesize   

\begin{center}
\begin{longtable}{l c r l l l c}

\caption{The XSL first-year sample.}\label{slib}\\

\hline
\hline

Star name	&RA (J2000)	&DEC (J2000)	&	MJD	&	Airmass	&	Sp. type	& Flux note	\\

\hline                                                  
\endfirsthead
{{\bfseries \tablename\ \thetable{} -- Continued.}} \\
\hline
\hline

Star name	&RA (J2000)	&DEC (J2000)	&	MJD	&	Airmass	&	Sp. type  & Flux note		\\

\hline                                                  
\endhead

HD57060	&	07:18:40.38	&	$-$24:33:31.3	&	55235.12	&	1.00	&	O7e...	& *  \\	
HD34816	&	05:19:34.52	&	$-$13:10:36.4	&	55467.38	&	1.03	&	B0.5IV	& *  \\	
HD96446	&	11:06:05.82	&	$-$59:56:59.6	&	55200.33	&	1.25	&	B2IIIp	& \\	
HD224926	&	00:01:49.45	&	$-$03:01:39.0	&	55113.00	&	1.80	&	B7III-IV	& \\	
HD34797	&	05:19:18.31	&	$-$18:30:34.4	&	55235.06	&	1.02	&	B8/B9IV:	& \\	
HD196426	&	20:37:18.38	&	$+$00:05:49.1	&	55408.26	&	1.18	&	B8IIIp	& \\	
HD358	&	00:08:23.26	&	$+$29:05:25.6	&	55113.05	&	2.29	&	B8IVmnp...	& *\\	
HD128801	&	14:38:48.09	&	$+$07:54:40.4	&	55438.00	&	1.86	&	B9	& \\	
HD163641	&	17:56:55.98	&	$+$06:29:15.8	&	55457.08	&	1.72	&	B9III	& \\	
HD175640	&	18:56:22.66	&	$-$01:47:59.5	&	55410.23	&	1.39	&	B9III	& \\	
HD27295	&	04:19:26.10	&	$+$21:08:32.3	&	55178.20	&	1.54	&	B9IV	& \\	
HD147550	&	16:22:38.90	&	$-$02:04:47.5	&	55438.05	&	1.37	&	B9V	& \\	
HD164257	&	18:00:07.32	&	$+$06:33:14.1	&	55412.09	&	1.17	&	A0	& \\	
HD194453	&	20:25:25.43	&	$+$06:38:30.5	&	55408.23	&	1.21	&	A0	& \\	
HD204041	&	21:25:51.58	&	$+$00:32:03.6	&	55407.28	&	1.16	&	A1IV	& \\	
HD72968	&	08:35:28.20	&	$-$07:58:56.3	&	55596.33	&	1.72	&	A1spe...	& \\	
HD174240	&	18:49:37.19	&	$+$00:50:10.3	&	55395.20	&	1.14	&	A1V	& \\	
HD2857	&	00:31:53.79	&	$-$05:15:42.9	&	55113.03	&	1.59	&	A2	& \\	
HD193281	&	20:20:27.88	&	$-$29:11:50.0	&	55408.20	&	1.01	&	A2III	& \\	
HD190073	&	20:03:02.51	&	$+$05:44:16.7	&	55409.24	&	1.28	&	A2IVe	& \\	
HD28978	&	04:34:08.27	&	$+$05:34:07.1	&	55162.12	&	1.37	&	A2Vs	& \\	
HD38237	&	05:44:30.59	&	$+$04:20:21.0	&	55461.29	&	1.81	&	A3	& \\	
HD163346	&	17:55:37.51	&	$+$02:04:29.8	&	55457.07	&	1.50	&	A3	& \\	
HD174966	&	18:53:07.83	&	$+$01:45:19.7	&	55395.21	&	1.16	&	A3	& \\	
HD18769	&	03:01:54.14	&	$+$26:27:44.5	&	55162.11	&	1.63	&	A3m	& \\	
HD19445	&	03:08:25.59	&	$+$26:19:51.4	&	55162.08	&	1.79	&	A4p	& \\	
HD172230	&	18:38:54.95	&	$+$06:16:14.8	&	55395.19	&	1.19	&	A5	& \\	
HD16031	&	02:34:11.05	&	$-$12:23:03.5	&	55162.04	&	1.14	&	F0V	& \\	
HD29391	&	04:37:36.13	&	$-$02:28:24.8	&	55178.22	&	1.15	&	F0V	& *\\	
HD284248	&	04:14:35.52	&	$+$22:21:04.3	&	55162.10	&	1.82	&	F2	& \\	
HD167278	&	18:14:33.65	&	$+$00:10:32.9	&	55457.09	&	1.54	&	F2	& \\	
HD205202	&	21:33:02.84	&	$+$30:21:35.1	&	55407.26	&	1.75	&	F2	& \\	
HD170756	&	18:30:16.24	&	$+$21:52:00.6	&	55466.02	&	1.63	&	F4Ibpv	& \\	
G029-023	&	23:19:40.45	&	$+$03:22:16.7	&	55410.33	&	1.14	&	F5	& \\	
HD61064	&	07:37:16.69	&	$-$04:06:39.5	&	55507.36	&	1.09	&	F6III	& \\	
HD160365	&	17:38:57.85	&	$+$13:19:45.3	&	55357.22	&	1.27	&	F6III	& \\	
HD196892	&	20:40:49.38	&	$-$18:47:33.3	&	55408.27	&	1.09	&	F6V	& \\	
HD4813	&	00:50:07.59	&	$-$10:38:39.6	&	55150.03	&	1.08	&	F7IV-V	& \\	
HD19019	&	03:03:50.82	&	$+$06:07:59.9	&	55162.09	&	1.23	&	F8	& \\	
G20-15	&	17:47:27.97	&	$-$08:46:47.7	&	55395.12	&	1.04	&	F8	& \\	
HD175805	&	18:56:58.07	&	$+$02:27:42.1	&	55410.24	&	1.53	&	F8	& \\	
G188-22	&	21:43:57.12	&	$+$27:23:24.0	&	55407.26	&	1.63	&	F8	& \\	
HD217877	&	23:03:57.27	&	$-$04:47:41.5	&	55409.29	&	1.06	&	F8V	& \\	
HD52298	&	06:57:45.44	&	$-$52:38:54.5	&	55461.32	&	1.60	&	F8VFe-1CH-0.5	& \\	
HD157089	&	17:21:07.06	&	$+$01:26:35.0	&	55357.21	&	1.11	&	F9V	& \\	
HD45282	&	06:26:40.77	&	$+$03:25:29.8	&	55240.16	&	1.38	&	G0	& \\	
HD161770	&	17:47:46.08	&	$-$09:36:18.5	&	55357.24	&	1.04	&	G0	& \\	
HD188262	&	19:53:45.93	&	$+$16:46:41.1	&	55409.21	&	1.41	&	G0	& \\	
HD200081	&	21:01:22.42	&	$-$02:30:50.4	&	55407.28	&	1.16	&	G0	& \\	
G187-40	&	21:21:57.75	&	$+$27:27:10.4	&	55408.25	&	1.64	&	G0	& \\	
HD52973	&	07:04:06.53	&	$+$20:34:13.1	&	55586.10	&	1.48	&	G0Ibv	& *\\	
HD216219	&	22:50:52.15	&	$+$18:00:07.6	&	55409.28	&	1.36	&	G0IIp	& \\	
HD39587	&	05:54:22.98	&	$+$20:16:34.2	&	55586.09	&	1.42	&	G0VCH+M	& *\\	
HD345957	&	20:10:48.16	&	$+$23:57:54.5	&	55409.23	&	1.62	&	G0Vw	& \\	
HD13043	&	02:07:34.27	&	$-$00:37:02.7	&	55408.31	&	1.52	&	G2V	& \\	
HD17072	&	02:40:40.09	&	$-$69:13:58.8	&	55119.14	&	1.54	&	G2w...	& \\	
HD163810	&	17:58:38.45	&	$-$13:05:49.6	&	55458.13	&	2.18	&	G3V	& \\	
G169-28	&	16:50:11.47	&	$+$22:18:50.0	&	55412.05	&	1.48	&	G3V	& \\	
HD8724	&	01:26:17.59	&	$+$17:07:35.1	&	55176.05	&	1.35	&	G5	& \\	
HD204155	&	21:26:42.91	&	$+$05:26:29.9	&	55407.29	&	1.24	&	G5	& \\	
HD179821	&	19:13:58.61	&	$+$00:07:31.9	&	55409.19	&	1.16	&	G5Ia	& \\	
HD193896	&	20:23:00.79	&	$-$09:39:17.0	&	55408.22	&	1.05	&	G5IIIa	& \\	
HD6229	&	01:03:36.46	&	$+$23:46:06.4	&	55110.15	&	1.58	&	G5IIIw	& \\	
HD18907	&	03:01:37.64	&	$-$28:05:29.6	&	55119.19	&	1.07	&	G5IV	& \\	
HD44007	&	06:18:48.53	&	$-$14:50:43.4	&	55240.15	&	1.19	&	G5IV:w...	& \\	
HD99648	&	11:27:56.24	&	$+$02:51:22.6	&	55589.26	&	1.21	&	G8Iab	& \\	
HD83212	&	09:36:19.95	&	$-$20:53:14.8	&	55637.16	&	1.04	&	G8IIIw...	& \\	
HD201626	&	21:09:59.27	&	$+$26:36:54.9	&	55408.24	&	1.61	&	G9p	& \\	
$\rm  [PHS2008]$ RGB512	&	05:10:57.03	&	$-$71:11:34.0	&	55408.39	&	1.83	&	K	& \\	
$\rm  [PHS2008]$ RGB522	&	05:11:22.58	&	$-$71:07:27.7	&	55439.35	&	1.64	&	K	& \\	
$\rm  [PHS2008]$ RGB533	&	05:13:12.67	&	$-$71:18:00.5	&	55429.35	&	1.77	&	K	& \\	
HD37828	&	05:40:54.65	&	$-$11:12:00.2	&	55224.08	&	1.03	&	K0	& \\	
HD173158	&	18:43:45.31	&	$+$05:44:14.6	&	55395.19	&	1.19	&	K0	& \\	
HD93813	&	10:49:37.49	&	$-$16:11:37.1	&	55636.02	&	1.53	&	K0/K1III	& *\\	
HD179870	&	19:13:53.59	&	$+$09:01:59.6	&	55409.18	&	1.25	&	K0IIb	& \\	
HD82734	&	09:33:12.46	&	$-$21:06:56.6	&	55591.36	&	1.33	&	K0III	& \\	
HD170820	&	18:32:13.11	&	$-$19:07:26.3	&	55410.22	&	1.25	&	K0III	& \\	
HD33299	&	05:10:34.98	&	$+$30:47:51.1	&	55235.01	&	1.78	&	K1Ib	& \\	
HD1638	&	00:20:28.31	&	$-$30:33:19.9	&	55113.01	&	1.48	&	K1III	& \\	
HD165438	&	18:06:15.20	&	$-$04:45:04.5	&	55413.09	&	1.07	&	K1IV	& \\	
HD190404	&	20:03:52.13	&	$+$23:20:26.5	&	55409.22	&	1.59	&	K1V	& \\	
HD25329	&	04:03:15.00	&	$+$35:16:23.8	&	55162.17	&	2.00	&	K1V...	& \\	
HD50877	&	06:54:07.95	&	$-$24:11:03.2	&	55474.37	&	1.09	&	K2.5Iab	& *\\	
HD63302	&	07:47:38.53	&	$-$15:59:26.5	&	55636.14	&	1.20	&	K2Iab	& \\	
HD19787	&	03:11:37.76	&	$+$19:43:36.0	&	55180.07	&	1.42	&	K2III	& *\\	
HD37763	&	05:31:53.01	&	$-$76:20:27.5	&	55465.18	&	2.32	&	K2III	& \\	
HD175545	&	18:55:51.45	&	$-$00:44:22.1	&	55395.22	&	1.16	&	K2III	& \\	
HD31421	&	04:56:22.28	&	$+$13:30:52.1	&	55202.06	&	1.35	&	K2IIIb	& *\\	
BS4432	&	11:30:18.89	&	$-$03:00:12.6	&	55589.27	&	1.13	&	K3.5III	& \\	
HD81797	&	09:27:35.24	&	$-$08:39:31.0	&	55586.12	&	1.44	&	K3II-III	& *\\	
HD65354	&	07:57:03.02	&	$-$34:22:42.4	&	55475.36	&	1.29	&	K3III	& *\\	
HD232078	&	19:38:12.07	&	$+$16:48:25.6	&	55409.21	&	1.42	&	K3IIp	& \\	
HD165195	&	18:04:40.07	&	$+$03:46:44.7	&	55413.08	&	1.14	&	K3p	& \\	
HD16160	&	02:36:04.89	&	$+$06:53:12.7	&	55162.06	&	1.28	&	K3V	& \\	
HD160346	&	17:39:16.92	&	$+$03:33:18.9	&	55357.23	&	1.14	&	K3V	& \\	
HD52005	&	07:00:15.82	&	$+$16:04:44.4	&	55508.36	&	1.32	&	K4Iab	& \\	
HD74088	&	08:38:27.86	&	$-$62:50:35.7	&	55475.37	&	1.58	&	K4III	& *\\	
BS4104	&	10:27:09.10	&	$-$31:04:04.0	&	55636.21	&	1.07	&	K4III	& \\	
BS3923	&	09:54:52.21	&	$-$19:00:33.7	&	55636.20	&	1.11	&	K5III	& \\	
HD114960	&	13:13:57.57	&	$+$01:27:23.2	&	55405.03	&	1.64	&	K5III	& *\\	
$\rm  [M2002]$ SMC052334	&	01:01:54.17	&	$-$71:52:18.7	&	55110.15	&	1.50	&	K7	& \\	
HD79349	&	09:11:43.04	&	$-$48:46:23.5	&	55590.36	&	1.43	&	K7IV	& \\	
 ISO-MCMS J004950.3-731116	&	00:49:50.14	&	$-$73:11:16.6	&	55110.11	&	1.58	&	M	& \\	
 ISO-MCMS J005059.4-731914	&	00:50:59.32	&	$-$73:19:13.4	&	55110.19	&	1.51	&	M	& \\	
 ISO-MCMS J005101.9-731607	&	00:51:01.94	&	$-$73:16:06.8	&	55110.21	&	1.51	&	M	& \\	
 ISO-MCMS J005304.7-730409	&	00:53:04.87	&	$-$73:04:09.5	&	55115.31	&	1.71	&	M	& \\	
 ISO-MCMS J005314.8-730601	&	00:53:14.65	&	$-$73:06:01.0	&	55115.33	&	1.81	&	M	& \\	
 ISO-MCMS J005332.4-730501	&	00:53:32.75	&	$-$73:04:58.2	&	55115.35	&	1.91	&	M	& \\	
 ISO-MCMS J005622.2-730334	&	00:56:22.80	&	$-$73:03:33.9	&	55114.10	&	1.59	&	M	& \\	
 ISO-MCMS J005714.4-730121	&	00:57:14.48	&	$-$73:01:21.3	&	55112.09	&	1.62	&	M	& \\	
 SHV0503595-691915	&	05:03:42.13	&	$-$69:15:12.2	&	55407.41	&	1.69	&	M	& \\	
 SHV0503595-691915	&	05:03:42.13	&	$-$69:15:12.2	&	55142.21	&	1.47	&	M	& \\	
 SHV0506368-681557	&	05:06:27.68	&	$-$68:12:03.7	&	55407.43	&	1.61	&	M	& \\	
 SHV0506368-681557	&	05:06:27.68	&	$-$68:12:03.7	&	55119.27	&	1.45	&	M	& \\	
 SHV0515461-691822	&	05:15:26.90	&	$-$69:15:07.5	&	55142.26	&	1.41	&	M	& \\	
 SHV0523357-692038	&	05:23:14.70	&	$-$69:17:57.8	&	55142.35	&	1.45	&	M	& \\	
 SHV0525543-692639	&	05:25:32.18	&	$-$69:24:09.2	&	55226.16	&	1.53	&	M	& \\	
 SHV0526364-693639	&	05:26:12.44	&	$-$69:34:13.0	&	55226.17	&	1.57	&	M	& \\	
 SHV0533015-720047	&	05:32:11.05	&	$-$71:58:48.8	&	55226.21	&	1.75	&	M	& \\	
 TLE Sgr I 55	&	17:59:29.65	&	$-$29:03:20.6	&	55412.06	&	1.01	&	M	& \\	
 TLE Sgr I 11	&	17:59:56.10	&	$-$28:53:28.0	&	55412.08	&	1.00	&	M	& *\\	
 TLE Sgr I 117	&	18:00:45.70	&	$-$29:11:05.0	&	55412.10	&	1.01	&	M	& \\	
 Cl* NGC 6522 Arp 4329	&	18:03:28.43	&	$-$29:58:41.7	&	55397.25	&	1.31	&	M	& \\	
2MASS J18033716-2954227 	&	18:03:37.16	&	$-$29:54:22.7	&	55393.14	&	1.01	&	M	& \\	
2MASS J18351799-3428093	&	18:35:17.99	&	$-$34:28:09.3	&	55356.40	&	1.42	&	M	& \\	
2MASS J18352206-3429112	&	18:35:22.06	&	$-$34:29:11.2	&	55356.17	&	1.12	&	M	& \\	
2MASS J18352834-3444085	&	18:35:28.34	&	$-$34:44:08.5	&	55393.05	&	1.20	&	M	& \\	
2MASS J18355679-3434481	&	18:35:56.79	&	$-$34:34:48.1	&	55355.41	&	1.45	&	M	& \\	
HD209290	&	22:02:10.27	&	$+$01:24:00.8	&	55408.28	&	1.13	&	M0.5V	& \\	
U Crt	&	11:12:45.30	&	$-$07:17:54.5	&	55351.96	&	1.05	&	M0e	& \\	
$\rm  [M2002]$ SMC046662	&	00:59:35.00	&	$-$72:04:06.6	&	55110.14	&	1.52	&	M0I	& \\	
 Cl* NGC 371 FTW R20	&	01:00:41.51	&	$-$72:10:37.1	&	55119.13	&	1.50	&	M0I	& \\	
CD-603636	&	11:36:34.84	&	$-$61:36:35.2	&	55586.13	&	2.04	&	M0Iab	& \\	
$\rm  [M2002]$ LMC170452	&	05:38:16.01	&	$-$69:10:11.2	&	55392.41	&	2.08	&	M1.5I	& \\	
HD35601	&	05:27:10.22	&	$+$29:55:15.8	&	55599.07	&	1.75	&	M1.5Ia0-Ia...	& \\	
CPD-573502	&	10:35:43.71	&	$-$58:14:42.3	&	55636.22	&	1.26	&	M1.5Iab	& *\\	
CD-603621	&	11:35:44.95	&	$-$61:34:41.0	&	55231.37	&	1.31	&	M1.5Iab:	& \\	
$\rm  [M2002]$ LMC162635	&	05:35:24.52	&	$-$69:04:03.4	&	55439.39	&	1.51	&	M1I	& \\	
HD98817	&	11:21:38.96	&	$-$60:59:28.2	&	55637.06	&	1.49	&	M1Iab	& \\	
BS4517	&	11:45:51.56	&	$+$06:31:45.7	&	55636.04	&	2.51	&	M1III	& \\	
$\rm  [M2002]$ LMC148035	&	05:30:35.53	&	$-$68:59:23.4	&	55240.04	&	1.40	&	M2.5I	& \\	
CD-314916	&	07:41:02.63	&	$-$31:40:59.1	&	55177.32	&	1.02	&	M2.5Iab:	& \\	
CD-314916	&	07:41:02.63	&	$-$31:40:59.1	&	55297.04	&	1.10	&	M2.5Iab:	& \\	
CM Car	&	09:47:55.57	&	$-$67:27:07.2	&	55393.01	&	2.26	&	M2e	& \\	
$\rm  [M2002]$ SMC083593	&	01:30:33.96	&	$-$73:18:41.7	&	55112.10	&	1.67	&	M2I	& \\	
$\rm  [M2002]$ LMC158646	&	05:33:52.17	&	$-$69:11:13.5	&	55429.42	&	1.52	&	M2I	& \\	
 HV2360	&	05:12:46.37	&	$-$67:19:37.9	&	55119.29	&	1.40	&	M2Ia	& \\	
 HD39801	&	05:55:10.31	&	$+$07:24:25.4	&	55116.34	&	1.22	&	M2Iab:	& *\\	
$\rm  [M2002]$ LMC150040	&	05:31:09.28	&	$-$67:25:54.9	&	55240.05	&	1.37	&	M3-M4	& \\	
$\rm  [M2002]$ LMC168757	&	05:37:36.82	&	$-$69:29:23.4	&	55439.40	&	1.51	&	M3?m4I...	& \\	
$\rm  [M2002]$ SMC055188	&	01:03:02.45	&	$-$72:01:53.1	&	55110.13	&	1.54	&	M3.5I	& \\	
HD101712	&	11:41:49.41	&	$-$63:24:52.4	&	55314.21	&	1.52	&	M3Iab	& \\	
HD101712	&	11:41:49.41	&	$-$63:24:52.4	&	55393.03	&	1.70	&	M3Iab	& \\	
BS4463	&	11:35:13.28	&	$-$47:22:21.3	&	55621.12	&	1.34	&	M3III	& \\	
HV2255	&	04:57:43.32	&	$-$70:08:50.3	&	55119.20	&	1.66	&	M4	& \\	
HV2255	&	04:57:43.32	&	$-$70:08:50.3	&	55408.37	&	1.86	&	M4	& \\	
 SHV0549503-704331	&	05:49:13.36	&	$-$70:42:40.7	&	55116.36	&	1.45	&	M4	& \\	
B86 133	&	18:03:45.47	&	$-$30:03:00.7	&	55397.28	&	1.56	&	M4	& \\	
R Cha	&	08:21:46.42	&	$-$76:21:18.5	&	55636.16	&	1.73	&	M4.5e	& *\\	
EV Car	&	10:20:21.61	&	$-$60:27:15.6	&	55636.19	&	1.27	&	M4.5Ia	& *\\	
$\rm  [M2002]$ LMC143035	&	05:29:03.48	&	$-$69:06:46.2	&	55240.03	&	1.40	&	M4I	& \\	
2MASS J18025277-2954335	&	18:02:52.78	&	$-$29:54:33.6	&	55439.21	&	2.30	&	M5	& \\	
2MASS J18040638-3010497	&	18:04:06.39	&	$-$30:10:49.7	&	55393.17	&	1.03	&	M5	& \\	
HV2446	&	05:20:01.57	&	$-$67:34:42.2	&	55142.31	&	1.37	&	M5e	& \\	
 SHV0452361-683928	&	04:52:26.48	&	$-$68:34:37.5	&	55119.23	&	1.53	&	M6	& \\	
 SHV0452361-683928	&	04:52:26.48	&	$-$68:34:37.5	&	55407.39	&	1.75	&	M6	& \\	
 SHV0501215-680112	&	05:01:15.35	&	$-$67:56:58.7	&	55407.39	&	1.74	&	M6	& \\	
 SHV0501215-680112	&	05:01:15.35	&	$-$67:56:58.7	&	55119.24	&	1.51	&	M6	& \\	
 SHV0510004-692755	&	05:09:40.52	&	$-$69:24:16.6	&	55423.39	&	1.60	&	M6	& \\	
 SHV0510004-692755	&	05:09:40.52	&	$-$69:24:16.6	&	55119.28	&	1.47	&	M6	& \\	
 SHV0518570-692207	&	05:18:36.62	&	$-$69:19:05.2	&	55142.29	&	1.41	&	M6	& \\	
 SHV0543367-695800	&	05:43:08.02	&	$-$69:56:45.7	&	55237.06	&	1.43	&	M6	& \\	
 OGLEII DIA BUL-SC01 1821	&	18:02:12.76	&	$-$30:02:12.3	&	55439.19	&	1.90	&	M6	& \\	
2MASS J18032525-2959483	&	18:03:25.26	&	$-$29:59:48.4	&	55397.24	&	1.25	&	M6	& \\	
2MASS J18042244-3000534	&	18:04:22.45	&	$-$30:00:53.5	&	55393.18	&	1.04	&	M6	& \\	
2MASS J18024611-3004509	&	18:02:46.11	&	$-$30:04:51.0	&	55439.20	&	2.11	&	M6.5	& \\	
BMB 162	&	18:03:43.00	&	$-$30:07:51.8	&	55397.29	&	1.62	&	M6.5	& \\	
BMB 300	&	18:04:27.20	&	$-$30:02:56.6	&	55463.16	&	2.59	&	M6.5	& \\	
GL866	&	22:38:33.73	&	$-$15:17:57.3	&	55408.29	&	1.02	&	M6(M7e?)	& \\	
 SHV0606101-724012	&	06:05:09.76	&	$-$72:40:35.2	&	55240.14	&	1.59	&	M7	& \\	
2MASS J18024572-3001120	&	18:02:45.73	&	$-$30:01:12.0	&	55439.20	&	1.99	&	M7	& \\	
GL644C	&	16:55:35.29	&	$-$08:23:40.1	&	55395.11	&	1.06	&	M7V	& \\	
GL752B	&	19:16:55.26	&	$+$05:10:08.1	&	55409.20	&	1.24	&	M8V	& *\\	
WX Psc	&	01:06:25.98	&	$+$12:35:53.1	&	55110.16	&	1.30	&	M9	& *\\	
2MASS J18042265-2954518	&	18:04:22.66	&	$-$29:54:51.8	&	55393.19	&	1.06	&	M9	& \\	
LHS2065	&	08:53:36.20	&	$-$03:29:32.1	&	55297.06	&	1.12	&	M9.0V	& \\	
LHS2065	&	08:53:36.20	&	$-$03:29:32.1	&	55297.07	&	1.14	&	M9.0V	& \\	
IRAS15060+0947	&	15:08:25.77	&	$+$09:36:18.2	&	55414.02	&	1.33	&	M9III	& \\	
IRAS14303-1042	&	14:32:59.89	&	$-$10:56:03.6	&	55437.99	&	1.46	&	Me	& \\	
 SHV0522380-691255	&	05:22:18.28	&	$-$69:10:10.4	&	55142.33	&	1.42	&	Ms	& \\	
 SHV0529467-693825	&	05:29:22.66	&	$-$69:36:13.0	&	55226.20	&	1.66	&	Ms	& \\	
HD64332	&	07:53:05.27	&	$-$11:37:29.4	&	55637.15	&	1.29	&	S	& \\	
 ISO-MCMS J004900.4-732224	&	00:49:00.33	&	$-$73:22:23.8	&	55110.08	&	1.67	&	C	& \\	
 ISO-MCMS J004932.4-731753	&	00:49:32.62	&	$-$73:17:52.3	&	55110.09	&	1.63	&	C	& \\	
 ISO-MCMS J005307.8-730747	&	00:53:07.65	&	$-$73:07:47.8	&	55116.12	&	1.54	&	C	& \\	
 ISO-MCMS J005422.8-730105	&	00:54:22.66	&	$-$73:01:05.7	&	55119.08	&	1.61	&	C	& \\	
 ISO-MCMS J005531.0-731018	&	00:55:30.91	&	$-$73:10:18.6	&	55119.09	&	1.58	&	C	& \\	
 ISO-MCMS J005638.9-730452	&	00:56:39.06	&	$-$73:04:53.0	&	55114.12	&	1.55	&	C	& \\	
 ISO-MCMS J005644.8-731436	&	00:56:44.78	&	$-$73:14:34.7	&	55119.11	&	1.55	&	C	& \\	
 ISO-MCMS J005700.7-730751	&	00:57:00.70	&	$-$73:07:50.6	&	55111.07	&	1.69	&	C	& \\	
 ISO-MCMS J005712.2-730704	&	00:57:12.15	&	$-$73:07:04.6	&	55111.08	&	1.66	&	C	& \\	
 ISO-MCMS J005716.5-731052	&	00:57:16.48	&	$-$73:10:52.8	&	55111.11	&	1.58	&	C	& \\	
 ISO-MCMS J010031.5-730724	&	01:00:31.51	&	$-$73:07:23.7	&	55111.12	&	1.56	&	C	& \\	
 SHV0504353-712622	&	05:03:55.97	&	$-$71:22:22.1	&	55119.26	&	1.55	&	C	& \\	
 SHV0518222-750327	&	05:16:49.73	&	$-$75:00:22.7	&	55142.28	&	1.57	&	C	& \\	
 SHV0520505-705019	&	05:20:15.02	&	$-$70:47:26.2	&	55142.32	&	1.45	&	C	& \\	
 SHV0525478-690944	&	05:25:28.22	&	$-$69:07:13.3	&	55142.36	&	1.46	&	C	& \\	
 SHV0528537-695119	&	05:28:27.73	&	$-$69:49:01.8	&	55226.19	&	1.63	&	C	& *\\	
 SHV0536139-701604	&	05:35:42.81	&	$-$70:14:16.3	&	55226.23	&	1.78	&	C	& \\	
HD70138	&	08:19:43.09	&	$-$18:15:52.8	&	55596.30	&	1.37	&	C	& \\	
HD76221	&	08:55:22.88	&	$+$17:13:52.6	&	55558.36	&	1.46	&	C	& *\\	
Y Hya	&	09:51:03.72	&	$-$23:01:02.3	&	55636.18	&	1.06	&	C	& *\\	
T Cae	&	04:47:18.92	&	$-$36:12:33.6	&	55142.19	&	1.11	&	CII	& \\	
RU Pup	&	08:07:29.83	&	$-$22:54:45.3	&	55636.15	&	1.18	&	CNv...	& *\\	
 OGLEII DIA BUL-SC26 0532	&	17:47:42.33	&	$-$35:19:57.3	&	55375.34	&	1.68	&	LPV	& \\	
 OGLEII DIA BUL-SC15 1379	&	17:48:25.03	&	$-$23:15:34.4	&	55395.13	&	1.00	&	LPV	& \\	
 OGLEII DIA BUL-SC15 2106	&	17:48:30.35	&	$-$23:05:23.5	&	55395.14	&	1.01	&	LPV	& \\	
 OGLEII DIA BUL-SC41 3443	&	17:52:13.90	&	$-$32:46:34.2	&	55375.36	&	1.85	&	LPV	& \\	
 OGLEII DIA BUL-SC41 3304	&	17:52:25.93	&	$-$32:49:08.1	&	55375.37	&	2.05	&	LPV	& \\	
 OGLEII DIA BUL-SC03 3941	&	17:53:32.56	&	$-$29:57:13.2	&	55395.15	&	1.01	&	LPV	& \\	
 OGLEII DIA BUL-SC24 0989	&	17:53:39.35	&	$-$33:06:03.0	&	55375.38	&	2.39	&	LPV	& \\	
 OGLEII DIA BUL-SC03 1890	&	17:53:45.24	&	$-$30:11:51.6	&	55438.06	&	1.05	&	LPV	& \\	
2MASS J17535707-2931427 	&	17:53:57.07	&	$-$29:31:42.8	&	55438.07	&	1.08	&	LPV	& \\	
 OGLEII DIA BUL-SC04 9008	&	17:54:27.94	&	$-$29:15:46.0	&	55439.13	&	1.34	&	LPV	& \\	
 OGLEII DIA BUL-SC04 4628	&	17:54:33.41	&	$-$29:44:03.9	&	55439.12	&	1.30	&	LPV	& \\	
 OGLEII DIA BUL-SC04 1709	&	17:54:49.07	&	$-$30:00:31.0	&	55439.14	&	1.37	&	LPV	& \\	
 OGLEII DIA BUL-SC22 1319	&	17:57:10.19	&	$-$31:02:16.8	&	55439.14	&	1.40	&	LPV	& \\	
 OGLEII DIA BUL-SC30 0707	&	18:01:46.14	&	$-$29:11:18.3	&	55412.10	&	1.01	&	LPV	& \\	
 OGLEII DIA BUL-SC01 0235	&	18:02:12.64	&	$-$30:22:35.0	&	55412.11	&	1.02	&	LPV	& \\	
 OGLEII DIA BUL-SC33 4149	&	18:05:39.41	&	$-$28:30:28.7	&	55408.12	&	1.01	&	LPV	& \\	
 OGLEII DIA BUL-SC36 2158	&	18:05:52.67	&	$-$28:11:36.8	&	55408.13	&	1.02	&	LPV	& \\	
 OGLEII DIA BUL-SC33 0357	&	18:05:56.37	&	$-$29:16:37.7	&	55408.14	&	1.04	&	LPV	& \\	
 OGLEII DIA BUL-SC19 2302	&	18:07:35.22	&	$-$27:15:34.3	&	55408.15	&	1.05	&	LPV	& \\	
 OGLEII DIA BUL-SC06 2525	&	18:07:51.30	&	$-$31:49:01.8	&	55408.16	&	1.07	&	LPV	& \\	
 OGLEII DIA BUL-SC19 2332	&	18:07:57.74	&	$-$27:15:51.8	&	55408.16	&	1.08	&	LPV	& \\	
2MASS J18080765-3142020	&	18:08:07.65	&	$-$31:42:02.0	&	55408.17	&	1.11	&	LPV	& \\	
 OGLEII DIA BUL-SC19 2948	&	18:08:16.20	&	$-$27:08:52.6	&	55408.18	&	1.13	&	LPV	& \\	
2MASS J18083220-3201531	&	18:08:32.20	&	$-$32:01:53.1	&	55408.19	&	1.17	&	LPV	& \\	
 OGLEII DIA BUL-SC16 1428	&	18:10:29.54	&	$-$26:29:00.1	&	55397.17	&	1.03	&	LPV	& \\	
 OGLEII DIA BUL-SC17 1595	&	18:11:33.55	&	$-$26:22:07.9	&	55397.17	&	1.02	&	LPV	& \\	
 OGLEII DIA BUL-SC13 0324	&	18:16:47.52	&	$-$24:20:32.4	&	55465.15	&	2.51	&	LPV	& \\	
 OGLEII DIA BUL-SC13 1542	&	18:17:06.47	&	$-$23:59:35.9	&	55466.14	&	2.23	&	LPV	& \\	
 OGLEII DIA BUL-SC08 1687	&	18:23:20.26	&	$-$21:35:54.0	&	55410.21	&	1.25	&	LPV	& \\	
 Kelu-1	&	13:05:40.20	&	$-$25:41:06.0	&	55352.15	&	1.26	&	L2	& \\	
 2MASS J15065441+1321060	&	15:06:54.41	&	$+$13:21:06.1	&	55438.02	&	1.93	&	L3	& *\\	
 2MASS J22244381-0158521	&	22:24:43.82	&	$-$01:58:52.1	&	55409.26	&	1.09	&	L4.5	& \\\hline

\end{longtable}
\end{center}
\normalsize



\end{document}